\documentclass[a4paper]{ws-article}
\usepackage{rotating}
\usepackage{epsfig}
\usepackage{amsfonts}
\usepackage{amssymb}
\newcommand{\dzero}{\mbox{D\O}}
\newcommand{\ppbar}{\mbox{$p\overline{p}$}}

\newcommand{\PT}{\mbox{$P_T$}}
\newcommand{\pt}{\mbox{$P_T$}}
\newcommand{\ET}{\mbox{$E_T$}}
\newcommand{\et}{\mbox{$E_T$}}

\newcommand{\met}{\mbox{\ensuremath{\not \!\! E_T}}}
\newcommand{\pet}{\mbox{\ensuremath{\not \!\! P_T}}}

\newcommand{\abseta}{\mbox{$\mid \eta \mid$}}

\newcommand{\gevp}{GeV/$c$}

\newcommand{\Gcs}{${\rm{GeV/}c^2}$}
\newcommand{\Tcs}{${\rm{TeV/}c^2}$}
\newcommand{\Gc}{${\rm{GeV/}c}$}
\newcommand{\Gcsit}{\ensuremath{\it{GeV/c}^2}}

\newcommand{\epemit}{\mbox{$e^+e^-$}}

\newcommand{\mh}{\mbox{$m_H$}}
\newcommand{\mA}{\mbox{$m_A$}}
\newcommand{\hplus}{\mbox{$H^{\pm}$}}
\newcommand{\mhplus}{\mbox{$m_{H^{\pm}}$}}
\newcommand{\hgg}{\mbox{$H \to  \gamma \gamma$}}
\newcommand{\hbb}{\mbox{$H \to  b \bar{b}$}}

\newcommand{\hzzsfourl}{\mbox{$H \to Z Z^{(*)} \rightarrow 4\ell$}}
\newcommand{\hzzfourl}{\mbox{$H \to Z Z \rightarrow 4\ell$}}
\newcommand{\hww}{\mbox{$H \to  W W $}}
\newcommand{\hwws}{\mbox{$H \to  W W^{(*)}$}}
\newcommand{\hwwsll}{\mbox{$H \to  W W^{(*)} \rightarrow \ell \nu \ell \nu $}}

\newcommand{\gamgam}{\mbox{$\gamma \gamma$}}

\newcommand{\ttbar}{\mbox{$t \overline{t} $}}
\newcommand{\bbbar}{\mbox{$b \overline{b} $}}
\newcommand{\ccbar}{\mbox{$c \overline{c} $}}
\newcommand{\qqbar}{\mbox{$q \overline{q} $}}

\newcommand{\fbs}{\mbox{$\rm{fb}^{-1}$}}

\newcommand{\fbsit}{\ensuremath{{\it fb}^{-1}}}

\newcommand{\lhigh}{\mbox{${\cal L} = 10^{34} \ \rm{cm}^{-2} \rm{s}^{-1}$}}

\newcommand{\chione}{\mbox{$\chi^0_1$}}
\newcommand{\chitwo}{\mbox{$\chi^0_2$}}
\newcommand{\chithree}{\mbox{$\chi^0_3$}}
\newcommand{\chifour}{\mbox{$\chi^0_4$}}

\newcommand{\slep}{\mbox{$\tilde{\ell}$}}

\newcommand{\sq}{\mbox{$\tilde{q}$}}
\newcommand{\sgl}{\mbox{$\tilde{g}$}}
\newcommand{\st}{\mbox{$\tilde{t}$}}
\newcommand{\sbot}{\mbox{$\tilde{b}$}}
\newcommand{\mzero}{\mbox{$m_0$}}
\newcommand{\mhalf}{\mbox{$m_{1/2}$}}
\newcommand{\tanb}{\mbox{$\tan \beta$}}
\newcommand{\matb}{\mbox{$(m_A,\tan \beta )$}}
\newcommand{\matbit}{\ensuremath{{\it (m}_A, {\it tan} \beta {\it )}}}
\newcommand{\tanbit}{\ensuremath{{\it tan} \beta}}

\newcommand{\tb}{\ensuremath{\tan\!\beta}}
\newcommand{\bb}{\ensuremath{b\bar{b}}}
\newcommand{\tautau}{\ensuremath{\tau^+\tau^-}}

\newcommand{\fbinv}{\ensuremath{\mathrm{fb^{-1}}}}
\newcommand{\nn}{\ensuremath{\nu\bar{\nu}}}
\newcommand{\hs}{\ensuremath{\mathit{\Phi}}}

\begin{document}

\markboth{V.~B\"uscher and K.~Jakobs}
{Higgs Boson Searches at Hadron Colliders}

%
%

\begin{center}
{\normalsize International Journal of Modern Physics Letters A Vol. 20 (2005)}\\[1cm]
{\Large Higgs Boson Searches at Hadron Colliders}\\[0.5cm]
{\footnotesize Volker B\"uscher and \ Karl Jakobs\\[0.3cm]
Physikalisches Institut, Universit\"at Freiburg \\
Hermann-Herder-Str. 3, 79104 Freiburg, Germany \\[0.4cm]
buescher@fnal.gov \\
karl.jakobs@uni-freiburg.de}\\[0.5cm]
{\large February 18, 2005}\\[0.5cm]
\end{center}



\begin{abstract}

The investigation of the dynamics responsible for electroweak 
symmetry breaking is one of the prime tasks of experiments at 
present and future colliders. Experiments at 
the Tevatron \ppbar\ Collider and at the CERN Large Hadron 
Collider (LHC) must be able to discover a Standard Model 
Higgs boson over the full mass range as well as Higgs bosons 
in extended models. In this review, the discovery potential 
for the Standard Model Higgs boson and for Higgs bosons in the Minimal 
Supersymmetric extension is summarized. Emphasis is put on 
those studies which have been performed recently within 
the experimental collaborations using a realistic simulation of 
the detector performance. This includes a discussion of the search 
for Higgs bosons using the vector boson fusion mode at the LHC, 
a discussion on the measurement of Higgs boson parameters 
as well as a detailed review of the MSSM sector for different
benchmark scenarios. The Tevatron part of the review also contains 
a discussion of first physics results from data taken
in the ongoing Run~II.

\end{abstract}


\section{Introduction}  

Over the next decade, hadron colliders will play an important role 
in the investigation of fundamental questions of particle physics. 
While the Standard Model of electroweak\cite{SM} and strong\cite{QCD}
interactions is in excellent agreement with the  
numerous experimental measurements, the dynamics responsible for
electroweak symmetry breaking are still unknown.
Within the Standard Model, the Higgs mechanism\cite{higgs}
is invoked to break the electroweak symmetry. A doublet of complex 
scalar fields is introduced of which a single neutral scalar particle, 
the Higgs boson, remains after symmetry breaking.\cite{higgs-hunter}
Many extensions of this minimal version of the Higgs sector have
been proposed, including a scenario with two complex Higgs doublets as
realized in the Minimal Supersymmetric Standard 
Model (MSSM).\cite{MSSM}

Within the Standard Model, the Higgs boson is the only particle that has 
not been discovered so far. The direct search
at the \epemit\ collider {\em LEP} has led to a lower bound on its mass of 
114.4~\Gcs.\cite{LEP-higgs-limit}
 Indirectly, high precision electroweak 
data constrain the mass of the Higgs boson via their sensitivity to 
loop corrections. Assuming the overall validity of the Standard Model, 
a global fit\cite{LEP-EWWG} to all electroweak data leads to 
$m_H = 114 ^{+69}_{-45}$~\Gcs. 
On the basis of the present theoretical knowledge, the Higgs sector 
in the Standard Model remains largely unconstrained. 
While there is no direct prediction for the mass of the Higgs boson,
an upper limit of $\sim$1\,\Tcs\ can be inferred from unitarity 
arguments.\cite{higgs-unitarity} 

Further constraints can be derived under the assumption that the Standard
Model is valid only up to a cutoff energy scale~$\Lambda$, beyond
which new physics becomes relevant. Requiring that the electroweak
vacuum is stable and that the Standard Model remains perturbative 
allows to set upper and lower bounds on the Higgs boson mass.\cite{higgs-constraints,higgs-const-2} 
For a cutoff scale of the order of the Planck mass, the Higgs boson mass is required 
to be in the range 130 $< \mh <$ 190~\Gcs. If new physics appears at lower mass scales, 
the bound becomes weaker, {\em e.g.}, for $\Lambda =$ 1~TeV 
the Higgs boson mass is constrained to be in the range 50 $< \mh <$ 800~\Gcs.

The Minimal Supersymmetric Standard Model contains two complex 
Higgs doublets, leading to five physical Higgs bosons after
electroweak symmetry breaking: three neutral (two CP-even $h, H$ and
one CP-odd $A$) and a pair of charged Higgs bosons $H^\pm$.
At tree level, the Higgs sector of the MSSM is fully specified 
by two parameters, generally chosen to be $m_A$, the mass of the 
CP-odd Higgs boson, and \tanb, the ratio of the vacuum expectation 
values of the two Higgs doublets. Radiative corrections modify the 
tree-level relations significantly. 
This is of particular interest for the mass of 
the lightest CP-even Higgs boson, which at tree level 
is constrained to be below the mass of the $Z$ boson. Loop corrections 
are sensitive to the mass of the top quark, to the mass of the scalar particles
and in particular to mixing in the stop sector. The largest values for 
the mass of the Higgs boson $h$ are reached for large mixing, 
characterized by large values of the mixing parameter 
$X_t : = A_t - \mu \cot \beta$, where $A_t$ is the trilinear coupling and $\mu$ is 
the Higgs mass parameter. If the full one-loop and 
the dominant two-loop contributions are included,\cite{heinemeyer,heinemeyer-2} the upper
bound on
the mass of the light Higgs boson $h$ is expected to be around 135~\Gcs.
While the light neutral Higgs boson may be difficult to distinguish from the 
Standard Model Higgs boson, the other heavier Higgs bosons 
are a distinctive signal of physics beyond the 
Standard Model. The masses of the heavier Higgs bosons $H, A$ and \hplus\ 
are often almost degenerate. 

Direct searches at LEP have given lower bounds of 92.9 (93.3)~\Gcs\ and 93.4 
(93.3)~\Gcs\ on the masses of the lightest CP-even Higgs boson $h$ and the CP-odd Higgs
boson $A$ within the $m_h$-max (no-mixing) scenario.\cite{lep-mssm} 
In those scenarios, the mixing parameter in the stop sector  
is set to values of $X_t$~=~2~\Tcs\ and $X_t$~=~0, respectively. 
Given the LEP results, the \tanb\ regions of 0.9 $< \tanb <$ 1.5 and  
0.4 $< \tanb <$ 5.6 are excluded at 95\% confidence level for the 
$m_h$-max and the no-mixing scenarios, respectively.\cite{lep-mssm}
However, it should be noted that the exclusions in \tanb\ depend
critically on the exact value of the top-quark mass. 
In the LEP analysis $m_t$ = 179.3~\Gcs\ has been assumed. With increasing 
top mass the theoretical upper bound on $m_h$ increases and hence the 
exclusion in \tanb\ decreases, {\em e.g.}, for $m_t$ of about 183~\Gcs\ 
or higher the exclusions in \tanb\ vanish. 

The charged Higgs boson mass is related to $m_A$ via the
tree-level relation $m^2_{H^{\pm}} = m_W^2 + m_A^2$ and is less
sensitive to radiative corrections.\cite{mssm-hplus} 
Direct searches for charged Higgs bosons in the decay modes 
$\hplus \to \tau \nu$ and $\hplus \to c s$
have been carried out at LEP, yielding a lower
bound of 78.6~\Gcs\ on \mhplus\ independent of the $\hplus \to \tau \nu$ 
branching ratio.\cite{lep-hplus}
At the Tevatron, the CDF and \dzero\ experiments have performed direct and 
indirect searches for the charged Higgs boson through the process
$\ppbar \to \ttbar$ with at least one top quark decaying via 
$t \to \hplus b$. These searches have excluded the small and large \tanb\
regions for \hplus\ masses up to $\sim$160~\Gcs.\cite{tev-hplus}
Other experimental bounds on the charged Higgs boson mass can be derived using
processes where the charged Higgs boson enters as a virtual particle. 
For example, the measurement of the $ b \to s \gamma$ decay rate
allows to set indirect limits on the charged Higgs boson mass,\cite{btosgamma}
which, however, are strongly model dependent.\cite{bsg-modeldep}

The high collision energy of the Fermilab Tevatron \ppbar\ 
collider and the CERN {\em Large Hadron Collider} (LHC) allow 
to extend the search for Higgs bosons into unexplored mass regions. 
In particular the experiments at the LHC have a large discovery 
potential for Higgs bosons in both the Standard 
Model and in the MSSM over the full parameter range. 
Should the Higgs boson be light, {\em i.e.}, have a mass in 
the range favoured by the precision electroweak measurements, also 
the experiments at the Tevatron will have 
sensitivity for discovery. The mass range accessible depends, however, 
critically on the integrated luminosity that can be collected.

In this article, the potential for Higgs boson searches 
at these hadron colliders is reviewed, focussing on the
investigation of the Higgs sectors in the Standard  
Model and in the MSSM. This subject is discussed in the literature in 
numerous notes and publications. For this review the main emphasis is
put on studies which have been performed recently 
within the experimental collaborations using a realistic
simulation of the detector performance.
This includes a discussion of the search 
for Higgs bosons using the vector boson fusion mode at the LHC, 
a discussion of the measurement of Higgs boson parameters at the 
LHC as well as a detailed review of
analyses within the MSSM for different
benchmark scenarios. For the Tevatron, first results based on the data 
taken in the ongoing Run~II are discussed in addition to a review 
of the Monte Carlo studies and projections. 

In Section 2, Higgs boson production and decay processes 
and the status of the calculation of higher order 
QCD corrections are presented. The experimental scenarios at the 
Tevatron and at the LHC are briefly discussed in Section 3. 
The current status and expected performance in the search for both a Standard 
Model Higgs boson
and MSSM Higgs bosons at the Tevatron is presented in Section 4. 
The LHC potential for discovery of a Standard Model Higgs boson, for 
the measurement of its parameters
and for the discovery of MSSM Higgs bosons
in various benchmark scenarios is summarized in Sections 5 to 7.

\section{Higgs Bosons at Hadron Colliders}

\subsection{Higgs boson production \label{s:higgs-prod}}
At hadron colliders, Higgs bosons can be produced via four different production 
mechanisms: 

\begin{itemize}
\item gluon fusion, $gg \to H$, which is mediated at lowest order by a heavy
quark loop; 
\item vector boson fusion (VBF), $ qq \to qq H $; 
\item associated production of a Higgs boson with weak gauge bosons, \\
$qq \to W/Z \ H$
(Higgs Strahlung, Drell-Yan like production); 
\item associated Higgs boson production with heavy quarks, \\
$gg, qq \to tt H$, $gg, qq \to bb H$  (and $gb \to b H$).
\end{itemize}

\begin{figure}
\begin{center}
\mbox{\epsfig{file=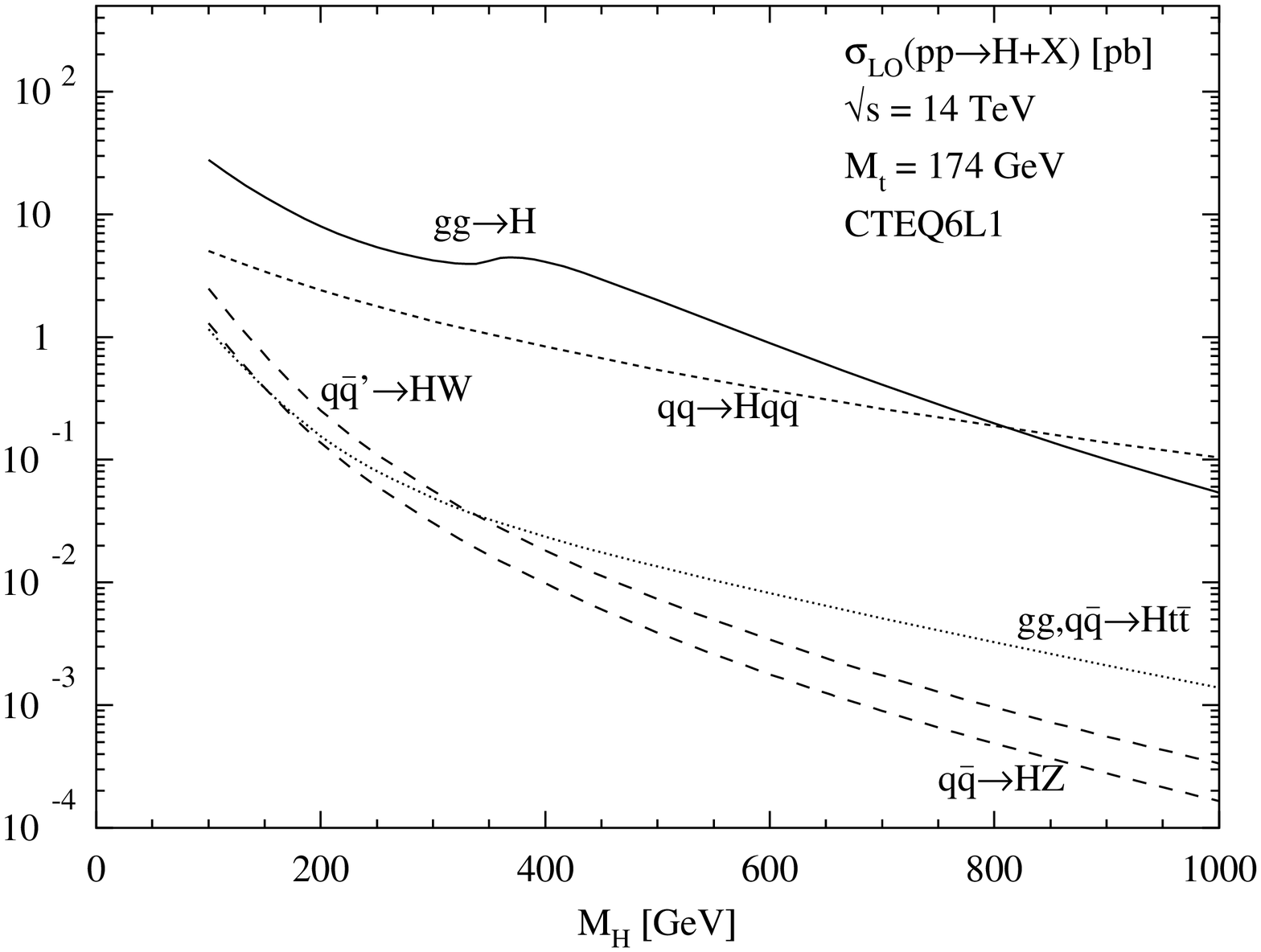,width=0.49\textwidth,clip=true}} 
\mbox{\epsfig{file=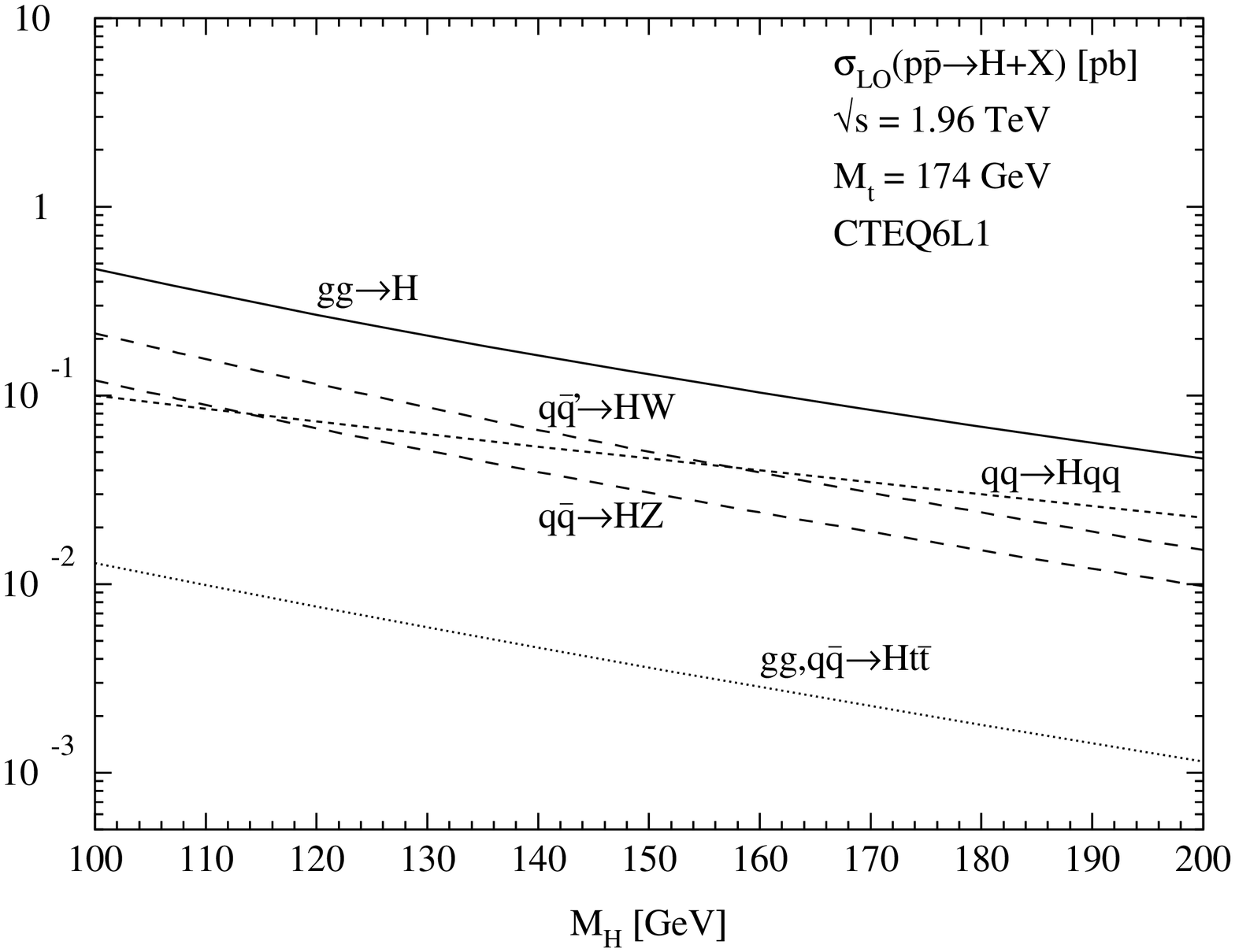,width=0.49\textwidth,clip=true}}
\end{center}
\caption{\it \footnotesize
Leading order production cross sections for a Standard Model Higgs boson 
as a function of 
the Higgs boson mass at the LHC 14 TeV 
pp collider (left) and at the 1.96 TeV Tevatron \ppbar\ collider (right). 
In the cross section calculation the CTEQ6L1 structure 
function parametrization has been used.  (The calculations have been 
performed by M.~Spira, Ref.~\protect\refcite{spira-xsect})}
\label{f:prod_xs}
\end{figure}

The lowest order production cross sections for the four different processes 
are shown in Fig.~\ref{f:prod_xs} for both the Tevatron and the LHC collider 
as a function of the Higgs boson mass.\cite{spira-xsect} 
At both colliders, the dominant production 
mode is the gluon-fusion process.
At the LHC, the vector boson fusion has the second largest cross section.
In the low mass region it amounts at leading order to about 20\% of the 
gluon-fusion cross section, whereas it reaches the same level for masses around 800~\Gcs.
At the Tevatron \ppbar\ 
collider, the contribution of the associated W/Z H production mode is more important
and, unlike at the LHC, Higgs boson searches heavily exploit this production mode. 
At the LHC, the associated $WH$, $ZH$ and $\ttbar H$ production
processes are relevant only for the search of 
a light Standard Model Higgs boson with a mass close to the LEP limit. 

For all production processes higher order QCD corrections have been 
calculated. In particular, significant progress has been made during the last 
two to three years in the calculation of QCD corrections for the gluon fusion 
and for the associated $\ttbar H$ and $\bbbar H$ production processes.

\begin{figure}
\begin{center}
\hspace*{-0.5cm}
\mbox{\epsfig{file=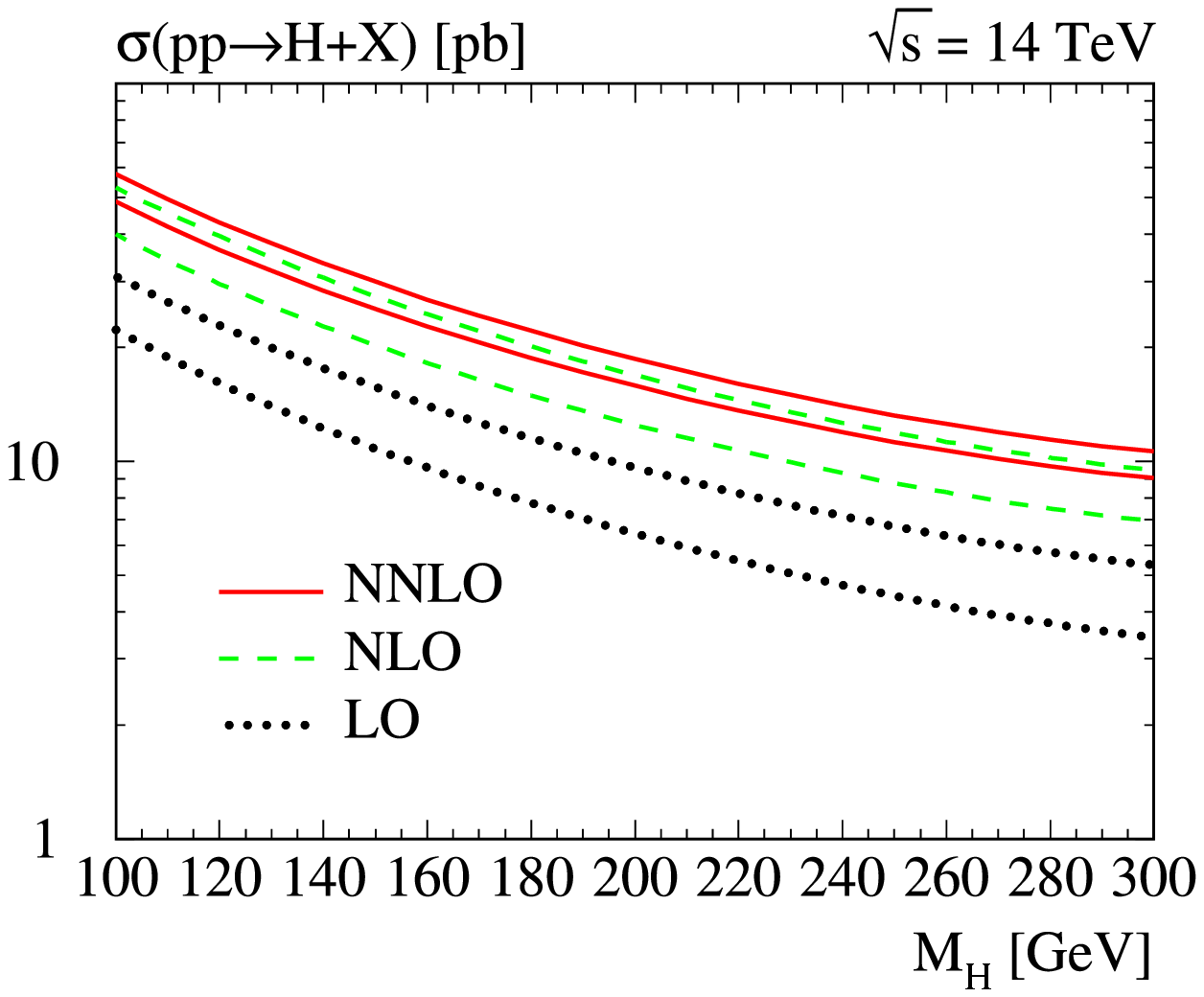,width=0.45\textwidth,clip=true}} \hspace*{0.6cm}
\mbox{\epsfig{file=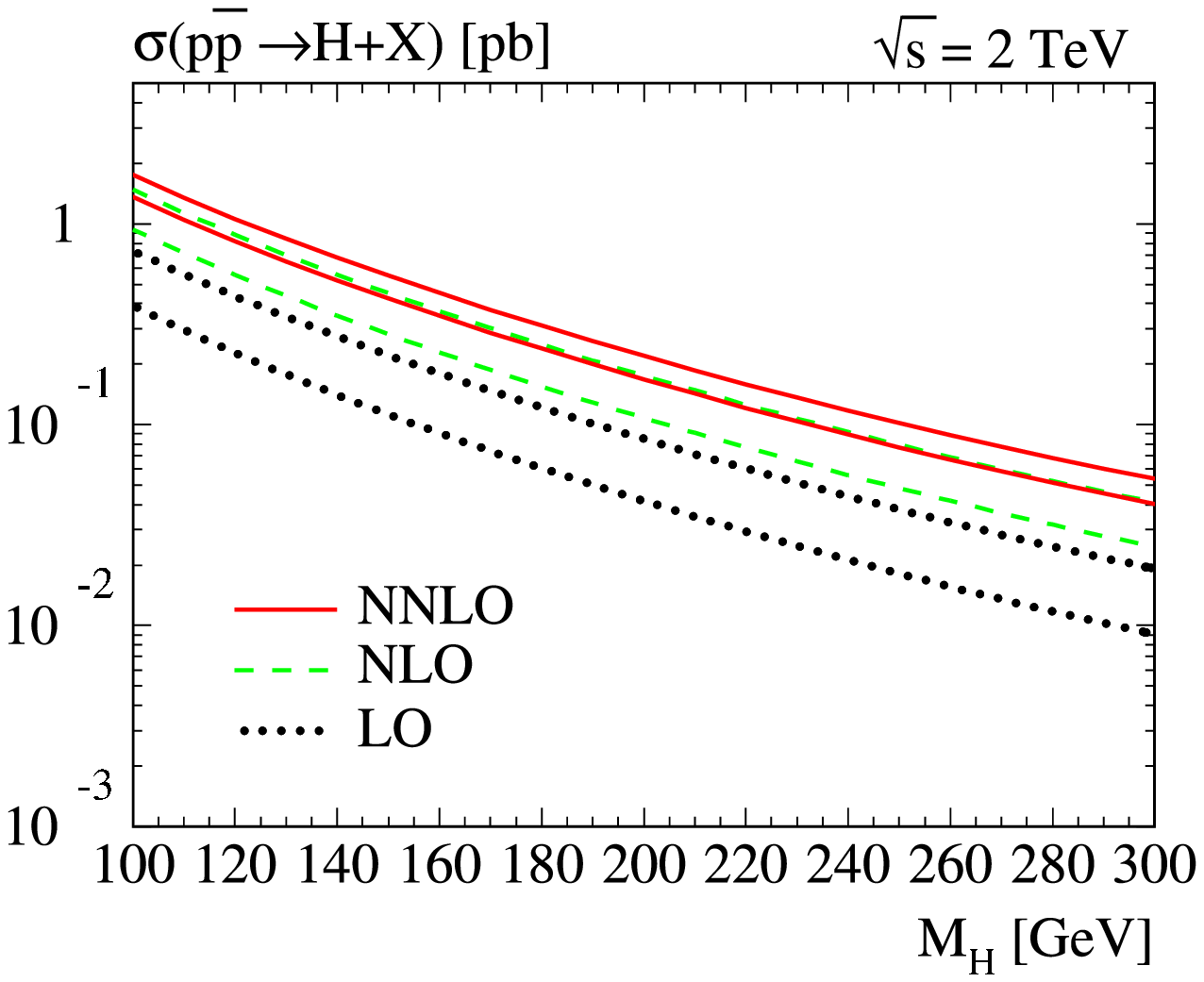,width=0.45\textwidth,clip=true}}
\end{center}
\caption{\it \footnotesize
Gluon-fusion production cross section for a Standard Model Higgs boson at 
the LHC 14 TeV pp collider (left) and at the Tevatron 2 TeV \ppbar\ collider (right) 
at leading (dotted), next-to-leading (dashed) and next-to-next-to-leading order
(solid). The upper (lower) curve of each pair corresponds to a choice of the 
renormalization and factorization scale of $\mu_{\it R} = \mu_{\it F} = {\it m}_{\it H} /{\it 2}$ 
($\mu_{\it R} = \mu_{\it F} = {\it 2 \ m_H}$) (from Ref.~\protect\refcite{harlander2}, 
see also Refs.~\protect\refcite{har-kil,anastasiou,rsn}). 
}
\label{f:xs_gg}
\end{figure}

Already more than ten years ago, the next-to-leading order (NLO) QCD corrections 
to the gluon-fusion process have been calculated and have been found to be 
large.\cite{gg-nlo} Their calculation appears challenging since already in the 
leading order diagram a massive one-loop triangle appears. The NLO 
calculation showed a significant increase of the predicted total cross 
section by about 50-100\%. These large corrections stimulated 
the calculation of the next-to-next-to-leading order (NNLO) corrections, 
to which many authors contributed\cite{gg-nnlo,har-kil,anastasiou,rsn} 
and which meanwhile has been completed in the heavy 
top-quark limit ($m_{t} \to \infty$). In this 
limit the calculation of the Feynman diagrams is considerably simplified. 
The approximation has been tested at NLO and has been found to agree within 
5\% with the full NLO calculation up to $\mh = 2 \ m_{t}$, 
if the NLO result obtained with ($m_{t} \to \infty$) is rescaled with
the ratio of
the LO cross sections calculated with a finite top mass and with ($m_{t} \to \infty$).  
For larger Higgs boson masses still 
a surprisingly good agreement is found, {\em e.g.}, the approximation deviates
from the exact result by only 10\% at $m_H$ = 1~\Tcs, 
if the same rescaling is applied.\cite{harlander} 
Based on these observations, it appears reasonable to apply the heavy 
top-quark approximation also at NNLO. The results for the total cross 
section show a modest increase between the NLO and NNLO calculation at the 
level of 10--20\%, indicating that a nicely converging
perturbative series seems to be emerging.
The results of the LO, NLO and NNLO cross section calculations 
are shown for both colliders in Fig.~\ref{f:xs_gg}.
In particular, a clear reduction of the uncertainty due to higher
order corrections has been observed, which is estimated to be about
15\% based on variations of the renormalization scale.\cite{harlander}

\begin{figure}
\begin{center}
\hspace*{-0.5cm}
\mbox{\epsfig{file=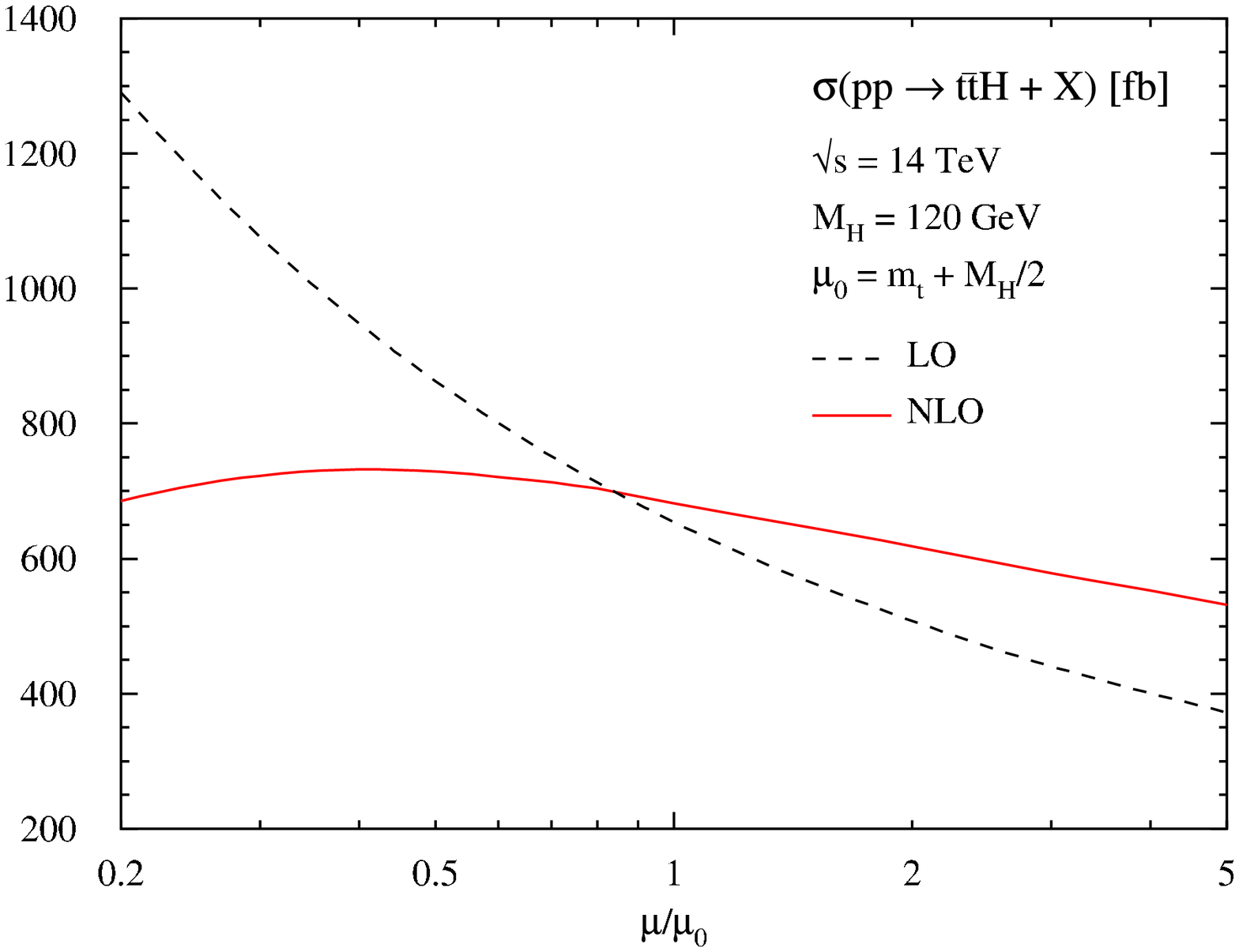,width=0.44\textwidth,clip=true}} \hspace*{0.8cm}
\mbox{\epsfig{file=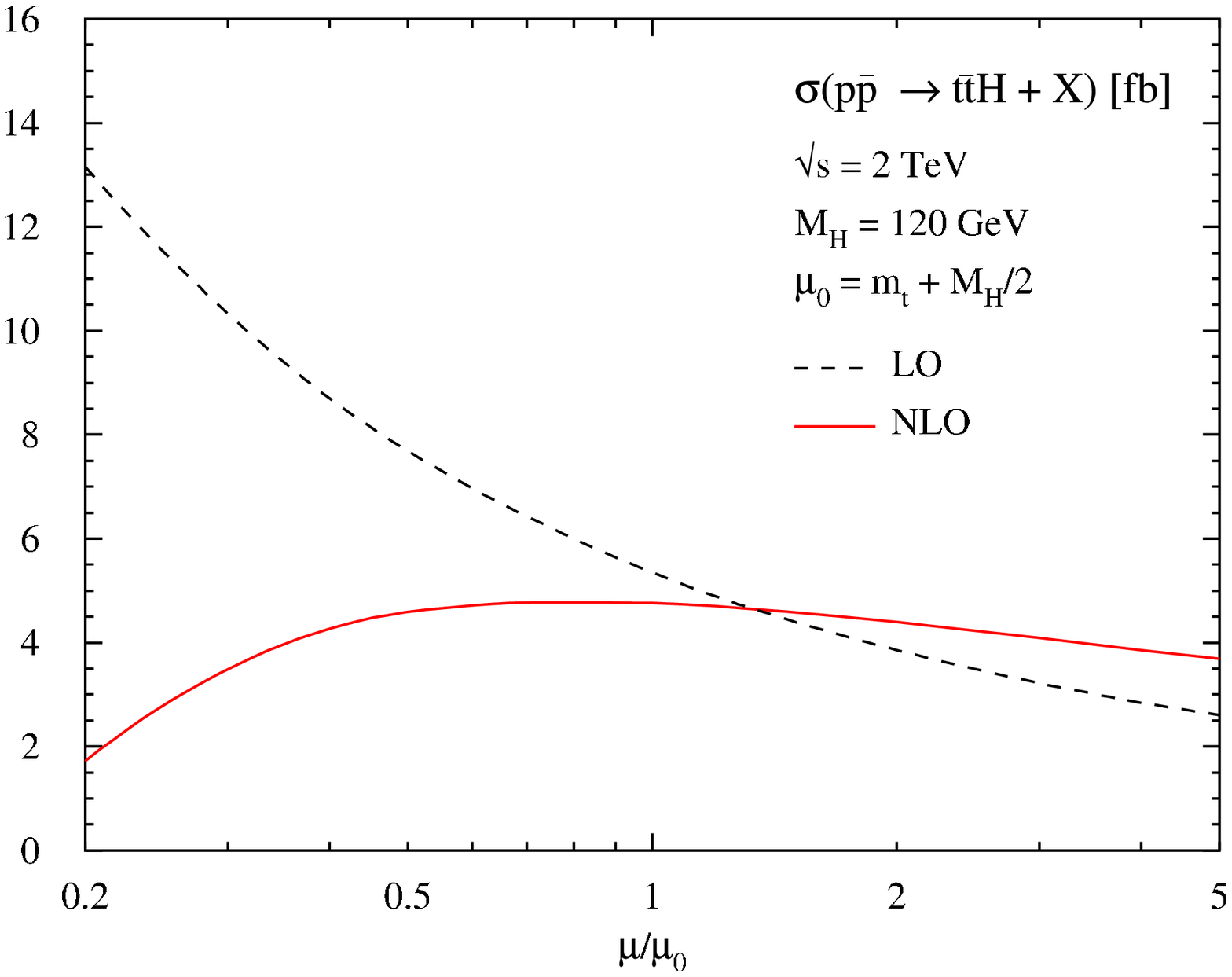,width=0.43\textwidth,clip=true}}
\end{center}
\caption{\it \footnotesize
Variation of the $\ttbar H$ production cross section at the LHC 14 TeV pp collider
(left) and at
the Tevatron 2 TeV \ppbar\ collider (right) 
with the renormalization and factorization scale~$\mu = \mu_{\it R} = \mu_{\it F}$, 
varied around the value $\mu_0 = m_t + m_H/2$
(from Refs.~\protect\refcite{tth-nlo} and \protect\refcite{tth-nlo2}). 
}
\label{f:tth-nlo}
\end{figure}

Another important and challenging theoretical calculation constitutes the 
NLO calculation of the QCD corrections to the associated $\ttbar H$ 
production. Results have been published recently for a finite top-quark 
mass.\cite{tth-nlo,tth-nlo2} Also in this case a dramatic 
reduction in the variation of the cross-section prediction with
the renormalization and factorization scale $\mu$ has been 
found,\cite{tth-nlo,tth-nlo2} as can be seen 
in Fig.~\ref{f:tth-nlo}, where the LO and NLO cross sections
are shown as a function of the scale for both the LHC and the Tevatron. 
For the choice $\mu = m_{t} + \mh/2$, for example, the NLO correction 
is found to be negative for the Tevatron, whereas a small 
increase of the cross section of the order of 20\% is found for the LHC. 

In the Standard Model, the cross section for producing a Higgs boson
in association with $b$ quarks is relatively small. 
However, in a supersymmetric theory with a 
large value of \tanb, the $b$-quark Yukawa coupling can be strongly enhanced 
and Higgs boson production in association with $b$ quarks becomes the dominant
production mechanism. Cross-section calculations including next-to-leading 
order corrections have been presented in two different 
approaches.\cite{Les-Houches-bb-higgs} 
In the so-called four-flavour scheme, no $b$ quarks 
are present in the initial state and the lowest order processes are the 
$gg \to \bbbar h$ and 
$ q \bar{q} \to \bbbar h$ tree level processes. Due to the splitting of 
gluons into \bbbar\ pairs and the small $b$-quark mass, the inclusive cross 
section is affected by large logarithms and the convergence 
of the perturbative expansion may be poor. Depending on the final state considered, 
the convergence can be improved by summing the logarithms to all orders in 
perturbation theory through the use of $b$-quark parton distributions, 
{\em i.e.}, moving to a five-flavour scheme.\cite{five-flavours} 
In this scheme, the lowest order inclusive process is 
$\bbbar  \to  h$. The first order correction to this process includes 
the process $gb \to b h$. 

In many analyses two high-\PT\ $b$ quarks 
are required experimentally to improve the signal-to-background ratio.
The leading subprocess in this region of phase
space is $gg \to \bbbar h$.  The  
relevant production cross section, implementing parton level cuts 
on the $b$ quarks that closely reproduce the experimental cuts,
have been computed at NLO in Refs.~\refcite{spira-h-bbbar} and 
\refcite{dawson-h-bbbar} for both the Tevatron and the LHC.
The NLO corrections modify the leading order predictions by less than 
30\% at the Tevatron and less than 50\% at the LHC. For a cut of 
$\PT > 20$~\Gc\ and $ | \eta | <$ 2.0 (2.5) for the $b$ quarks 
at the Tevatron (LHC) and using $\mu = ( 2 m_b + m_h) / 4$ as factorization 
and renormalization scale,  
the NLO corrections are negative for small Higgs boson masses around 
120~\Gcs\ and positive for large masses, with cross-over points
around 140~\Gcs\ at the Tevatron and around 300~\Gcs\ at the LHC.

For cases of one and no tagged b-jets in the final state, results for 
the relevant cross sections can be calculated in the four- and 
five-flavour schemes. The two calculations represent different
perturbative expansions of the same physics process and should agree at 
sufficiently high order. The NLO four-flavour result is obtained by 
integration over one of the $b$ quarks in the $gg \to \ \bbbar h$ 
calculation.\cite{spira-h-bbbar,dawson-h-bbbar} For the one-jet case, this calculation 
is found to agree within their respective scale uncertainties 
with the NLO calculation performed in the five-flavour 
scheme.\cite{willenbrock-bbbar} 
For the inclusive cross section (no tagged $b$ jets) 
the five-flavour calculation has been 
performed to NLO in Ref.~\refcite{five-flavour-nlo} and to NNLO in 
Ref.~\refcite{five-flavour-nnlo}. Again, within the large respective 
uncertainties, agreement between the NLO four- and NNLO five-flavour 
calculation is found for small Higgs boson masses, while for 
large Higgs boson masses the five-flavour scheme tends to yield 
larger cross sections. 
This represents major progress compared to several years 
ago, when large discrepancies between the NLO $bb \to h$ and the LO 
$gg \to \bbbar h$ calculation had been reported.\cite{tev-report,bbbar-LO}

Next-to-leading order calculations of the production cross sections are also 
available for the remaining two production mechanisms: $WH, ZH$ and $qqH$ production.
For the vector boson fusion process the NLO corrections are found 
to be moderate,\cite{vbf-nlo,nlo-mc-vbf} {\em i.e.}, at the level of 10\%. 
The NLO QCD corrections for the associated production of a Higgs boson with 
a vector boson can be derived from the Drell-Yan process and give a 30\% 
increase with respect to the leading order prediction.\cite{wh-nlo}
Recently, the NNLO QCD corrections for this process have been calculated.\cite{wh-nnlo}
For the Drell-Yan type corrections, a moderate increase of the NLO 
cross sections of up to 3\% (10\%) is found for the LHC (Tevatron) for a Higgs boson 
mass in the range 100 $< m_H <$ 300~\Gcs. For the $ZH$ associated production, 
additional gluon fusion contributions appear at NNLO (a triangular
diagram $gg \to Z^* \to ZH$ 
and a box diagram where the Z and H bosons are emitted from internal quark lines). 
These contributions have been found to be relevant only at the LHC and increase the 
$ZH$ production cross section in the low mass region by about 10\%.  

\begin{figure}
\begin{center}
\mbox{\epsfig{file=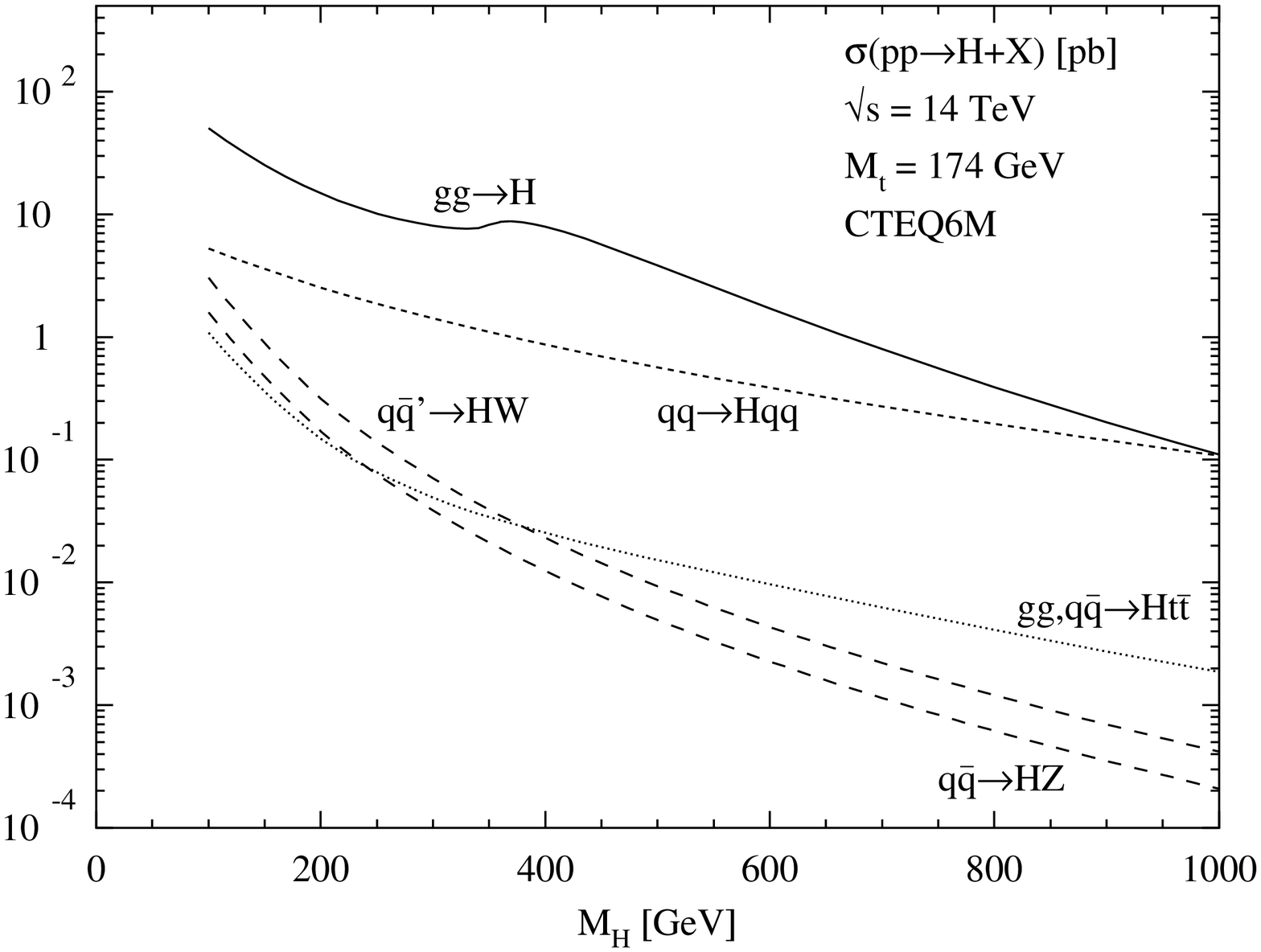,width=0.49\textwidth,clip=true}} 
\mbox{\epsfig{file=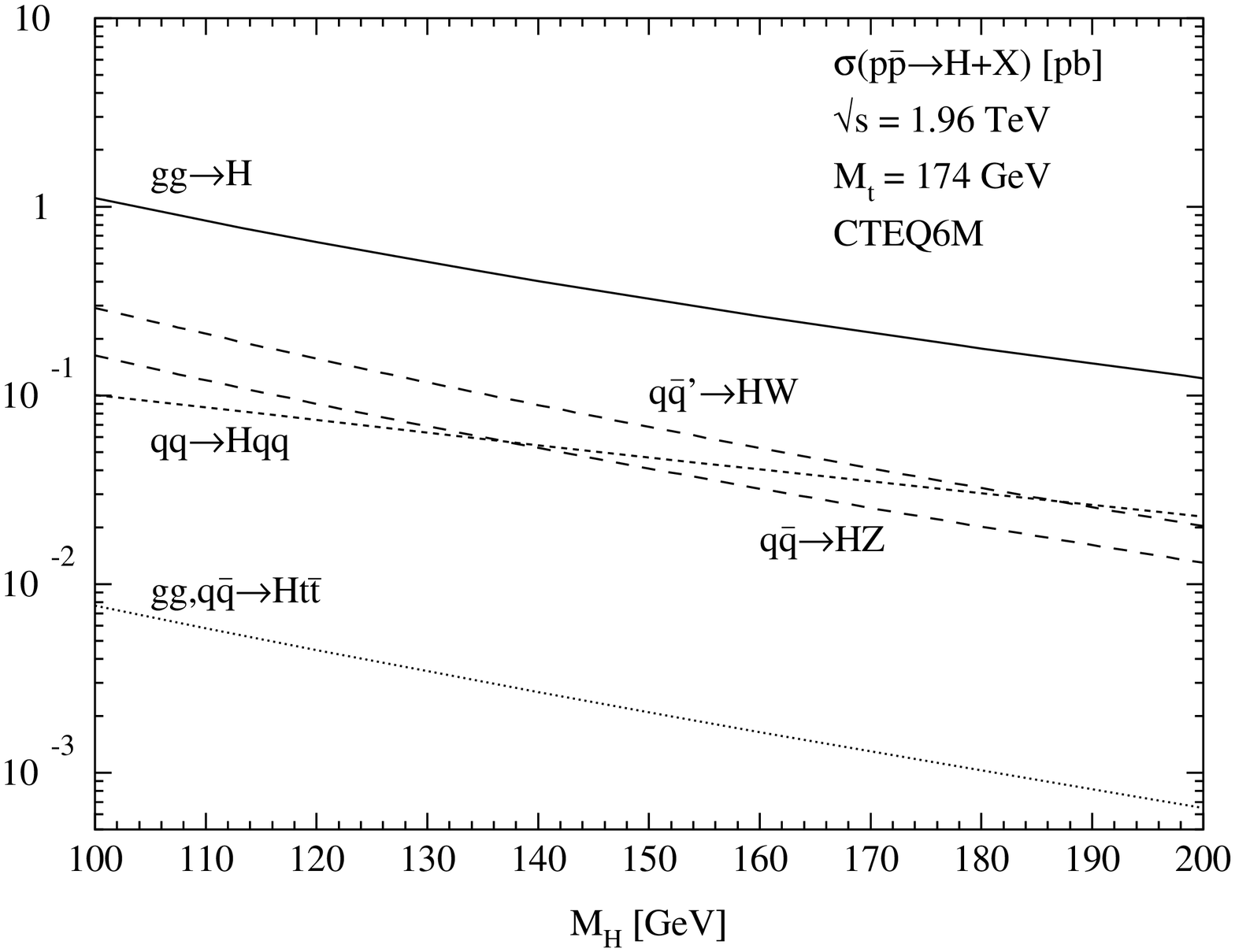,width=0.49\textwidth,clip=true}}
\end{center}
\caption{\it \footnotesize
NLO production cross sections for a Standard Model Higgs boson 
as a function of 
the Higgs boson mass at the LHC 14 TeV 
pp collider (left) and at the 1.96 TeV Tevatron \ppbar\ collider (right). 
In the cross section calculation the CTEQ6M structure 
function parametrization has been used.  (The calculations have been 
performed by M.~Spira, Ref.~\protect\refcite{spira-xsect})}
\label{f:prod_xs_nlo}
\end{figure}

The production cross sections for the four different processes including the NLO 
QCD corrections are shown in Fig.~\ref{f:prod_xs_nlo} for both the Tevatron and the 
LHC collider as a function of the Higgs boson mass.\cite{spira-xsect}

\subsection{Higgs boson decays}

\begin{figure}
\begin{center}
\epsfig{file=plots/smbr.epsi,height=0.45\textwidth,angle=-90} \hspace*{0.3cm}
\hspace*{0.5cm}
\epsfig{file=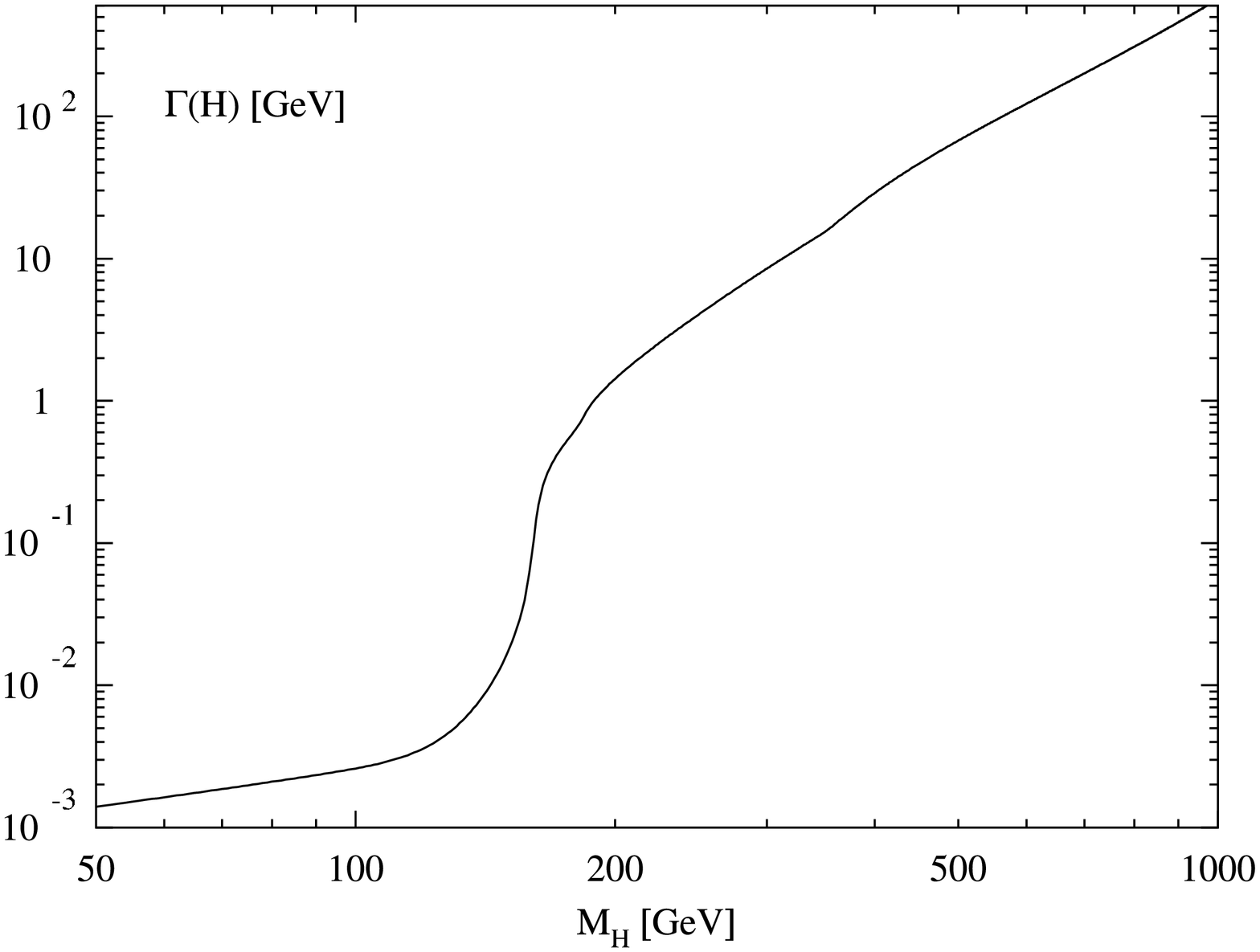,height=0.45\textwidth,angle=-90}
\end{center}
\caption{\it \footnotesize
Branching fractions (left) and total decay width (right) of the Standard Model Higgs 
boson as a function of Higgs boson mass 
(from Ref.~\protect\refcite{spira-decays}). 
}
\label{f:br_sm}
\end{figure}

The branching fractions of the Standard Model Higgs boson are shown in
Fig.~\ref{f:br_sm}(left) as a function of the Higgs boson mass. 
They have been calculated taking into account both electroweak and
QCD corrections.\cite{spira-decays} The latter include 
logarithmic corrections
which are sizeable in particular for decays into $b$ or $c$ quarks.
Large logarithms are resummed by using both the running quark
mass and the strong coupling constant evaluated at the
scale of the Higgs boson mass.
When kinematically accessible, decays of the Standard Model Higgs boson into
vector boson pairs $WW$ or $ZZ$ dominate over all other decay modes. 
Above the kinematic threshold, the branching fraction into \ttbar\ can
reach up to 20\%.
All other fermionic decays are only relevant for
Higgs boson masses below 2 $m_W$, with $H\to\bb$ dominating below 140~\Gcs.
The branching fractions for $H\to\tau\tau$ and $H\to gg$ both reach up
to about 8\% at Higgs boson masses between 100 and
120~\Gcs. Decays into two photons, which  are of interest due to their
relatively clean experimental signature, can proceed via fermion and $W$ loops
with a branching fraction of up to 2~$\cdot$~10$^{-3}$ at low Higgs 
boson masses.

Compared to the mass resolution of hadron collider experiments, the total decay 
width of the Standard Model Higgs boson is negligible at low masses and
becomes significant only above the threshold for decays into $ZZ$, as shown in 
Fig.~\ref{f:br_sm}(right). At $m_H$ = 1~\Tcs, the Higgs 
resonance is broad with a width of about 600~\Gcs.
In this mass regime, the Higgs field is coupling strongly, 
resulting in large NNLO corrections. With increasing coupling, the
peak position of the Higgs boson resonance does not increase beyond a
saturation value close to 1~\Tcs, as found in both perturbative and
non-perturbative calculations.\cite{binoth-gamma-h}

\begin{figure}
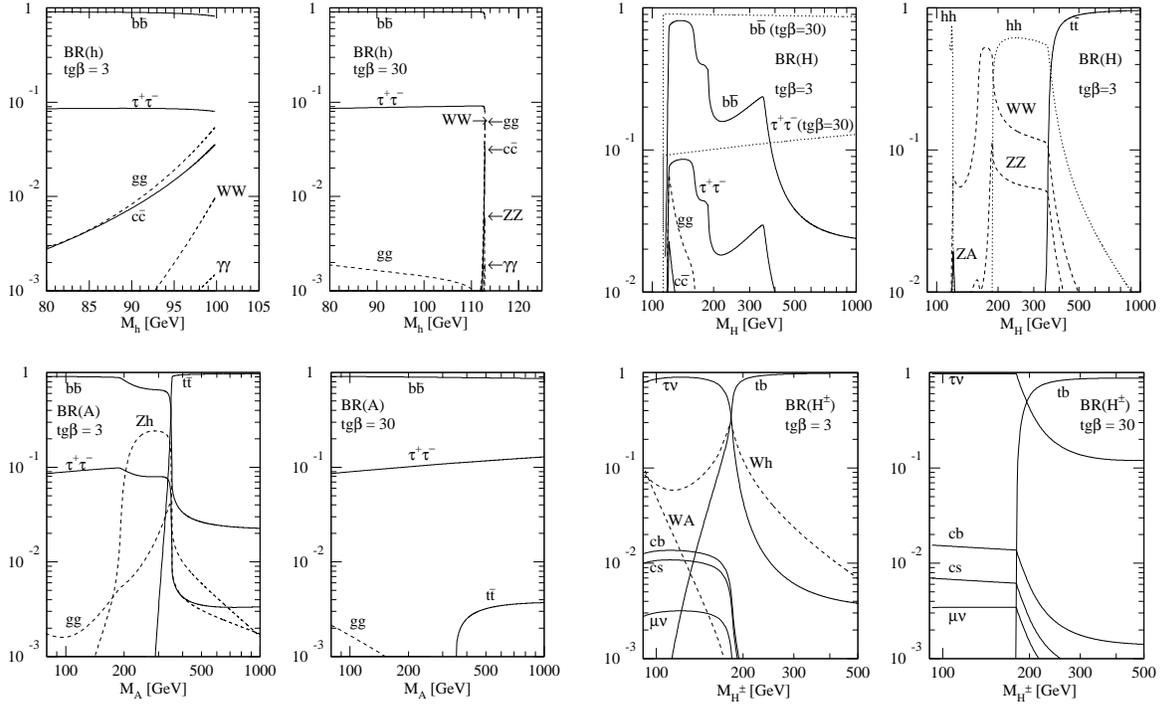

\begin{center}
\epsfig{file=plots/br.l.epsi,height=0.465\textwidth,angle=-90} \hfill
\epsfig{file=plots/br.h.epsi,height=0.48\textwidth,angle=-90}\\[0.4cm]
\epsfig{file=plots/br.a.epsi,height=0.48\textwidth,angle=-90} \hfill
\epsfig{file=plots/br.c.epsi,height=0.48\textwidth,angle=-90}
\end{center}
\caption{\it \footnotesize
Branching fractions of the MSSM Higgs bosons as a function of
their masses for tan$\beta$=3 and tan$\beta$=30 assuming a SUSY mass scale of
1~TeV/c$^2$ and vanishing mixing (from Ref.~\protect\refcite{spira-decays}). 
}
\label{f:br_mssm}
\end{figure}

Within the MSSM, branching fractions of five physical Higgs bosons have to be considered
as a function of their masses as well as \tb\ and the masses of the SUSY particles.
In Fig.~\ref{f:br_mssm} the branching fractions of all MSSM Higgs
bosons are shown assuming that supersymmetric particles are heavy 
enough to be neglected.\cite{spira-decays}
The neutral Higgs bosons decay dominantly into \bb\ and \tautau\ at large \tb\ or for masses 
below 150~\Gcs\ (up to 2 $m_t$ for the CP-odd Higgs boson~A). 
Decays into $WW$, $ZZ$ and photons are generally suppressed by  
kinematics as well as the Higgs couplings and become relevant only in the 
decoupling limit $m_A\to\infty$, where the light CP-even Higgs
boson~$h$ effectively behaves  
like a Standard Model Higgs boson while all other MSSM Higgs bosons are heavy.
Charged Higgs bosons preferably decay into $tb$ if accessible. For masses below $m_t+m_b$, 
the decay $H^{\pm}\to\tau\nu$ dominates, with small contributions from
$H^{\pm}\to cb$ and $H^{\pm}\to cs$.
In addition, both charged and heavy neutral Higgs bosons can decay into lighter Higgs 
bosons: $H^{\pm}\to Wh$, $WA$ as well as $H\to hh$,$AA$,$ZA$ and $A\to Zh$.
Generally, the branching fractions of these decay modes are
significant only at small \tb.

Supersymmetric particles can influence the phenomenology of Higgs decays
either via loop effects or, if light enough, as viable decay modes.
In particular, Higgs boson decays to charginos, neutralinos and third generation sfermions
can be relevant if kinematically allowed. With R-parity
conserved, Higgs boson decays into the lightest neutralino are not
directly detectable (``invisible decay'').

\newcommand{\pbinv}{\ensuremath{\mathrm{pb^{-1}}}}

\section{Experimental Scenarios}

\subsection{The Tevatron Run II}
After a successful first period of data taking until 1996 (Run I), the 
Fermilab accelerator complex has been upgraded to provide collisions
with increased luminosity and a centre-of-mass energy of 1.96~TeV. 
In 2001 the second phase (Run II) of the experimental program started.
After an initial period with low luminosity, peak luminosities of up to 
1.0 $\cdot$ 10$^{32}$ $\rm{cm}^{-2}\rm{s}^{-1}$ have been reached in
Summer 2004. A further significant increase in luminosity is expected
after the commissioning of the recycler, a new ring that will be used
for accumulation and cooling of antiprotons. With this, the
accelerator is expected to deliver an integrated luminosity of up to
8~\fbs\ by 2009. 

Both Tevatron experiments CDF and \dzero\ have undergone major upgrades to 
meet the requirements of the Run~II physics program as well as the
higher luminosity and collision rates of the upgraded
accelerator. In the design, particular emphasis has been placed on
achieving an efficient identification  
of leptons and b-jets as well as on providing good jet and missing
energy measurements. 
The CDF and D\O\ detectors are described in detail in
Refs.~\refcite{cdf-detector} and \refcite{d0-detector}. 
Only a brief overview is presented in the following.

The CDF tracking system consists of silicon detectors and a drift chamber
situated inside a solenoid that provides a 1.4~T magnetic field coaxial with
the beam.
The silicon microstrip detector has eight cylindrical
layers of mostly double-sided silicon, distributed in radius between
1.5~cm and 28~cm. The system is read out in about 700.000 channels and
can provide three-dimensional 
precision tracking up to pseudorapidities of 2.0.
Outside of the silicon detectors and for pseudorapidities less than
1.0, charged particles are detected with up to 96 hits per track by
the central outer tracker, an open-cell drift chamber with alternating
axial and 2$^\circ$ stereo superlayers with 12 wires each. 
Just inside the solenoid, a scintillator-based time-of-flight detector
allows particle identification with a timing resolution of about 100~ps.

The electromagnetic (hadronic) calorimeters are lead-scintillator
(iron-scintillator) sampling calorimeters, providing coverage up to
pseudorapidities of 3.6 in a segmented projective tower geometry.
Proportional wire and scintillating strip detectors situated at a
depth corresponding to the electromagnetic shower maximum provide
measurements of the transverse shower profile.
In addition, an early energy sampling is obtained using 
preradiator chambers positioned between the solenoid coil and the
inner face of the central calorimeter.
Outside of the calorimeter and behind additional steel absorbers, a
multi-layer system of drift chambers and 
scintillation counters allows detection of muons for pseudorapidities
up to 1.5.

The tracking system of the D\O\ detector consists of a silicon vertex
detector and 
a scintillating fibre tracker, situated inside a superconducting coil
providing a 2~T magnetic field.
The D\O\ silicon tracker has four cylindrical layers of mostly
double-sided microstrip detectors covering 2.7~cm up to 9.4~cm in
radius, interspersed with twelve disk detectors in the 
central region and two large disks in either
forward region. The full system
has about 800.000 channels and provides three-dimensional precision
tracking up to pseudorapidities of 3.0. 
The volume between the silicon tracker and the superconducting coil
is instrumented with eight cylindrical double layers of
scintillating fibres. Each layer has axial and stereo fibres (stereo
angle $\pm 3^{\circ}$) with a diameter of 835~$\mu$m, that are 
read out using solid-state photodetectors (Visible Light Photon Counters,
VLPCs).

The D\O\ calorimeter is a Liquid Argon sampling calorimeter with
uranium absorber (copper and steel for the
outer hadronic layers) with hermetic coverage up to 
pseudorapidities of 4.2. 
Signals are read out in cells of projective towers with four
electromagnetic, at 
least four hadronic layers and a transverse segmentation of 0.1 in
both azimuth and pseudorapidity.
The granularity is increased to 0.05 for the third EM layer,
roughly corresponding to the electromagnetic shower maximum.
To provide additional sampling of energy lost in dead material,
scintillator-based detectors are placed in front of the calorimeter
cryostats (preshower detectors) and between the barrel and end-cap cryostats
(intercryostat detector). The preshower
detectors consist of three layers of scintillator strips with VLPC readout
providing, in addition to the energy measurement, a precise
three-dimensional position measurement for electromagnetic showers.

The D\O\ muon system consists of three layers of drift tubes and
scintillators, with toroid magnets situated between the first and
second layer to allow for a stand-alone muon momentum measurement. 
Scintillator pixels are used for triggering and rejection of out-of-time
backgrounds in both the central and forward regions. 
Proportional drift tubes 
are stacked in three or four decks per layer in the central region. 
Tracking of muons in the forward region is accomplished by using decks of
mini drift tubes in each layer, allowing muons to be
reconstructed up to pseudorapidities of 2.0.
The muon system is protected from beam-related backgrounds by
shielding around the beampipe using an iron-polyethylene-lead
absorber.

Both CDF and D\O\ detectors are read out using a three-level trigger system which
reduces the event rate from 2.5~MHz to about 50~Hz. This includes
programmable hardware triggers at Level~1 that provide basic track,
lepton and jet reconstruction, secondary vertex or impact parameter
triggers at Level~2 as well as a PC-based quasi-offline event
reconstruction at Level~3.

After commissioning, calibration and alignment, about 500~\pbinv\ of
physics quality data have been collected by each experiment between April
2002 and July 2004. Physics results based on the analysis of up to 200~\pbinv\ have
been presented at the Summer Conferences 2004.

\subsection{LHC experiments}

The {\em Large Hadron Collider (LHC)} is presently being constructed 
as a proton-proton collider with a centre-of-mass energy of 14 TeV at CERN. 
This machine will open up the possibility to explore the TeV energy range, 
which plays a key role in the investigation of the electroweak 
symmetry breaking. Two experiments, 
ATLAS and CMS,  have been designed and optimized 
as general purpose $pp$ detectors, capable of running at high 
luminosity ($\lhigh$) and detecting a variety of 
final-state signatures. For details on the detector concepts, 
the reader is referred to the Technical Proposals and Technical 
Design Reports.\cite{atlas-tp,cms-tp} Only a brief summary of the main 
design features is given in the following. 

Both experiments use a superconducting solenoid around the inner 
detector cavity to measure the track momenta. 
Pattern recognition, momentum and vertex measurements are achieved 
with a combination of high-resolution silicon pixel and strip detectors. 
In the ATLAS experiment, tracking is performed in a 2 T magnetic field.
Electron identification is enhanced with a 
continuous straw-tube tracking detector with transition radiation
capability in the outer part of the tracking volume.
Due to the presence of silicon strip and pixel detectors, both
experiments will be able to perform $b$ quark tagging using impact
parameter measurements and the reconstruction of secondary vertices. 

The calorimeter of the ATLAS experiment consists of an inner barrel cylinder and
end-caps, using the intrinsically radiation resistant Liquid Argon (LAr)
technology. Over the full length, this calorimeter is surrounded by a novel
hadronic calorimeter using iron as absorber and scintillating tiles as active material.
The barrel part of the LAr calorimetry
is an electromagnetic {\em accordion} calorimeter,\cite{atlas-tp}
with a highly granular first sampling (integrated preshower detector).
In the end-cap region (1.5$< | \eta | <$ 3.2) the LAr technology is used for both 
electromagnetic and hadronic calorimeters. In order to achieve a good 
jet and missing transverse energy (\met) measurement, the 
calorimeter coverage is extended down to $\abseta < 4.9$ using a 
special forward LAr calorimeter with rod-shaped electrodes in a tungsten matrix.
The CMS calorimeter system\cite{cms-tp}
consists of a high resolution lead tungstate ($PbWO_4$) crystal
calorimeter, complemented by a hadronic copper-scintillator sandwich
calorimeter. Both the electromagnetic and a large fraction of the
hadronic calorimeter are located inside the coil of the solenoid. 
The choice of the detection technique is to a large extent motivated by
the search for the Higgs boson in the $\hgg$ decay channel, which requires
both an excellent electromagnetic energy and angular resolution. 

In addition to the inner solenoid, the ATLAS detector has 
large superconducting air-core
toroids consisting of independent coils arranged outside 
the calorimetry. This allows a stand-alone
muon momentum measurement with three stations
of high-precision tracking chambers at the inner and outer radius and
in the middle of the air-core toroid. In the CMS experiment 
the magnetic flux of the large solenoid is returned through 
a 1.8 m thick saturated iron return yoke (1.8 T) which is 
instrumented with muon chambers. A single magnet thus provides 
the necessary bending power for precise
tracking in the inner detector and in the muon spectrometer.  
The magnetic field in the central cavity, in
which the inner detectors are located, is 4 T.  

It is assumed that at the LHC an initial
luminosity of $10^{33}$~cm$^{-2}$~s$^{-1}$, 
hereafter called {\em low luminosity}, can be achieved at the 
startup, which is expected for the year 2007.
This value is supposed to increase during the first 
two to three years of operation to the design luminosity
of 10$^{34}$ cm$^{-2}$ s$^{-1}$, hereafter called {\em high luminosity}.
Integrated luminosities of 10~fb$^{-1}$ and 100~fb$^{-1}$
should therefore be collected at the LHC after about 
one and four years of data taking, respectively.

\section{Search for Higgs Bosons at the Tevatron}
As the Run~II luminosity increases, the Tevatron experiments CDF and
D\O\ will start reaching sensitivity to production of low-mass Higgs
bosons beyond the LEP limits. For Standard Model Higgs bosons decaying to \bb, the
production in association with $W$ or $Z$ bosons is the most promising
channel. In the mass range between $\sim$150 and $\sim$180~\Gcs, 
Higgs bosons produced via
gluon fusion might be observable in their decays to $WW$. Given the
current projections for the integrated Tevatron luminosity of about 8~\fbinv\ by
2009, a 5$\sigma$ discovery of a Standard Model Higgs boson will be
very difficult to achieve. Nevertheless, combining results from all search channels 
should provide sensitivity for exclusion of Higgs boson production at the 95\% confidence
level up to Higgs boson masses of 180~\Gcs. 

In the following subsections the current status and projections
for the most important channels at Tevatron Run~II are summarized.
When available, calculations of signal and background cross sections
beyond leading order have been used. Systematic errors are taken into
account unless specifically noted otherwise.
Over the last couple of years, the original Monte Carlo studies presented in
Ref.~\refcite{tev-report} have  
been cross-checked and refined using Run~II data and 
hit-based GEANT\cite{geant} simulation.\cite{tev-hss} In most channels, first
preliminary results from the analysis of about 200~\pbinv\ of data
exist and are summarized in the following.

In view of the reduced luminosity expectations for Run~II, challenging
search channels, such as $\ttbar H$ and diffractive Higgs
production,\cite{tev-goldstein,tev-report} are expected to provide
little sensitivity and will therefore not be discussed further.

\subsection{\label{s:tev-vh} Associated production}
The production of Higgs bosons in association with vector bosons can
be searched for in all leptonic decays of $W$ and $Z$: $W\to \ell \nu$,
$Z\to\nn$ and $Z\to \ell\ell$ (with $\ell$ = $e$, $\mu$, $\tau$).
Sensitivity studies based on Monte Carlo simulation of the
detector performance throughout the course
of Run~II exist.\cite{tev-report,tev-hss} 
For $WH$ production, these studies are compared to first preliminary
results of searches in Run~II data corresponding to integrated 
luminosities of 162~\pbinv\ (CDF) and 174~\pbinv\ (D\O).\cite{tev-wh}

Final states compatible with the $WH$/$ZH$ signature can be selected by
requiring leptons and/or missing \ET\ as well as two b-tagged jets. 
After offline cuts, the trigger efficiencies for events involving 
charged leptons are very close to 100\%, as measured in Run~II data.
The channel $ZH\to\nn \bb$ represents a challenge for the trigger
systems of both experiments because of high rates due to the QCD
background. Nevertheless, using a set of inclusive triggers exploiting 
calorimeter information at Level~1 combined with jet-, lepton- and 
impact-parameter-triggers at Level~2, D\O\ estimates a trigger efficiency of
$>$90\% for events surviving offline analysis cuts.\cite{tev-hss}

Backgrounds involving light-quark jets (u,d,s) are suppressed using b-tagging,
which for both experiments has an efficiency of about 50\%, as measured
in data for central jets with \pt$>$40~\gevp\ and for a mistag rate for
light-quark jets of less than 0.5\%.\cite{tev-hss}
The b-tagging performance in the forward region has been estimated
using Monte Carlo simulation; it is expected to improve further after
including silicon stand-alone tracking algorithms using the forward
silicon detectors. 

After the b-tagging requirements and additional topological cuts,
backgrounds in channels involving charged leptons are entirely
dominated by physics backgrounds from $W$\bb, $Z$\bb, $WZ$ and \ttbar\ production. For
$ZH\to\nn \bb$, a significant amount of background from QCD jet production 
remains in addition. 
To improve the signal-to-background ratio further, the Higgs boson mass 
has to be reconstructed with the best possible resolution.
Currently, a relative jet energy resolution of 13.9\% has been
achieved by D\O, as measured in Run~II data for central jets at
\et = 55~GeV.\cite{tev-hss} It is expected that this can be improved by
30\% due to more sophisticated jet reconstruction algorithms as well
as refinements  
in jet energy calibration, including a calibration of the \bb\ mass
reconstruction using the $Z\to\bb$ signal. 

\begin{figure}
\begin{center}
\epsfig{file=./plots/WHpexp115.epsi,width=0.49\textwidth} \hfill
\epsfig{file=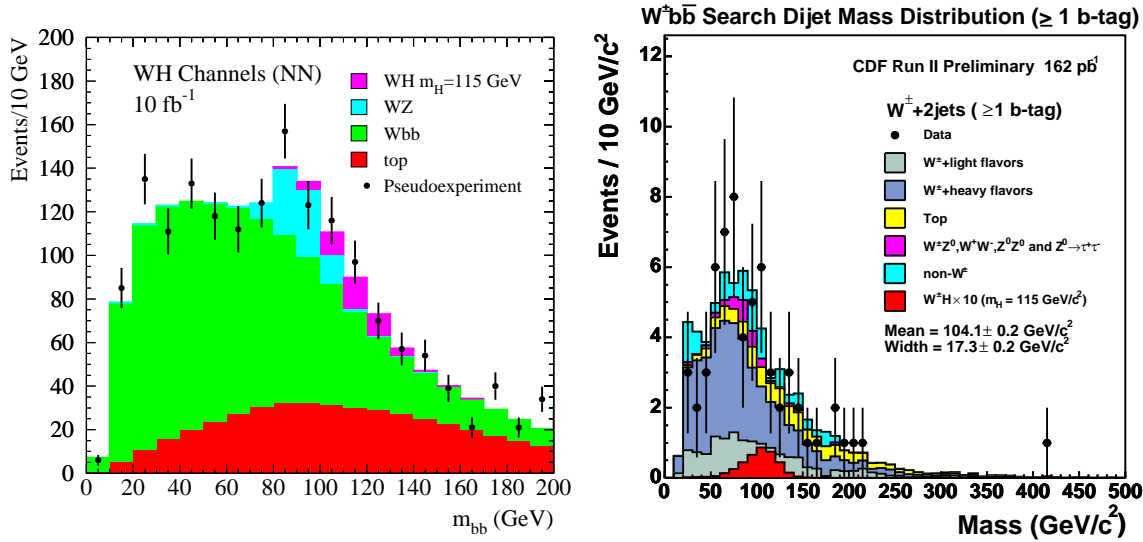,width=0.49\textwidth}
\end{center}
\caption{\it Invariant \bb\ mass spectrum in search for WH production in full simulation 
of Tevatron data corresponding to an integrated luminosity of 10~$fb^{-1}$ (left, from
Ref.~\protect\refcite{tev-hss}) and in CDF Run~II
data corresponding to an integrated luminosity of 162~$pb^{-1}$ in
comparison with signal and background expectation (right, from
Ref.~\protect\refcite{tev-wh}).} 
\label{f:tev-wh}
\end{figure}

A Higgs signal is then searched for as an excess in the \bb\ mass
spectrum, as shown in Fig.~\ref{f:tev-wh}(left) 
for full simulation of Tevatron data corresponding to an integrated luminosity of 10\fbs.
Given the Higgs event yields (3 $WH$ events selected per \fbinv), this 
requires precise knowledge of the backgrounds over the entire mass
range. While the normalization of the background can be obtained from
a fit outside of the signal region, the shape and relative normalization 
of the \bb\ mass spectrum of the various background components has to be 
known to allow extrapolation below the Higgs peak. Procedures to
obtain this information from data are outlined in
Ref.~\refcite{tev-hss} and typically involve a measurement of the
shape of the dijet mass spectrum in background-enriched samples, which
is then extrapolated to the final signal sample using a mixture of
Monte Carlo and data-driven methods.

\begin{figure}
\begin{center}
\epsfig{file=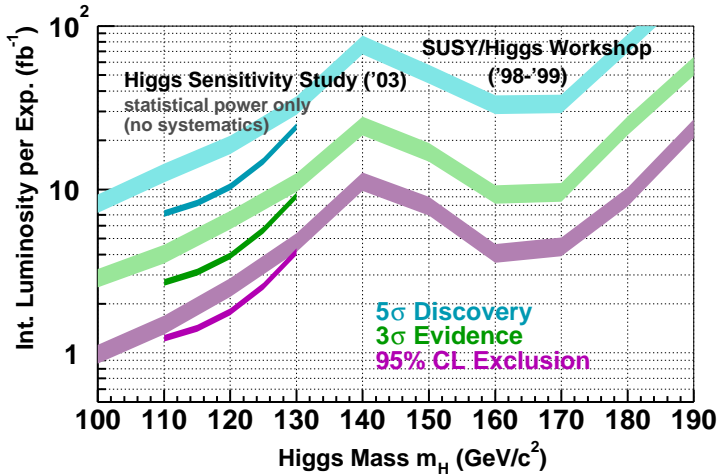,width=0.65\textwidth}
\end{center}
\caption{\it 
The integrated luminosity required per experiment for a 95\% C.L.
exclusion, a 3$\sigma$ and a 5$\sigma$  
discovery of a Standard Model Higgs boson at the Tevatron as a function of the 
Higgs boson mass. Thin lines show recent estimates based
on full simulation verified with Run~II data\protect\cite{tev-hss}
for the combination of searches for associated production with a
vector boson with no systematic uncertainties included.
Thick lines indicate results of an earlier study\protect\cite{tev-report} with fast
simulation which includes the combination with searches for $gg\to H\to WW$ (see
Section~\protect\ref{s:tev_hww}). 
For the latter, the line thickness indicates the impact of systematic uncertainties.} 
\label{f:tev-higgs}
\end{figure}

The \bb\ mass spectrum as measured in Run~II data
corresponding to an integrated luminosity of 162~\pbinv\ is shown 
in Fig.~\ref{f:tev-wh}(right) after all cuts. In this version of the
analysis only one jet is required to be b-tagged.
No evidence for $WH$ production is observed in current Run~II searches
by CDF and D\O, allowing to set an upper limit on the product of
cross section and branching fraction
$\sigma$($WH$)$\times$BR($H\to$\bb) of 5~pb for a Higgs boson mass of
120~\Gcs.\cite{tev-wh} 
Due to the small amount of integrated luminosity that has been
collected so far, the limit is still more than an order of magnitude
higher than the Standard Model expectation. 
In Fig.~\ref{f:tev-higgs} the luminosity required to observe (or exclude) 
a Standard Model Higgs boson is shown as a function of mass.
This estimate 
assumes a 30\% improvement in jet energy resolution 
and anticipates a number of enhancements to the current analysis, including the use
of forward b-tagging and multivariate methods as well as an increase
in lepton acceptance by extending the analysis to the forward region
and using isolated tracks to identify leptons. Systematic errors have
not been taken into account.  
After combining all channels and both experiments, a sensitivity at
the 95\% C.L. for m$_H$ = 120~\Gcs\ is expected to be achieved with an
integrated luminosity of 1.8~\fbinv\ per experiment. 
Evidence for a signal at 
the 3$\sigma$ (5$\sigma$) level will require 4~\fbinv\ (10~\fbinv) per
experiment for the same Higgs boson mass. 

\subsection{Search for $H\to WW$ \label{s:tev_hww}}

Higgs boson decays into two $W$ bosons are the dominant decay mode for
masses above 140~\Gcs.   
The relatively clean signature of two leptonic $W$~decays 
allows a search for this decay in the gluon-fusion channel. While the
production cross section in this channel is higher compared to the
associated production, the suppression due to the branching fractions of
the leptonic $W$~decays limits the event yield to only about 4 events per
\fbinv\ for $H \to WW \to ee,e\mu,\mu\mu + \met$ with $m_H$ = 160~\Gcs. 

\begin{figure}
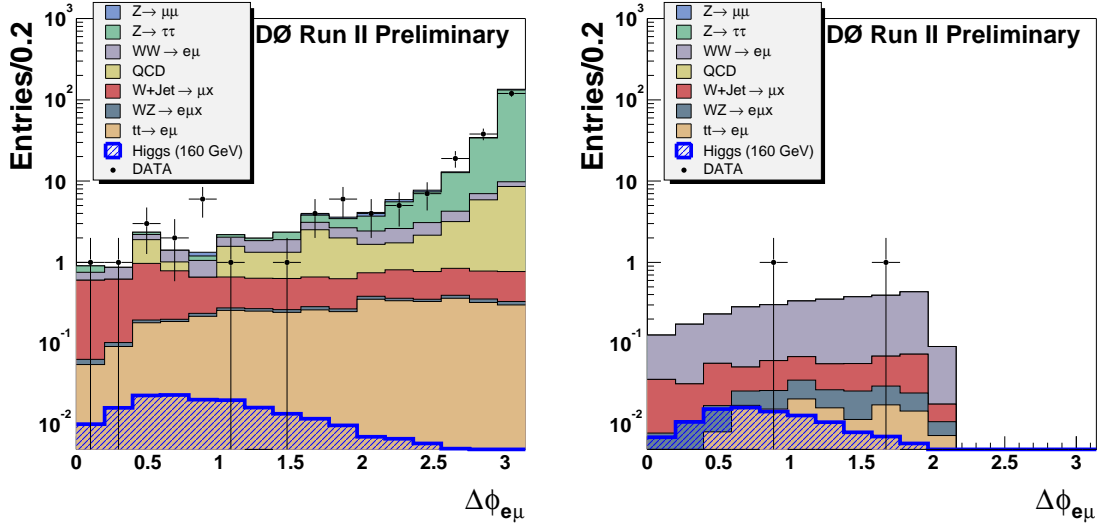

\begin{center}
\epsfig{file=./plots/H03F02a.epsi,width=0.45\textwidth,clip=true} \hspace{0.5cm}
\epsfig{file=./plots/H03F02b.epsi,width=0.45\textwidth,clip=true}
\end{center}
\caption{\it 
Difference in azimuthal angle of the reconstructed electron and muon
at an early stage of the selection (left) and after all cuts (right)
in the search for H$\to$WW$\to$e$\nu\mu\nu$ in D\O\ Run~II data 
corresponding to an integrated luminosity of 176~$pb^{-1}$
(from Ref.~\protect\refcite{tev-hww}).
} 
\label{f:tev-hww-dphi}
\end{figure}

\begin{figure}
\begin{center}
\mbox{\epsfig{file=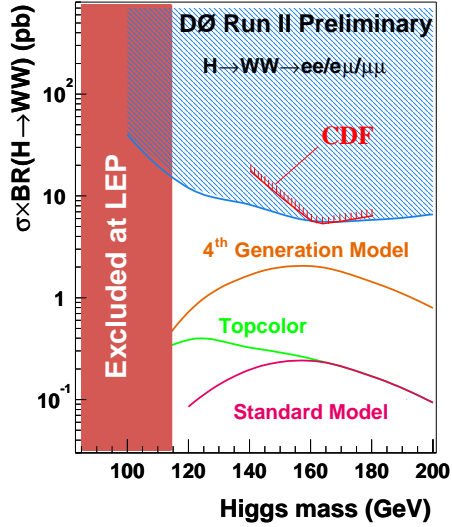,width=6.0cm}}
\end{center}
\caption{\it 
Upper limit (at 95\% C.L.) on the cross section $\sigma \cdot BR(H\to
WW)$ set by the D\O\ Collaboration using Run~II data corresponding to
an integrated luminosity of 176~$pb^{-1}$, with the limit set by the
CDF Collaboration (184~$pb^{-1}$) superimposed (from
Ref.~\protect\refcite{tev-hww}). The limits are compared to
expectations from Standard Model Higgs production and alternative models.} 
\label{f:tev-hww}
\end{figure}

Both Tevatron collaborations have started analyzing their Run~II data
in search for a $H\to WW$ 
signal.\cite{tev-hww} So far, efficiencies of up to 15--20\% have been achieved for the 
dilepton plus \met\ final states with electrons or muons. The background
is dominated by $WW$ production, which remains after selection cuts with
a cross section of about 25~fb for the sum of all three analysis channels.
Further separation of signal and $WW$ events is
possible using the difference in azimuthal angle~$\Delta\phi$ between the two charged
leptons.\cite{dittmar}
Due to spin correlations, $\Delta\phi$ tends to be small for
decays of a spin-0 resonance (see Fig.~\ref{f:tev-hww-dphi} and
Section \ref{s:lhc-vbf}).  
Both CDF and D\O\ observe no significant excess of events in Run~II
data corresponding to an integrated luminosity of 184~\pbinv\ and 
176~\pbinv,
respectively. Limits on the production 
cross section of $H\to WW$ 
have been set as a function of the Higgs boson mass as shown in
Fig.~\ref{f:tev-hww}. 
For a mass of 160~\Gcs, cross sections larger
than 5.6~pb have been excluded at 95\% C.L., which is still more than
an order of magnitude higher than the expectation within the Standard
Model. 

The performance of these analyses is consistent with the
expectations based on the fast simulation, which projected a total
background of 30.4~fb at an efficiency of 18.5\% for a Higgs boson
mass of 150~\Gcs.\cite{tev-report} Based on this projection,
sensitivity at the 95\% C.L. to a Standard Model Higgs boson with
masses between 160 and 170~\Gcs\ will be reached with an integrated
luminosity of 4~\fbinv\ per experiment (10~\fbinv\ for a 3$\sigma$
sensitivity), as shown in Fig.~\ref{f:tev-higgs}. These estimates
include contributions from the vector boson fusion channel and Higgs
boson production in association with vector bosons.
The latter is best searched for using a like-sign dilepton
selection.\cite{tev-www-pheno,tev-report}
First results in this channel have been reported by the CDF collaboration, 
but are not yet competitive with the gluon-fusion channel.\cite{tev-www}

In models beyond the Standard Model, the rate of $H\to WW$ events can be 
enhanced due to larger production cross sections (models with heavy 4th 
generation quarks) or due to an increase in branching fraction 
(Topcolor models\cite{topcolor}). 
In the former case, the gluon-fusion process is enhanced due to 
loop-diagrams involving heavy quarks by a factor of 
about 8.5 within the mass range of interest at the Tevatron, with only 
a mild dependence on the heavy quark mass.\cite{4gen}
The latter class of models also predicts an enhanced branching fraction 
for $H\to\gamma\gamma$, which can be searched for with diphoton analyses 
to increase the Tevatron sensitivity at low Higgs boson masses.\cite{tev-report}
First Run~II results from D\O\ in this channel exist, but improve only
marginally on existing limits from LEP and Run~I.\cite{tev-diphoton}

\subsection{Neutral Higgs bosons in supersymmetry}
Sensitivity to a low-mass Higgs boson is of particular interest within
supersymmetric extensions of the Standard Model, which predict the
existence of at least one neutral Higgs boson~\hs = $h$, $H$, $A$ with a mass below
$\sim$135~\Gcs. Searches for the Standard Model Higgs boson produced in
association with a vector boson (cf. Section~\ref{s:tev-vh}) can be
interpreted within SUSY parameter space.\cite{tev-report}
In addition, the enhancement of the Higgs coupling to \bb\ at large \tb\ 
results in sizeable cross sections for two search channels that are
inaccessible within the Standard Model: the production of Higgs bosons in
association with one or more $b$ quarks\cite{willenbrock-bbbar} as well as the 
gluon-fusion channel $gg\to \hs$ with the subsequent 
decay $\hs\to\tau\tau$. 

\subsubsection{$\hs b(b)\to bbb(b)$}
The D\O\ collaboration has analyzed a Run~II dataset 
corresponding to an integrated luminosity of 131~\pbinv, 
collected with multijet triggers optimized for the $\hs\bb\to\bb\bb$
signal.\cite{tev-bbh-d0} 
Requiring two jets with transverse momenta~\pt$>$25~\gevp\ and a third
jet with \pt$>$15~\gevp, this trigger consumed less than 4~Hz of
Level-3 bandwidth at instantaneous luminosities of \mbox{4.0 $\cdot$
10$^{31}$ $\rm{cm}^{-2}\rm{s}^{-1}$}, while 
maintaining a signal efficiency of about 70\% after offline cuts.
The offline analysis
requires at least three b-tagged jets with \pt$>$15~\gevp. Depending on
the Higgs mass hypothesis, the \pt\ cuts for the two leading jets are
tightened to values between 35 and 60~\gevp\ to optimize for best
expected sensitivity.  

\begin{figure}
\begin{center}
\epsfig{file=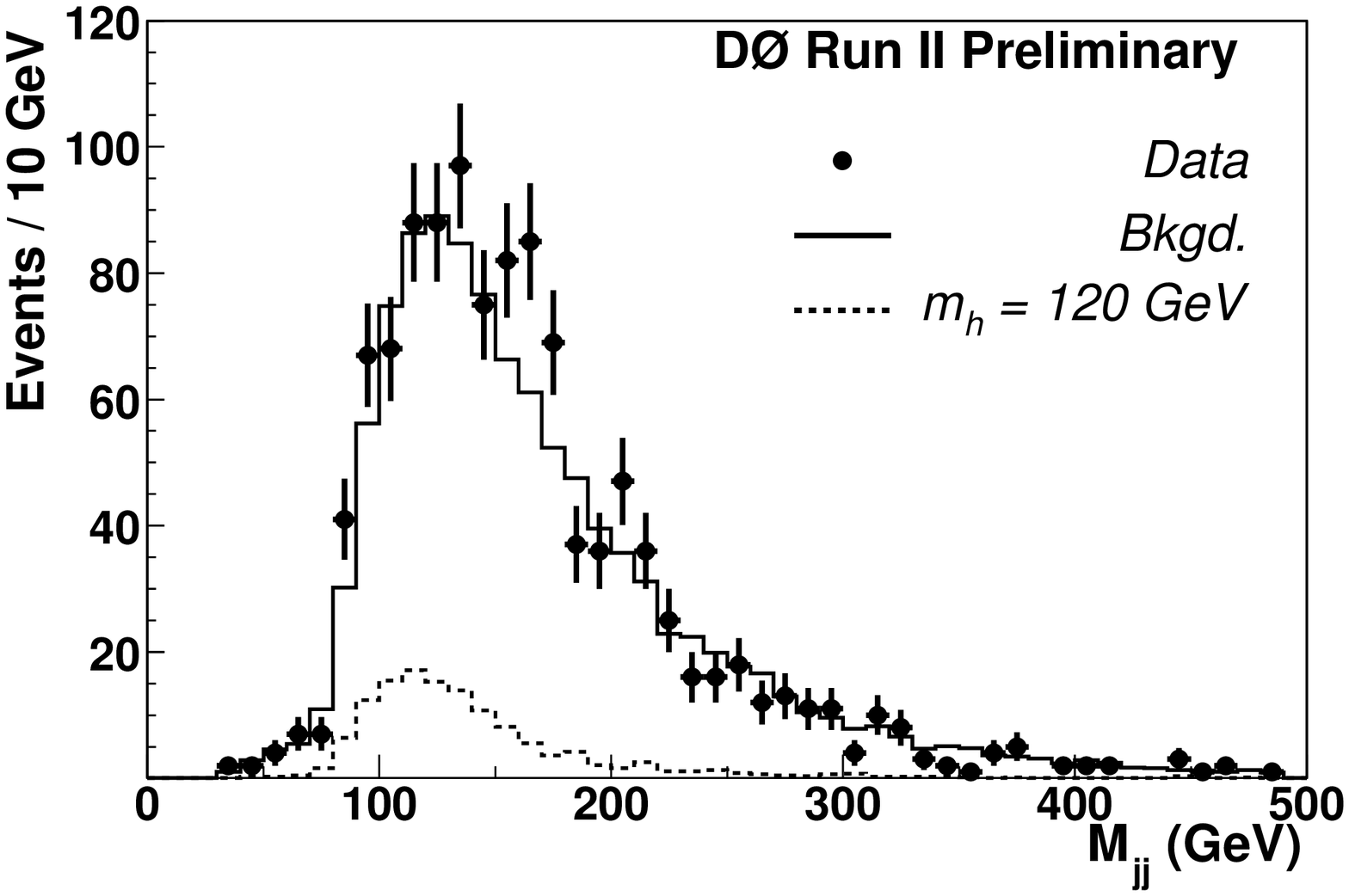,width=0.49\textwidth}
\epsfig{file=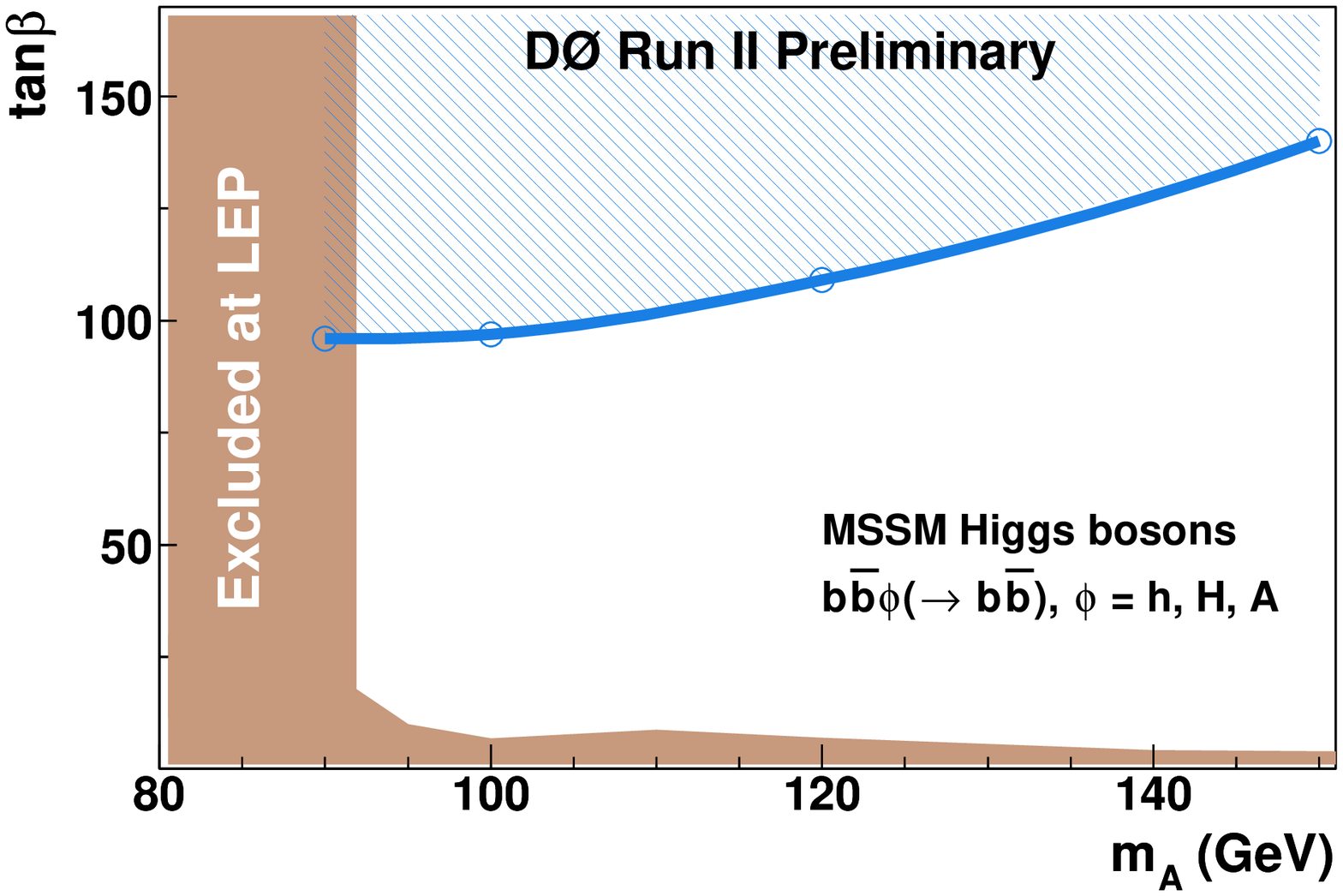,width=0.49\textwidth}
\end{center}
\caption{\it 
(Left) Invariant mass spectrum of the two leading jets in the 3-jet sample
with three b-tags after final cuts in the D\O\ search for bh
production using Run~II data corresponding to an
integrated luminosity of 131~$pb^{-1}$; 
the dashed histogram shows the distribution for a Higgs signal with
$m_H$ = 120~\Gcs and \tanbit=100;
(Right) Regions in
($m_A$,tan$\beta$) excluded by this analysis at 95\% C.L. in
comparison with LEP limits (from Ref.~\protect\refcite{tev-bbh-d0}). 
}
\label{f:tev-bbh}
\end{figure}

The background at this stage is dominated by multijet production with
$b$ quarks. The signal can be searched for as a
peak in the invariant mass spectrum of the two leading 
jets, which is shown in Fig.~\ref{f:tev-bbh} in comparison with the
expectation from background and a Higgs signal with m$_\hs$=120~\Gcs. 
The shape of the dijet mass spectrum in background is obtained from a
multijet sample with two b-tagged jets, which is expected to have
negligible contamination from signal, by weighting events using b-tag
fake rates measured in data as a function of jet \pt\ and~$\eta$.  
The background is then normalized by fitting this shape to the mass
spectrum outside the signal region. 

No evidence for production of neutral Higgs bosons~$h$, $H$, $A$ in
association with b-jets has been observed, and limits on the
production cross section have been set at 95\% C.L.
This limit has been translated into
an exclusion region in the \matb-plane under the assumption that the
production cross section is proportional to $\tan^2\beta$.
However, it should be noted that in the large \tanb\
region, higher order corrections to the bottom-Yukawa coupling are large.
The resulting limit is shown in Fig.~\ref{f:tev-bbh} in comparison
with existing limits from LEP.
The D\O\ Run~II limit is significantly worse than the 
limit published in 2001 by the CDF collaboration, which was based on
the analysis of Run~I data corresponding to an
integrated luminosity of 91~\pbinv.\cite{tev-bbh-cdf}
Detailed comparisons of both results indicate that the apparent
loss of sensitivity observed by D\O\ 
can be traced back to the use of more recent cross section
calculations and fits of parton distribution functions, 
which cause a significant reduction in signal acceptance and cross section compared 
to the CDF Run~I analysis.\cite{tev-bbh-comparison}

For an integrated luminosity of 5~\fbinv\ and after combining results
from both Tevatron experiments, the reach in \tb\ within the
$m_h$-max\ scenario (see Section~\ref{s:mssm-benchmarks})
will be extended down to about \tb = 25 for m$_A$ = 120~\Gcs\ at the 95\% C.L.,
but deteriorates quickly with increasing m$_A$.  

\subsubsection{$\hs\to\tau\tau$}
In addition to the dominant decay mode 
$\hs\to\bb$, a light supersymmetric
Higgs boson can be searched for in its decay to $\tautau$.
This decay mode is of particular interest both for SUSY scenarios that favour
suppressed couplings of Higgs bosons to $b$ quarks as well as for the large
\tb\ region, where the channels $\hs b(b)\to \tau\tau b(b)$ and
$gg\to \hs\to\tau\tau$ provide a viable complement to the search for
$\hs b(b)\to bbb(b)$.

Both
Tevatron experiments have demonstrated the ability to reconstruct
hadronic tau decays in Run~II data by measuring the $Z\to\tau\tau$
cross section.\cite{tev-ztautau}
The CDF collaboration has analyzed Run~II data corresponding to an
integrated luminosity of 195~\pbinv\ in search
for $gg\to \hs\to\tau\tau$ with one tau decaying leptonically to
electron or muon and the other tau decaying into
hadrons.\cite{tev-Htautau-cdf} 
The hadronic tau decay is reconstructed as one or more tracks pointing
to a narrow energy deposition in the calorimeter.  
Background from jets misreconstructed
as tau objects is further suppressed using cuts on track multiplicity,
mass and isolation of the tau candidate. 
The selection then 
requires one such tau candidate in addition to an isolated electron or muon.
After topological cuts using the
transverse momenta of the lepton and the 
hadronic tau candidate as well as the transverse missing energy, the sample is
dominated by irreducible background from $Z\to\tau \tau$ with a fraction
of about 90\%. Events from $\hs\to\tau\tau$ are selected with an  
efficiency, including branching fractions, of about 0.8\% (0.6\%) in
the electron (muon) channel for $m_{\hs}$ = 130~\Gcs.

\begin{figure}
\begin{center}
\epsfig{file=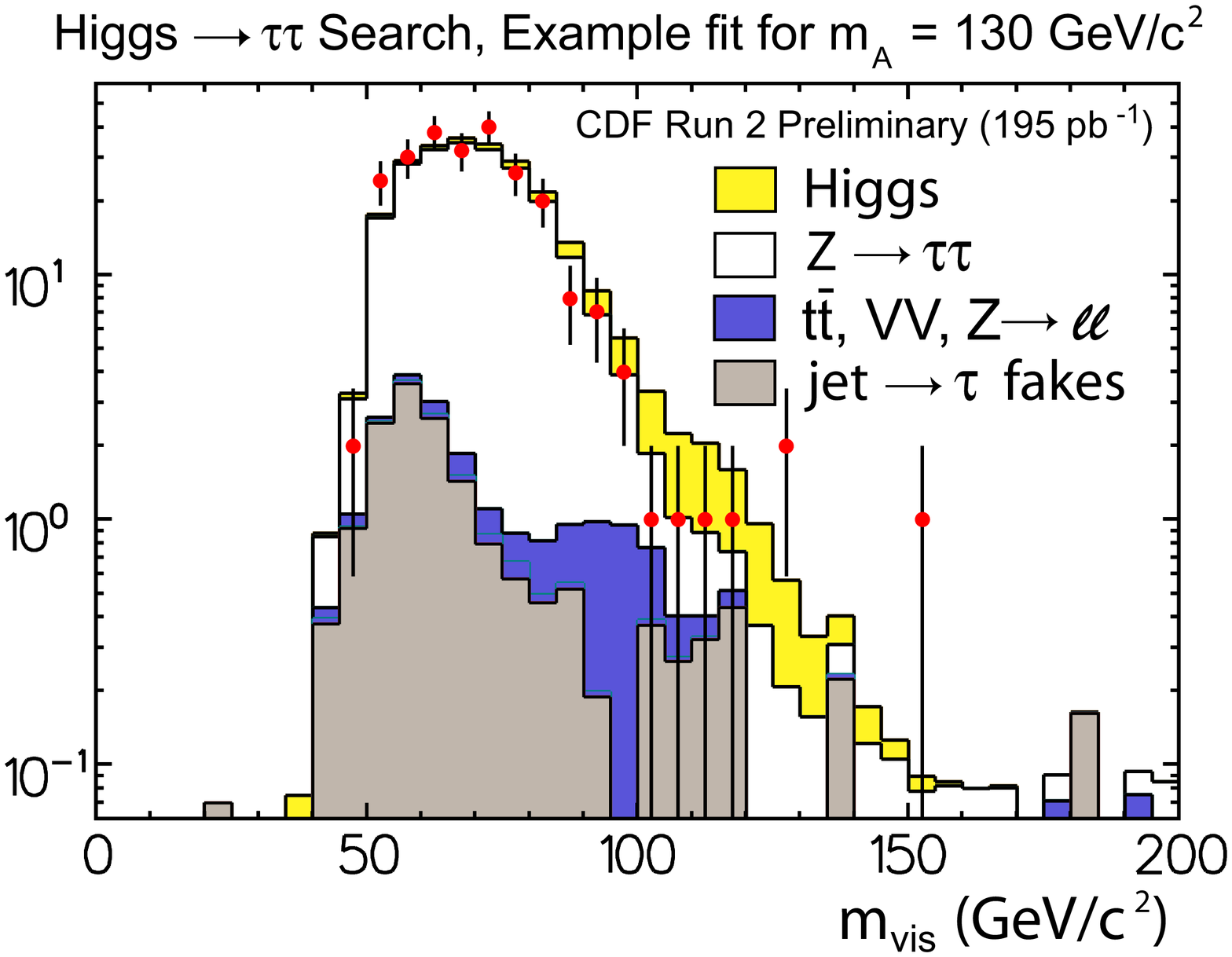,width=0.47\textwidth}
\epsfig{file=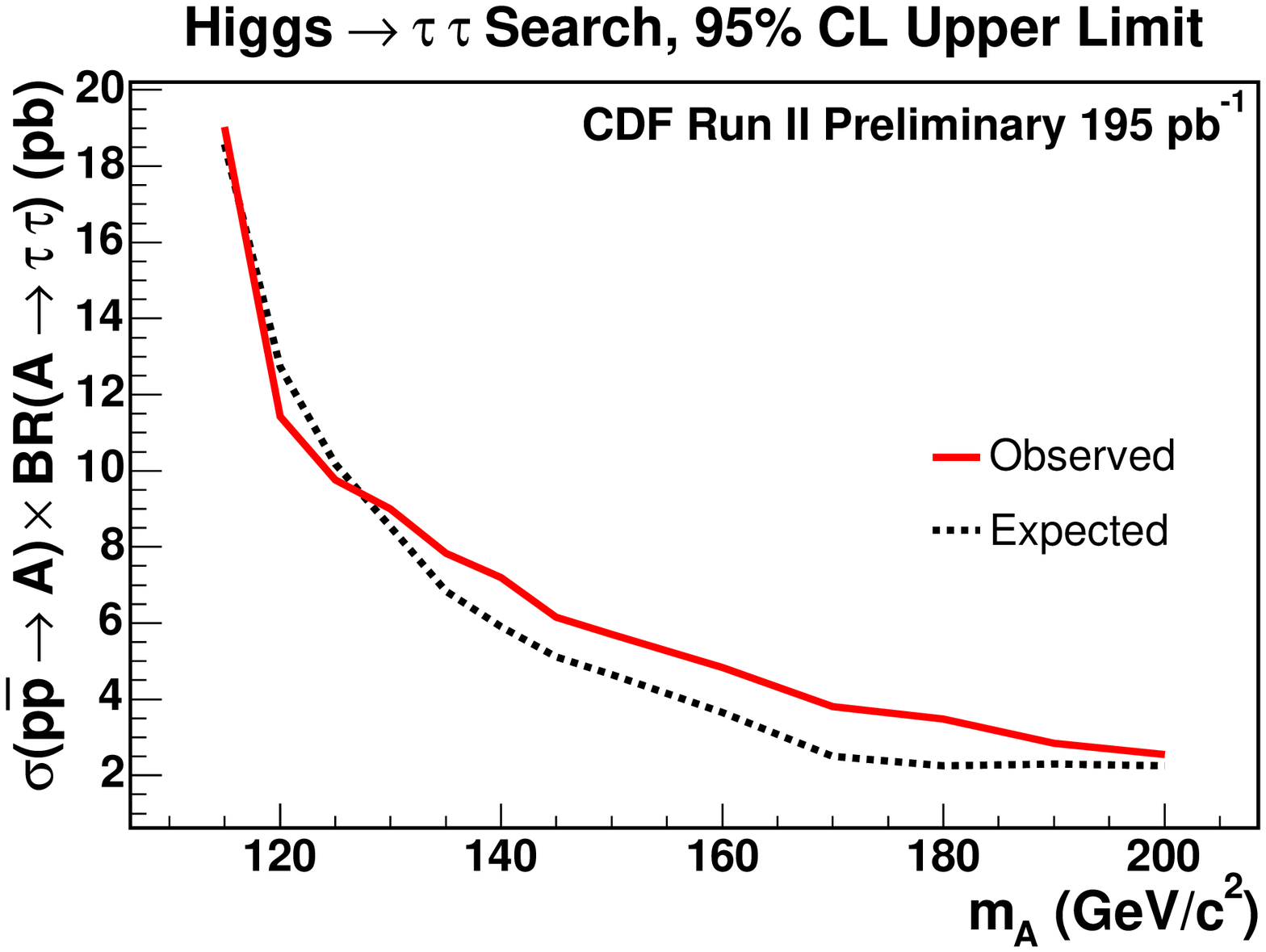,width=0.51\textwidth}
\end{center}
\caption{\it 
(Left) Distribution of the visible mass after all cuts of the CDF search
for $\hs\to\tau \tau$ in 195~$pb^{-1}$ of Run~II data and for a Higgs
boson signal with $m_\hs$=130~GeV/$c^2$ on top of the Standard Model
background; (Right) Upper limit (at 95\% C.L.) on 
the cross section $\sigma \cdot BR(\hs\to \tau\tau)$  in comparison with
the expected limit (from Ref.~\protect\refcite{tev-Htautau-cdf}).  
}
\label{f:tev-htautau}
\end{figure}

Separation of signal events from the $Z\to\tau \tau$ background is possible
by reconstructing an event mass~$m_{vis}$  using the momentum vectors
of the lepton and tau candidate as well as the missing transverse energy vector.
In Figure~\ref{f:tev-htautau}(left) 
the distribution of $m_{vis}$ is shown for data, backgrounds and a potential Higgs signal.
No evidence for an excess of events with respect to the Standard Model prediction
has been observed.
Using a binned likelihood fit of this distribution, a limit on the production
cross section of $\hs\to \tau\tau$ has been extracted as displayed in
Fig.~\ref{f:tev-htautau}(right) as a function of the Higgs boson mass.

\subsubsection{Combined reach}
\begin{figure}
\begin{center}
\epsfig{file=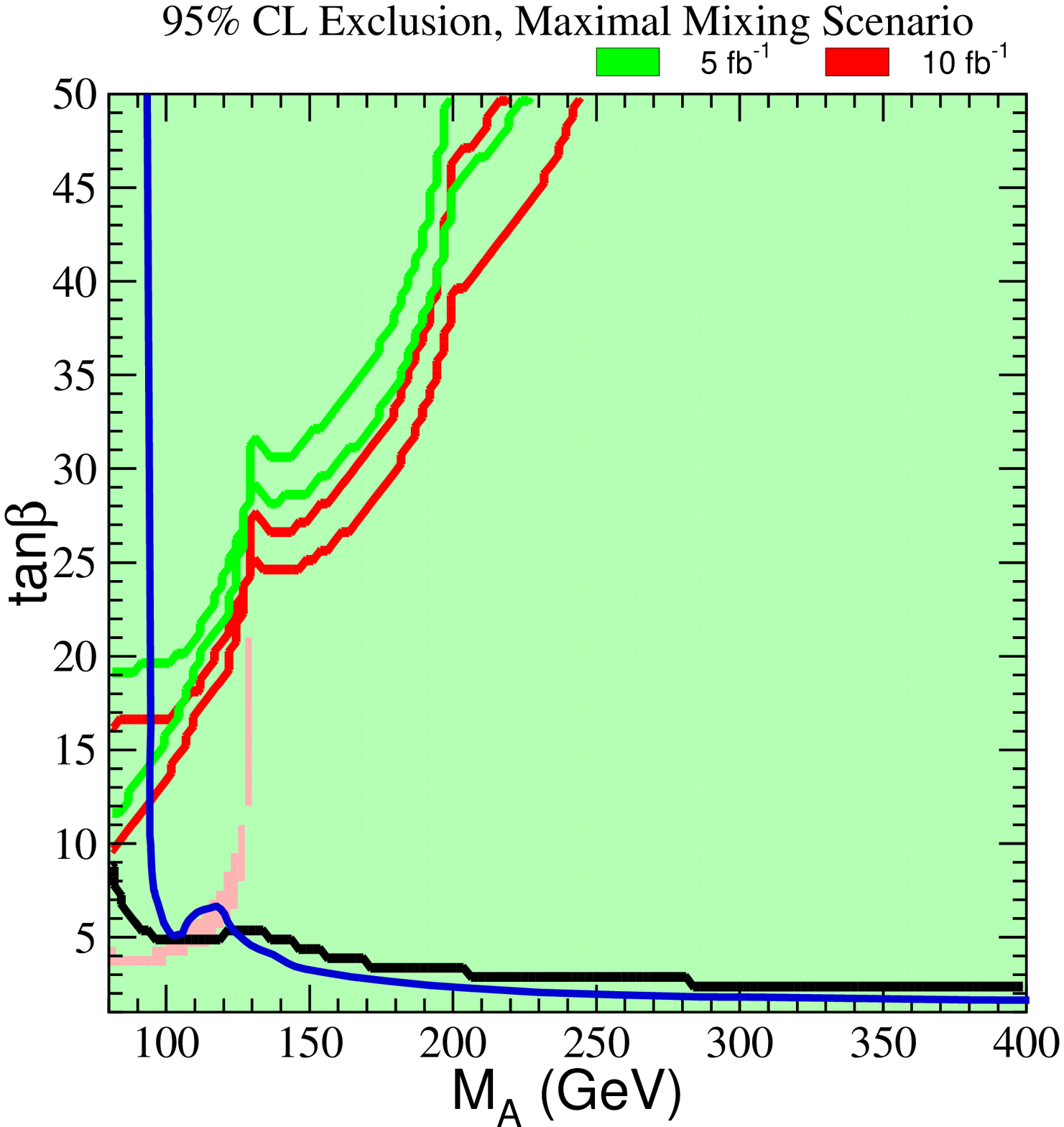,width=0.46\textwidth}
\epsfig{file=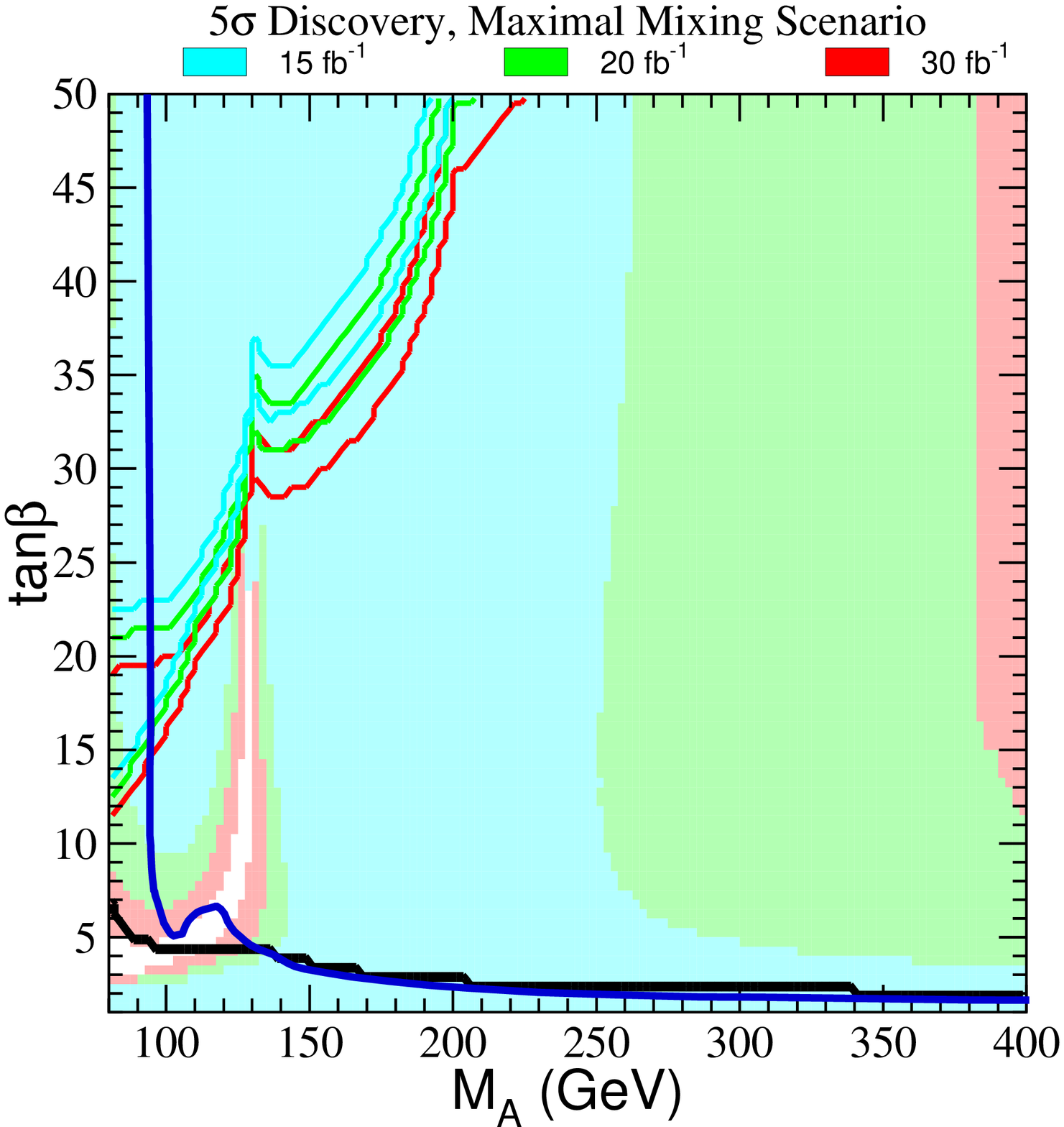,width=0.46\textwidth}
\end{center}
\caption{\it 
Luminosity required for exclusion at 95\% C.L. (left) or 5$\sigma$
discovery (right) of a SUSY Higgs boson as a function of m$_{\it A}$ and
\tanbit\ within the $m_{\it h}$-max\ scenario (taken from
Ref.~\protect\refcite{tev-report} and modified to include the most recent
LEP2 limit\protect\cite{lep-mssm}). Shaded regions indicate the
reach of $WH$ and $ZH$ searches, the region above the diagonal lines are
accessible to searches for $\hs b(b)$, the dark line indicates the LEP2 limit.} 
\label{f:tev-susyreach}
\end{figure}

Combining dedicated searches for Higgs bosons at large \tb\ with 
searches for production of Higgs bosons in association with vector
bosons, sensitivity at 95\% C.L. to MSSM Higgs bosons within the 
$m_h$-max\  scenario (see Section~\ref{s:mssm-benchmarks}) can be
achieved independent of \tb, 
as shown in Fig.~\ref{f:tev-susyreach}.\cite{tev-report}
However, within this challenging scenario, a 5$\sigma$ discovery will 
not be possible at Tevatron Run~II for most of the \matb-plane.

\subsection{Charged Higgs bosons}
Models with an extended Higgs sector predict charged Higgs
bosons~$H^\pm$ or, in the case of additional 
Higgs triplets, also
doubly-charged Higgs bosons~$H^{\pm\pm}$. For masses smaller than 
$m_t$-$m_b$, singly-charged Higgs bosons can be produced in top quark
decays. 
In this case, charged Higgs bosons can be searched for either as an
excess of events with $t\to H^+b$ or as a decrease in top-quark
branching fraction for $t\to W^+b$. Limits on the mass of charged
Higgs bosons as a function of \tb\ have been set at Run~I using both
strategies.\cite{tev-hplus} 
The reach of the Run~I analyses has been projected to
Run~II employing a parametrized detector simulation: from measuring
the branching fraction of leptonic top quark decays with 2~\fbinv\ of
data, charged Higgs boson masses of 80-140 \Gcs\ can be probed at 95\%
C.L. for \tb\ from 20-30, respectively. 
Within supersymmetric models, this allows a test of the large \tb\
region complementary to the search for $\hs b(b)$ and $\hs\to\tau\tau$.

\begin{figure}
\begin{center}
\epsfig{file=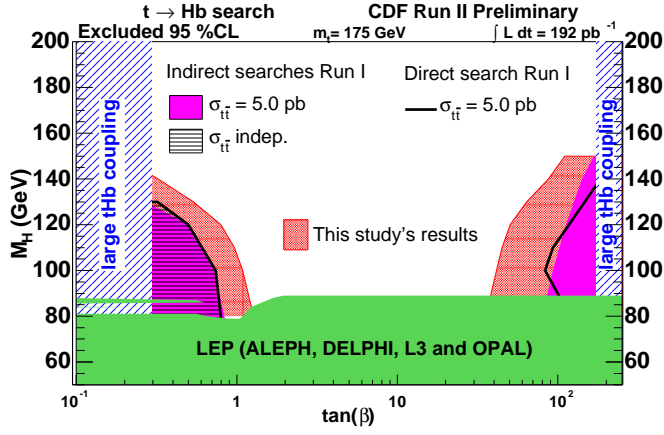,width=0.60\textwidth}
\end{center}
\caption{\it 
Regions in the MSSM ($m_{H^{\pm}}$,tan$\beta$)-plane excluded by the
CDF Run~II \ttbar\ 
cross section measurements, in comparison with regions excluded at LEP
and Run~I; Tevatron results do not apply for non-perturbative $Htb$
couplings at small and large \tanb\ (from Ref.~\protect\refcite{tev-hplus-run2}).}
\label{f:tev-hpm}
\end{figure}

Preliminary Run~II results of \ttbar\ cross-section measurements by
the CDF collaboration have been interpreted to obtain limits on the
branching fraction of top quark decays into charged Higgs bosons.\cite{tev-hplus-run2}
This analysis is based on cross-section measurements in the dilepton,
lepton plus jets and lepton plus hadronic-tau channel using Run~II data
corresponding to an integrated luminosity of 192~\pbinv. Assuming that
charged Higgs bosons decay only into $cs$, $\tau\nu$ or $Wb\bar{b}$, limits in
the ($m_{H^{\pm}}$,tan$\beta$)-plane have been derived at tree level
within the MSSM, as shown in Fig.~\ref{f:tev-hpm}.

Production of doubly charged Higgs bosons can provide particularly
striking signatures in left-right symmetric models,\cite{lr-models} where 
BR($H^{\pm\pm}\to
\ell^\pm \ell^\pm$) is expected to be 100\%. 
The CDF (D\O) collaborations have analyzed Run~II data corresponding to an
integrated luminosity of 240~\pbinv\ (107~\pbinv) to search for 
$H^{\pm\pm}$ production in like-sign dilepton events.\cite{tev-hpp} Requiring
two acoplanar, isolated electrons or muons, no excess of events has been
observed at high dilepton masses. For left-handed (right-handed)
$H^{\pm\pm}$, the CDF collaboration sets a lower mass 
limit of 135~\Gcs\ (112~\Gcs) for BR($H^{\pm\pm}\to \ell^\pm \ell^\pm$) = 1. 

The searches described above loose sensitivity once
the leptons are produced with significant impact parameter due to a
non-negligible lifetime of the Higgs boson. Since
the lifetime of doubly charged Higgs bosons depends on the unknown
Yukawa coupling, scenarios with large $H^{\pm\pm}$ decay lengths have
to be considered  
as well. The CDF collaboration has searched for long-lived doubly charged Higgs bosons that 
decay outside the detector in Run~II data corresponding to an
integrated luminosity of 206~\pbinv.\cite{tev-hpp-stable}
Experimentally, these events can be identified requiring two isolated high-\PT\
tracks associated with hits
in the muon detectors and large energy loss in the tracking detectors.
Efficiencies of about 30\% have been achieved with negligible
backgrounds from Standard Model sources. No events have been observed
in the data, which translates 
into a lower limit on the $H^{\pm\pm}$ mass of 134~\Gcs.

\subsection{Summary of Higgs searches at the Tevatron}
With an integrated Run~II luminosity of the order of 8~\fbinv, both
Tevatron experiments combined are expected to reach 
sensitivity to the production of Standard Model Higgs 
bosons up to masses of 180~\Gcs\ at 95\% C.L. Within the context of 
Supersymmetry, most of the parameter space can be tested at 95\%
C.L. by a combination of searches for Higgs bosons produced in
association with vector bosons or $b$ quarks. 

These estimates are based on analyses employing full detector
simulation and reconstruction that has been tuned to Run~II data.
Preliminary results from searches using Run~II data 
corresponding to an integrated luminosity of about 200~\pbinv\ 
are available in the most important channels.
The performance of the analyses is consistent with the 
projections, even though a number of assumptions about future improvements to 
reconstruction algorithms still remain to be verified. 
Due to the small amount of integrated luminosity that has been collected so far,
the current results are not yet sensitive 
to Higgs boson production.
Even with the full Run~II luminosity and after combining all channels
and both experiments, a 5$\sigma$ 
discovery will most likely not be possible at the Tevatron, leaving
it to the LHC to explore and discover Higgs bosons over
the full parameter range of both Standard Model and Supersymmetry.

\pagebreak

\section{The Search for a Standard Model Higgs Boson at the LHC \label{s:higgs-lhc}}

The Standard Model Higgs boson is searched for at the LHC in various decay 
channels, the choice of which is given by the signal rates and the 
signal-to-background ratios in the different mass regions. 
The search strategies and background rejection methods have been established 
in many studies over the past ten years.\cite{physics-tdr,cms-higgs}
Originally, inclusive final states 
have been considered, among them the 
well established \hgg\ and $H \to Z Z^{(*)} \to \ell \ell \ \ell \ell$
decay channels. More exclusive channels have been considered in the low mass 
region by searching for Higgs boson decays in \bbbar\ or $\gamma \gamma$ 
in association with a lepton from a decay of an accompanying $W$ or $Z$ boson 
or a top quark. 
The search for a Standard Model 
Higgs boson in the intermediate mass region using the vector boson
fusion mode and exploiting forward jet tagging, which had 
been proposed in the literature several years ago,\cite{zeppenfeld,zeppenfeld-tau,zeppenfeld-ww} 
has meanwhile been studied by the experimental collaborations.
In the following, a brief summary of the Standard Model Higgs boson 
discovery potential at the LHC is given. After a discussion of the 
inclusive analyses, more exclusive final states are discussed, ordered 
according to the different production processes. 

Despite the enormous progress in the calculation of higher order 
QCD corrections over the past few years (s. Section \ref{s:higgs-prod}), 
the LHC physics performance generally has been evaluated by using 
Born-level predictions for both signals and backgrounds.
Since the higher order QCD corrections (K-factors) are not known 
yet for all background processes this approach was considered to be 
more consistent and conservative. In the following K-factors are ignored, 
unless otherwise stated. 

The non-diffractive inelastic proton-proton cross section has been assumed to be
70~mb. This leads on average to a superposition of 2.3 or 23 
minimum bias events on top of the hard collision at low or high 
luminosity respectively. These so called  
pile-up contributions have been included for both low and high luminosity.

Physics processes have mainly been simulated with the PYTHIA Monte 
Carlo program,\cite{pythia} 
including initial- and final-state radiation, hadronisation and 
decays. Although many results have been obtained 
using a fast simulation, 
all key performance characteristics have been evaluated with a full
GEANT\cite{geant} simulation, both at low and high luminosity.\cite{physics-tdr,cms-higgs}

\subsection{Inclusive Higgs boson searches \label{s:higgs-incl}}

Several important channels for Higgs boson discovery at the LHC 
have been discussed extensively in the literature. 
Among those channels are the \hgg\ decay mode, the gold plated decay channel 
\hzzsfourl\ as well as the decay channel \hwwsll. 
If no additional particles except the Higgs boson decay products are searched 
for, the production via gluon fusion provides 
the largest contribution to the signal event yields.

\subsubsection{\hgg\ decays}
\begin{figure}[hbtn]
\begin{center}
\mbox{\epsfig{file=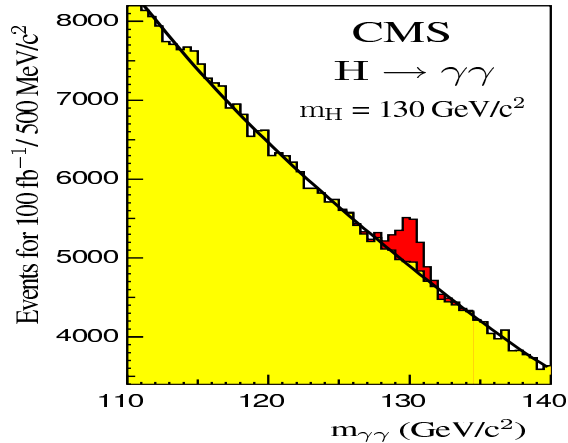,height=6.0cm}}
\caption{\small \it 
Reconstructed $\gamma \gamma$ invariant mass distribution of a  
\hgg\ signal (dark) with m$_{\it H}$~=~130~\Gcsit\ and the background (light) for an integrated 
luminosity of 100~\fbsit\ in the CMS experiment (from Ref.~\protect\refcite{cms-higgs}).
} 
\label{f:h_gamgam}
\end{center}
\end{figure}

The decay $\hgg$ is a rare decay mode, which is only detectable in a limited
Higgs boson mass region between 80 and 150~\Gcs,
where both the production cross section and the decay
branching ratio are relatively large. 
Excellent energy and angular resolution are required to observe the
narrow mass peak above the irreducible prompt $\gamgam$ continuum.
In addition, there is a large reducible background resulting from direct photon 
production or from two-jet production via QCD processes. 
Using a realistic detector simulation, it has been 
demonstrated\cite{physics-tdr,cms-higgs} that the required rejection can be achieved and
that the residual jet-jet and $\gamma$-jet backgrounds
can be brought to the level of approximately 20\% of the
irreducible $\gamgam$ background over the mass range relevant 
to the $\hgg$ search. For an integrated luminosity of 100~\fbs, a
Standard Model Higgs boson in the  
mass range between 105 and 145~\Gcs\ can be observed in the \hgg\ channel in the 
ATLAS experiment with a significance of more than 5$\sigma$.\cite{physics-tdr} 
The discovery range in the CMS experiment is slightly larger, due to the better 
$\gamma \gamma$ mass resolution.\cite{cms-higgs} 
As an example of signal reconstruction above background, the expected
signal from a Higgs  
boson with a mass of 130~\Gcs\ in the CMS experiment is shown in Fig.~\ref{f:h_gamgam}, 
assuming an integrated luminosity of 100~\fbs.

The $\gamma \gamma$ decay mode has also been studied for the case where the 
Higgs boson is accompanied by a high-\PT\ jet.\cite{cms-ggjet,physics-tdr} 
A better signal-to-background ratio is found than in the inclusive case.
However, sizeable uncertainties on the background estimates from Monte Carlo calculations 
prevent a precise estimate of the discovery significance at present.

\subsubsection{$H \to Z Z^{(*)}$ decays}
The decay channel $H \to Z Z^* \to \ell \ell \ \ell \ell$ provides a rather 
clean signature in the intermediate mass region 115~\Gcs\ $< m_H < 2 \ m_Z$. 
In addition to the irreducible backgrounds from $ZZ^*$ and $Z \gamma^*$ production, 
there are large reducible backgrounds from \ttbar\ and $Z \bbbar$ production. 
Due to the large production cross section, the \ttbar\ events
dominate at production level, whereas the $Z \bbbar$ events contain a
genuine $Z$ boson in the final state and are therefore more difficult to reject.
In addition, there is background 
from $ZZ$ continuum production,
where one of the $Z$ bosons decays into a 
$\tau$ pair, with subsequent leptonic decays of the $\tau$ leptons, and the other 
$Z$ decays into an electron or muon pair. It has been shown that in both LHC 
experiments the reducible backgrounds can be suppressed well below the level of 
the irreducible background from  $ZZ^* \to  4 \ell $. Calorimeter and track isolation together 
with impact parameter measurements can be used to achieve the necessary 
background rejection.\cite{physics-tdr,cms-higgs}
Assuming an integrated luminosity of 30\,\fbs, the $H \to Z Z^* \to  4 \ell$
signal can be observed with a significance of more 
than 5$\sigma$ in the mass range 130 $< m_H <$ 180~\Gcs, except for a narrow region around 170~\Gcs,
where the branching ratio is suppressed due to the opening up of the $WW$ decay 
mode (see Fig.~\ref{f:br_sm}).

It has also been studied whether $Z$-decay modes involving $b$ quarks can be used to enhance 
the signal significance.\cite{weizmann}
To take advantage of the fact that the branching ratio of
$Z\to\bbbar$ is approximately five times larger than that of $Z\to \ell \ell$, a search for 
$ H \to Z Z^* \to \bbbar\ \ \ell^+ \ell^-$ has been considered, with
the on-shell $Z$ boson decaying into  
a \bbbar\ pair and the other decaying into a pair of electrons or muons.
Since the large backgrounds  from \ttbar\ and $Z \bbbar$ production contain a genuine pair 
of $b$ quarks, the discovery potential of a Higgs boson in that channel is marginal. Assuming 
an integrated luminosity of 30\,\fbs, a signal significance of only 2.7$\sigma$ has been 
estimated with a signal-to-background ratio of about 0.3 for a Higgs boson 
with $m_H$ = 150~\Gcs.

\begin{figure}[hbtn]
\begin{center}
\mbox{\epsfig{file=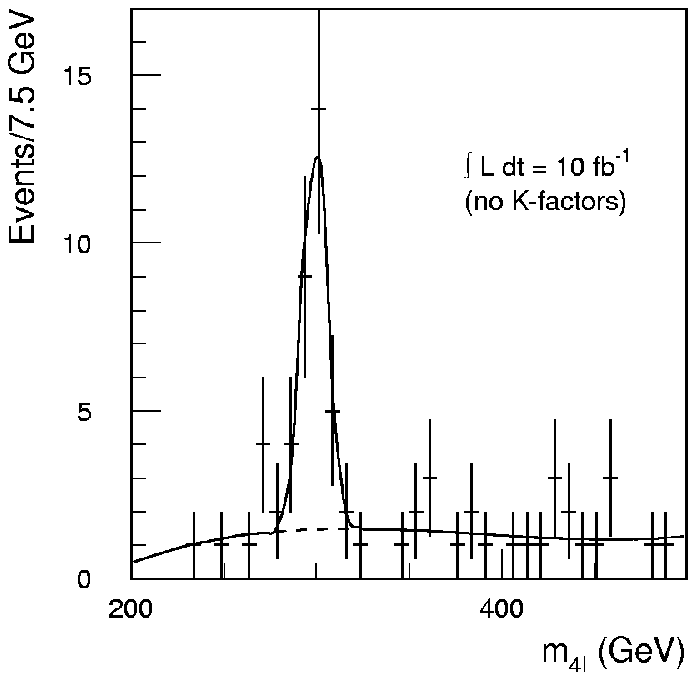,height=5.5cm}}
\hspace*{0.5cm}
\mbox{\epsfig{file=./plots/h_ww_atlas.epsi,height=5.5cm}}
\caption{\small \it 
(Left) Expected $H \to \ ZZ \to 4 \ell$ signal above the background 
for $m_H$ = 300~\Gcsit\ and for an integrated luminosity 
of 10~\fbsit\ in the ATLAS experiment (from Ref.~\protect\refcite{physics-tdr}).
(Right) Transverse mass distribution for the summed 
$H \to W W^* \to \ell \nu \ \ell \nu$ signal (m$_{\it H}$~=~150~\Gcsit) and total background, 
for an integrated luminosity of 30~\fbsit. The distribution for the background alone is 
also shown separately. The shaded histogram represents the contributions from 
the $Wt$ and \ttbar\ backgrounds. The dashed lines indicate the selected signal region
(from Ref.~\protect\refcite{physics-tdr}).}
\label{f:htowwzz}
\end{center}
\end{figure}

For Higgs boson masses in the range 
180~\Gcs\ $< m_H \lesssim$~700~\Gcs, the 
$H \to ZZ \to 4 \ell$ decay mode is the most reliable one
for the discovery of a Standard Model Higgs boson at the LHC. The expected background, which is dominated by the continuum production of $Z$ boson pairs, is smaller than the signal. In this mass range 
the Higgs boson width grows rapidly with increasing $m_H$, and
dominates over the experimental 
mass resolution for $m_H >$ 300~\Gcs. The momenta of the final-state leptons are high and their
measurement does not put severe requirements on the detector
performance. The 
$H \to ZZ \to  4 \ell$ signal would be
easily observable above the 
$ZZ \to \ 4 \ell$ continuum background after less than one year 
of low luminosity operation for 200 $< m_H <$ 600~\Gcs.\cite{mohn-stugu} 
It should be mentioned that the interference effect between the resonant signal 
and the non-resonant background, as discussed in Ref.~\refcite{baur-glover}, has not 
been taken into account. 
As an example of signal reconstruction above background, 
the expected signal from a Higgs boson with $m_H$~=~300~\Gcs\ is shown 
in Fig.~\ref{f:htowwzz}(left) for an integrated luminosity of only 10~\fbs\ in the ATLAS 
experiment.

For larger values of $m_H$, the Higgs boson signal becomes very 
broad and the signal rate drops rapidly. In the high mass region, the decay modes 
$H \to  Z Z \to \ell \ell \ \nu \nu$ and 
$H \to  Z Z \to \ell \ell \ j j$  provide additional discovery 
potential\cite{physics-tdr,cms-higgs} and allow to extend the 5$\sigma$-discovery range 
up to $\sim$ 1~\Tcs.

\subsubsection{$H \to W W^{(*)}$ decays  \label{s:hww-inclusive}}
For Higgs boson masses around 170~\Gcs, for which the $ZZ^*$ branching ratio 
is suppressed, the discovery potential can be enhanced by searching for the 
\hwwsll\ decay.\cite{dittmar,higgs-ww-atlas-cms} 
In this mode it is not possible to reconstruct a 
Higgs boson mass peak. Instead, an excess of events above the expected 
backgrounds can be observed and used to establish the presence of a 
Higgs boson signal. Usually, the transverse mass computed from the leptons and 
the missing transverse momentum, 
$m_T = \sqrt{2 \ \PT^{\ell \ell} \met \ ( 1 - \cos \Delta \varphi)}$,
is used to discriminate between signal and background. 
The $WW$, \ttbar\ and single-top production processes constitute 
severe backgrounds and the signal significance depends 
critically on their absolute knowledge. 
After relaxing cuts, the Monte Carlo predictions 
for those backgrounds can, however, be normalized to the data in 
regions where only a small fraction of signal events is expected. 
As an example, 
the distribution of the transverse mass is shown in Fig.~\ref{f:htowwzz}(right) for the 
sum of signal plus background and for the background alone in the ATLAS experiment, assuming 
$m_H$ = 170~\Gcs\ and an integrated luminosity of 30~\fbs. 
Under the 
assumption that the total background is known with an uncertainty of $\pm$5\%, 
a Higgs boson signal can be extracted with a significance of more than 
5$\sigma$ for $\sim$150 $< m_H <$ 190~\Gcs\ for an integrated luminosity of 
30\,\fbs.

For Higgs boson masses beyond $\sim$800~\Gcs\ the decay mode
$ H \rightarrow W W \rightarrow \ell \nu \ jj$ provides additional 
discovery potential.\cite{physics-tdr,cms-higgs} The
branching ratio is about 150 times larger than
for the $H \to ZZ \to \ 4 \ell$ decay channel, thus providing the largest possible
signal rate with at least one charged lepton in the final state. An
important contribution to the production cross section of such a heavy
Higgs boson is the vector boson fusion process, resulting in the production of two
final state quark jets in the high rapidity regions in association with
the Higgs boson (see Section \ref{s:lhc-vbf}).
This production mechanism provides an important additional signature for
isolating the signal from the background.
 
The experimental challenge in the reconstruction of these events is
twofold: (i) the reconstruction of high-$\PT$ W decays to two jets, and
(ii) the tagging of the forward jets in the presence of large pile-up.
As an example, for a Higgs boson mass of 1~\Tcs\ and an integrated luminosity of
100\,\fbs, about 60 signal events above a background of 20 events are
expected. However, there is no substantial difference between the shapes
of the signal and background distributions and some years of running may
be needed before a signal could be established.

\subsection{Higgs boson searches using vector boson fusion \label{s:lhc-vbf}} 
In recent studies
it has been demonstrated that, not only in the high mass but also in the 
intermediate mass range, the discovery potential can be significantly increased 
by performing a
search for Higgs boson production in the vector boson fusion 
mode.\cite{zeppenfeld,zeppenfeld-tau,zeppenfeld-ww,atlas-vbf,cms-vbf,cms-higgs} 
Although the contribution to the cross section in the 
intermediate mass range amounts at leading order only to about 20\% of the 
total production cross section, the additional event characteristics can 
be exploited to suppress the large backgrounds. 
In vector boson fusion events, the Higgs boson is accompanied by two jets in the 
forward regions of the detector, originating from the 
initial quarks that emit the vector bosons. On the other hand, central jet 
activity is suppressed due to the lack of colour exchange between the initial 
state quarks. This is in contrast to most background processes, where 
colour flow appears in the $t$-channel.
Jet tagging in the forward region of the detector together with a veto 
of jet activity in the central region are therefore powerful tools 
to enhance the signal-to-background ratio. 

\begin{figure}[hbtn]
\begin{center}
\begin{minipage}{6.0cm}
\mbox{\epsfig{file=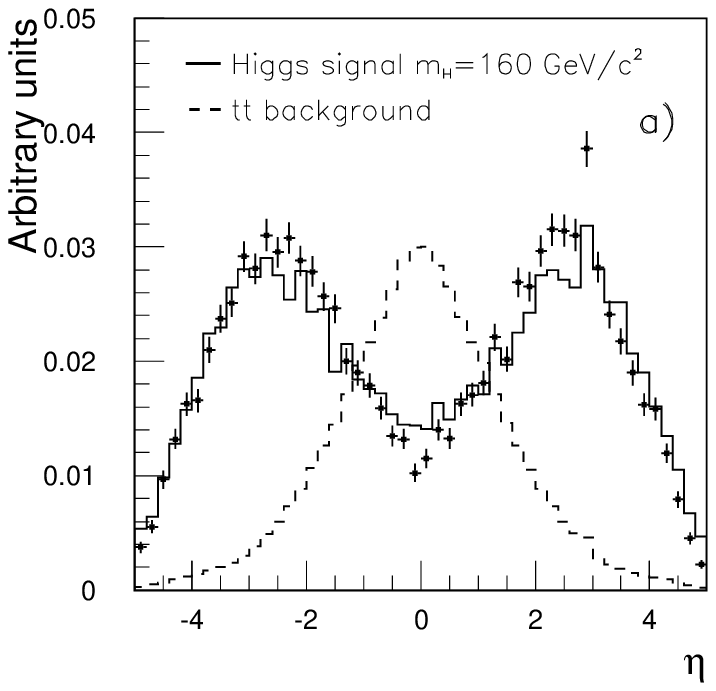,height=5.4cm}}
\end{minipage}
\begin{minipage}{6.0cm}
\mbox{\epsfig{file=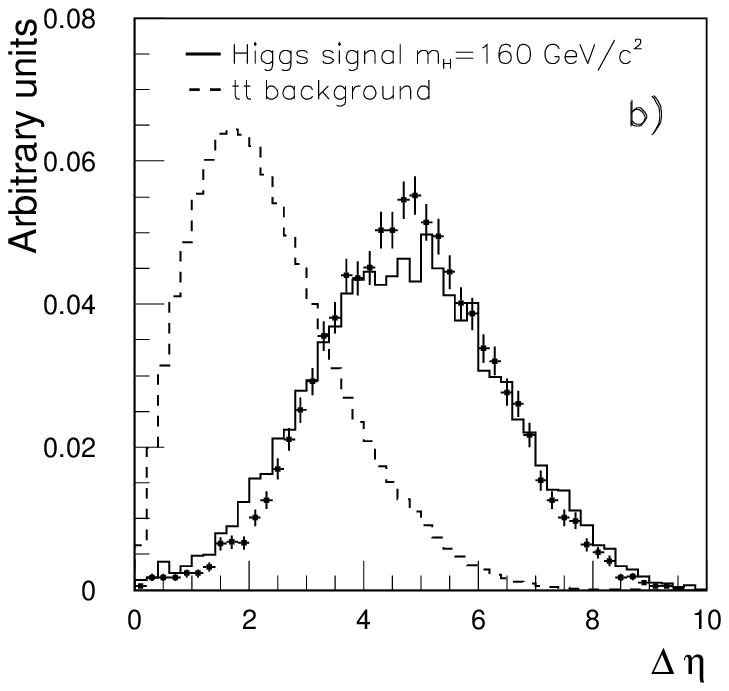,height=5.4cm}}
\end{minipage}
\caption{\small \it (Left) Pseudorapidity distribution of the tag jets in signal 
events with $m_H = 160$~\Gcsit\ and for \ttbar\ background events. 
(Right) Separation $\Delta \eta$ between the tag jets
for the same types of events. 
The full histograms show the distributions at parton level, the dots 
represent the distributions at
reconstruction level, after the tagging algorithm 
has been applied. The corresponding distributions for jets identified as tag 
jets in \ttbar\ events are superimposed as dashed histograms. All distributions 
are normalized to unity (from Ref.~\protect\refcite{atlas-vbf}).}
\label{f:tagjets}
\end{center}
\end{figure}

The performance of the detectors for forward jet tagging has been studied in a
detailed simulation.\cite{atlas-vbf,cms-higgs} In the study presented
in  
Ref.~\refcite{atlas-vbf}, the two tag jets are searched for over 
the full calorimeter coverage $(| \eta | < 4.9)$. 
The jets with
the highest \PT\ in the positive and negative regions of pseudorapidity
are taken to be the tag jet candidates. Each jet is required to have 
a transverse energy of at least 20~GeV. 
The pseudorapidity distribution and the separation $\Delta \eta$ between 
the two tag jets, as found from the parton-level information, 
is shown in Fig.~\ref{f:tagjets}
for signal events with $m_H =$  160~\Gcs. 
For the tagging algorithm described above, distributions at
parton and reconstruction level are in good agreement.
For comparison, the corresponding distributions for 
tag jets as reconstructed in \ttbar\ background events are 
superimposed on the figure. From these distributions it can be concluded that 
a large pseudorapidity separation can be used for the
discrimination between signal and backgrounds from QCD production. 

\begin{figure}[hbtn]
\begin{center}
\begin{minipage}{6.0cm}
\mbox{\epsfig{file=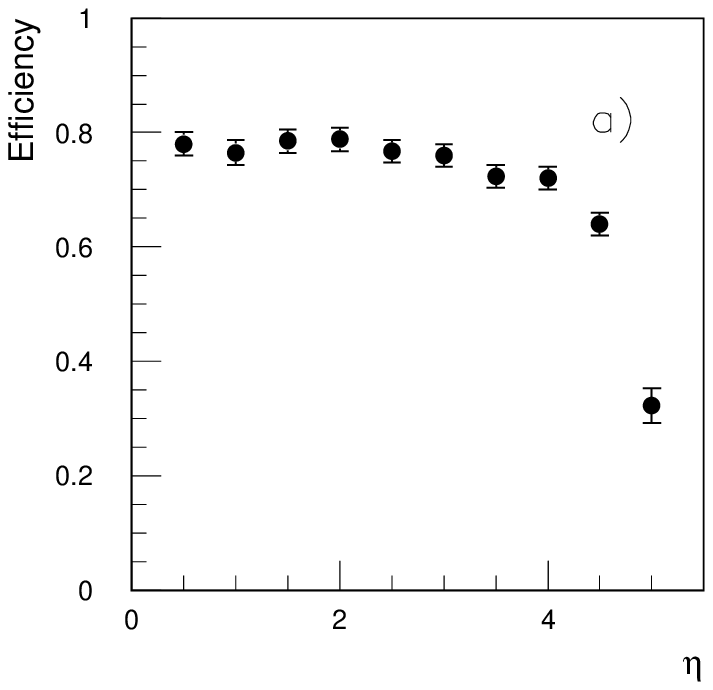,height=5.4cm}}
\end{minipage}
\begin{minipage}{6.0cm}
\mbox{\epsfig{file=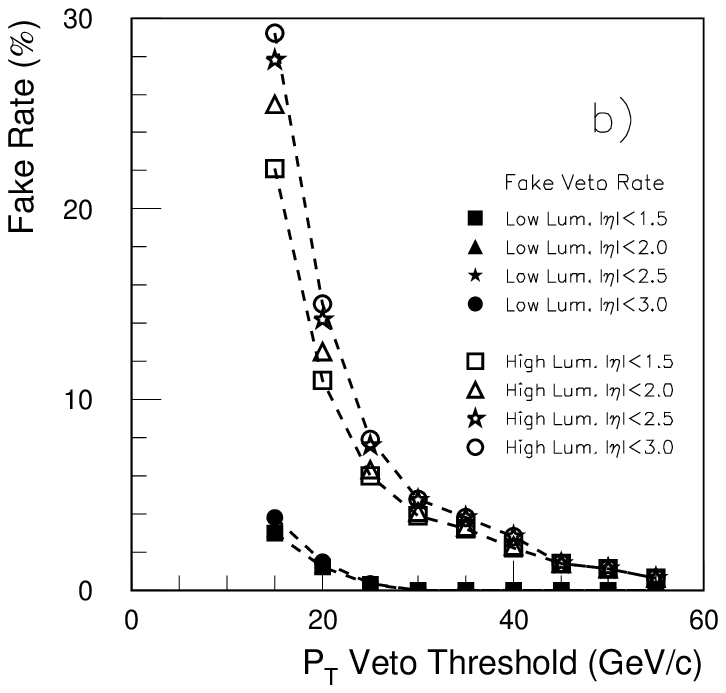,height=5.4cm}}
\end{minipage}
\caption{\small \it 
(Left) Efficiency for reconstructing a tag jet with $\PT > 20$~GeV/c which originates
from a parton with $\PT > 20$~GeV/c as a function of pseudorapidity $\eta$
of the parton.
(Right) Probability for finding at least one jet from pile-up events in 
central rapidity intervals
in the ATLAS detector as a function of jet \PT\ threshold.
The dashed curves connect the points for pseudorapidity itervals 
$| \eta |<$1.5 and $| \eta |<$3.0 for low and high LHC luminosities.}
\label{f:jetveto}
\end{center}
\end{figure}

The detailed simulation performed in Ref.~\refcite{pisa-tagging} has 
demonstrated that tag jets can be reliably reconstructed in the ATLAS 
detector. The efficiency for reconstructing a tag jet 
in signal events with \PT\ above 20 \Gc\ is shown 
in Fig.~\ref{f:jetveto}(left) as a function of pseudorapidity $\eta$.

Generally, the fast simulation package of the ATLAS 
detector\cite{ATLFAST} 
provides a sufficiently good description of the tagging efficiency. 
Differences between the fast and full simulation have been found in the 
transition regions between different calorimeters and at very forward 
rapidities. These differences have been parametrized as a function 
of \PT\ and $\eta$ to incorporate them in the fast 
simulation.\cite{pisa-tagging} 

As pointed out above, a veto against jets in the central region
will be an important tool to suppress background from QCD processes. 
It should be noted that a reliable estimate of the jet veto efficiency is 
difficult to obtain. Present estimates are based on leading-order
parton shower
Monte Carlos and might be affected by sizeable uncertainties. It is important 
to explore new Monte Carlo approaches\cite{sherpa}
and to compare their predictions to Tevatron data.  

At the LHC, jets in the central region can be produced also by pile-up events. 
In the full simulation study it has been found that after applying a 
threshold cut on the calorimeter cell energies of 0.2~GeV at low and 1.0~GeV at high luminosity, 
the veto rate due to fake jets from pile-up events can be kept at a low level, provided that \PT\
thresholds of 20~\Gc\ at low and 30~\Gc\ at high luminosity are used for the 
jet definition. The results of this study are presented in 
Fig.~\ref{f:jetveto}(right),
where the efficiency to find a jet from pile-up events in different 
intervals of central rapidity is 
shown as a function of the jet \PT-threshold for low and high luminosity.

The Higgs boson search in the vector boson fusion mode for 
various final states is discussed in the following.

\subsubsection{$ qqH \to qq WW^{(*)}$}
\begin{figure}
\center{
\begin{minipage}{6.0cm}
\mbox{{\epsfig{file=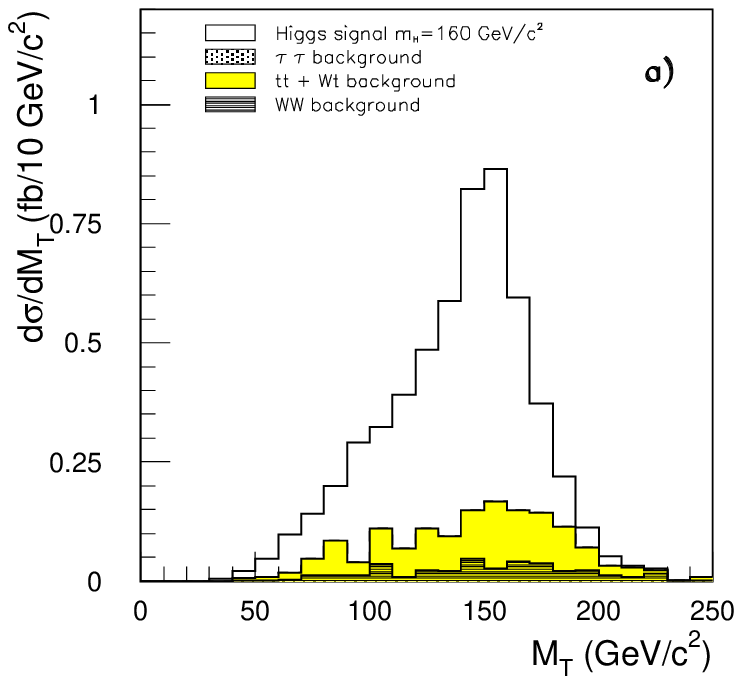,height=6.0cm}}}
\end{minipage}
\begin{minipage}{6.0cm}
\mbox{{\epsfig{file=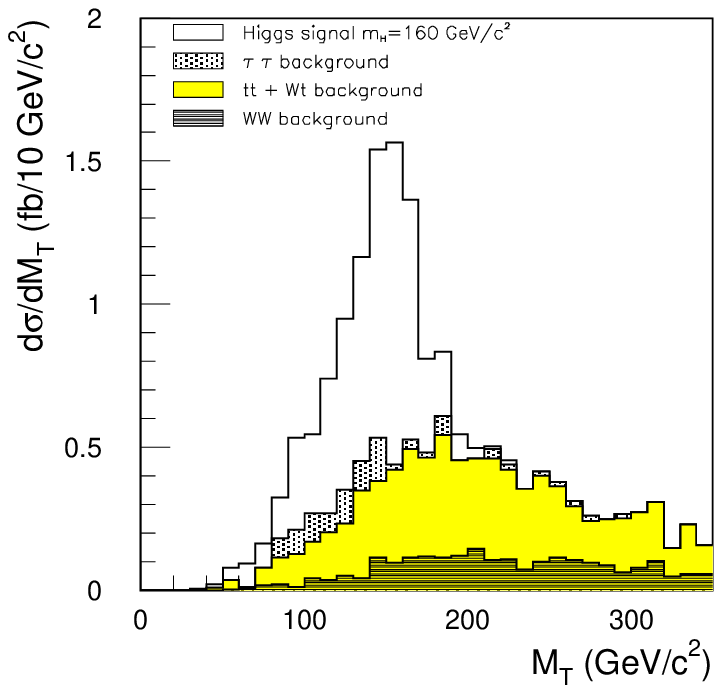,height=6.0cm}}}
\end{minipage}
}
\caption{\it 
(Left) Distribution of the transverse mass M$_{\it T}$ for a Higgs boson with a mass of 160~\Gcsit\  
and the backgrounds in the $e\mu$-channel after all cuts. 
(Right) The same, after relaxing kinematical cuts on the reconstructed leptons. 
The accepted cross sections  
$d \sigma / d M_T$ (in fb/10~\Gcsit) including all efficiency and acceptance factors 
are shown in both cases (from Ref.~\protect\refcite{atlas-vbf}).}
\label{f:vbf_WW}
\end{figure}

According to Monte Carlo studies, the 
LHC experiments have a large discovery potential in the 
$ \hwws \rightarrow \ell^+ \ell^- \pet $ decay 
mode.\cite{atlas-vbf-ww,cms-vbf-ww} 
The additional signatures
of tag jets in the forward and of a low jet activity in the
central regions of the detector allow for a significant reduction of
the background. This results in a better signal-to-background ratio compared to  
the inclusive $H \rightarrow WW^{(*)}$ channel, which is dominated by the 
gluon fusion process. As a consequence, 
the signal sensitivity is less affected
by systematic uncertainties on the predictions of the background. 
As an example, the reconstructed transverse mass distribution 
for a Higgs boson signal with a
mass of 160~\Gcs\ is shown on top of the backgrounds from \ttbar, $Wt$, 
and $WW$ production in Fig.~\ref{f:vbf_WW}(left). 
This figure demonstrates the better signal-to-background ratio as compared to the inclusive case shown in Fig.~\ref{f:htowwzz}(right). 
Since neutrinos appear 
in the final state, the transverse mass, defined as 
\[ 
M_T = \sqrt{(E_T^{\ell \ell} + E_T^{\nu \nu})^2 - ( \vec{p}_T^{\ \ell \ell} + 
\vec{\pet} )^2 }
\]
is used for the mass reconstruction. For Higgs boson masses below
$\sim 2 \ m_W$, the $W$ bosons are mostly at rest in the Higgs boson
centre-of-mass system, resulting in $m_{\ell\ell} = m_{\nu \nu}$, 
such that for both the dilepton and the neutrino system the 
transverse energy can be calculated as\cite{zeppenfeld}   
\[ 
E_T^{\ell \ell} = \sqrt{(P_T^{\ell \ell})^2 + m^2_{\ell \ell}} 
\hspace*{1.0cm} {\rm and} \hspace*{1.0cm} 
E_T^{\nu \nu} = \sqrt{(\pet)^2  + m^2_{\ell \ell}} . 
\]

After all cuts, a signal cross section in the $e \mu$ channel
for a Higgs boson with a mass of 160~\Gcs\ of the order of 4.6 fb is 
expected above a total background expectation of 1.2 fb. 
Due to this large signal-to-background ratio, this channel alone has 
a good discovery potential for a Higgs boson with a mass  
around 160~\Gcs\ and is not very
sensitive to systematic uncertainties on the background. 
This compares favourably with the  $gg \rightarrow WW^{(*)}$ channel discussed 
in Section \ref{s:hww-inclusive}. 
It is interesting to note that an application of looser kinematical  
cuts on the final-state leptons allows for a better discrimination between 
the signal and background shape.\cite{atlas-vbf} 
The corresponding transverse mass distribution
is shown in Fig.~\ref{f:vbf_WW}(right). In this case, the background extends to higher 
$M_T$ values and a background normalization outside the signal region 
is possible. 

\begin{figure}[hbtn]
\begin{center}
\mbox{\epsfig{file=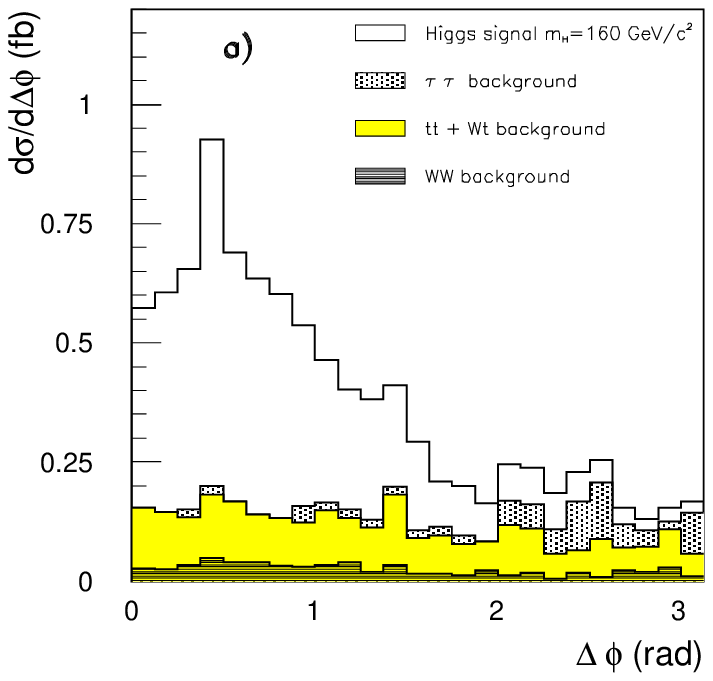,width=0.44\textwidth}} 
\mbox{\epsfig{file=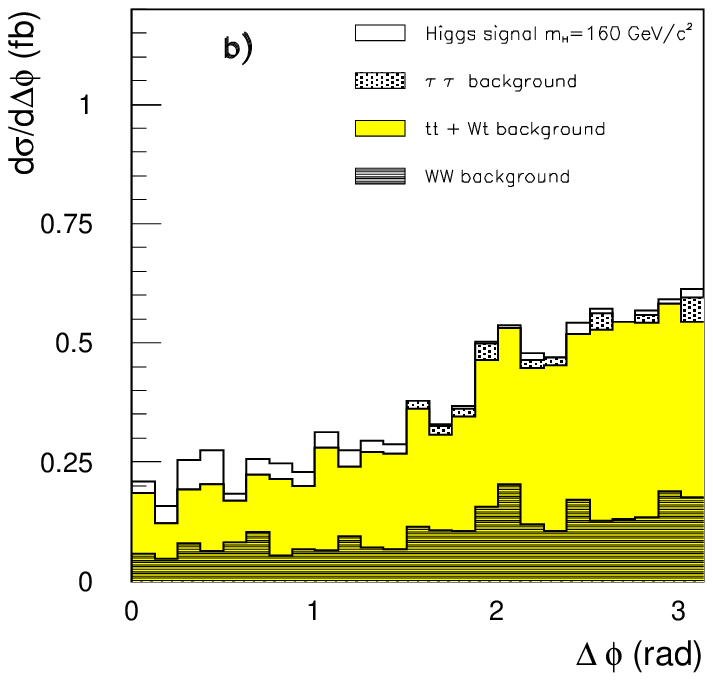,width=0.44\textwidth}}
\caption{\small \it 
Distributions of the azimuthal opening angle $\Delta \phi$ 
between the two leptons for events in the signal region (M$_{\it T} <$ 175~\Gcsit)
(left) and events outside the signal region (M$_{\it T} >$ 175~\Gcsit) (right). 
The accepted cross sections  
$d \sigma / d \Delta \phi$ (in fb/0.13 rad) including all efficiency and 
acceptance factors are shown in both cases (from Ref.~\protect\refcite{atlas-vbf}).}
\label{f:vbf-delphi}
\end{center}
\end{figure}

The presence of a signal can also be demonstrated in the distribution of 
$\Delta \phi$, the difference in azimuthal angle between the two
leptons in the final state.
This distribution is shown in Fig.~\ref{f:vbf-delphi} for $e\mu$ final
states passing the looser cuts, {\em i.e.},  
without cuts on the spatial separation of the leptons.
Depending on $M_T$
the event sample has been split into two subsamples: 
(a) in the so called signal sample ($M_T < 175$~\Gcs) and (b) in a control 
sample ($M_T > 175$~\Gcs). For events in the signal region 
(Fig.~\ref{f:vbf-delphi}(left)), a pronounced 
structure at small $\Delta \phi$ is seen, as expected for a spin-0 resonance. 
This behaviour is not present for events in the control  
sample (Fig.~\ref{f:vbf-delphi}(right)), where the \ttbar\ and $WW$ backgrounds are 
expected to dominate. Therefore, the unbiased $\Delta \phi$ distribution,
resulting from relaxed kinematical cuts, can be used for both a demonstration of the 
consistency of the signal with a spin-0 hypothesis and for an 
additional background normalization. This normalization can be performed in  
the high $\Delta \phi$ region using events directly below the peak.

In addition, it has been studied whether the larger hadronic branching ratio 
of the W boson can be exploited and the process 
$ qq \rightarrow qqH \rightarrow qqWW^{(*)} \rightarrow qq \ \ell \nu \ qq$ 
can be detected.\cite{pisa-vbf} 
The cross section times branching ratio is about
4.3 times larger than for the dilepton channel. However, since only one lepton is
present in the final state, $W$+jet production is a serious
additional background, with a cross section
more than two orders of magnitude larger than the signal cross section.
A detailed study has shown that after all cuts the signal rates 
and the signal-to-background ratio are 
much lower than the corresponding numbers in the 
dilepton channel. For a Higgs boson mass of 160~\Gcs\ and an integrated 
luminosity of 30\,\fbs, for example, 24 signal and 18 background events are 
expected.\cite{pisa-vbf} The estimate of the dominant $W$+jet background has been performed 
using a tree-level Monte Carlo generator.\cite{vecbos}
Given these results, this channel should not be
considered a Higgs boson discovery channel, but could be used to confirm an
observation of a Higgs boson with a mass around 160~\Gcs.

In addition, it has been shown in Ref.~\refcite{vbf-ww-highmass} that the 
$ qq \rightarrow qqH \rightarrow qqWW^{(*)} \rightarrow qq \ \ell \nu \ \ell \nu$
decay mode also increases the Higgs boson discovery potential in the high 
mass region between $\sim$300~\Gcs\ and $\sim$600~\Gcs.

\subsubsection{$ qqH \to qq \tau \tau$}
\begin{figure}[hbtn]
\begin{center}
\begin{minipage}{6.0cm}
\mbox{\epsfig{file=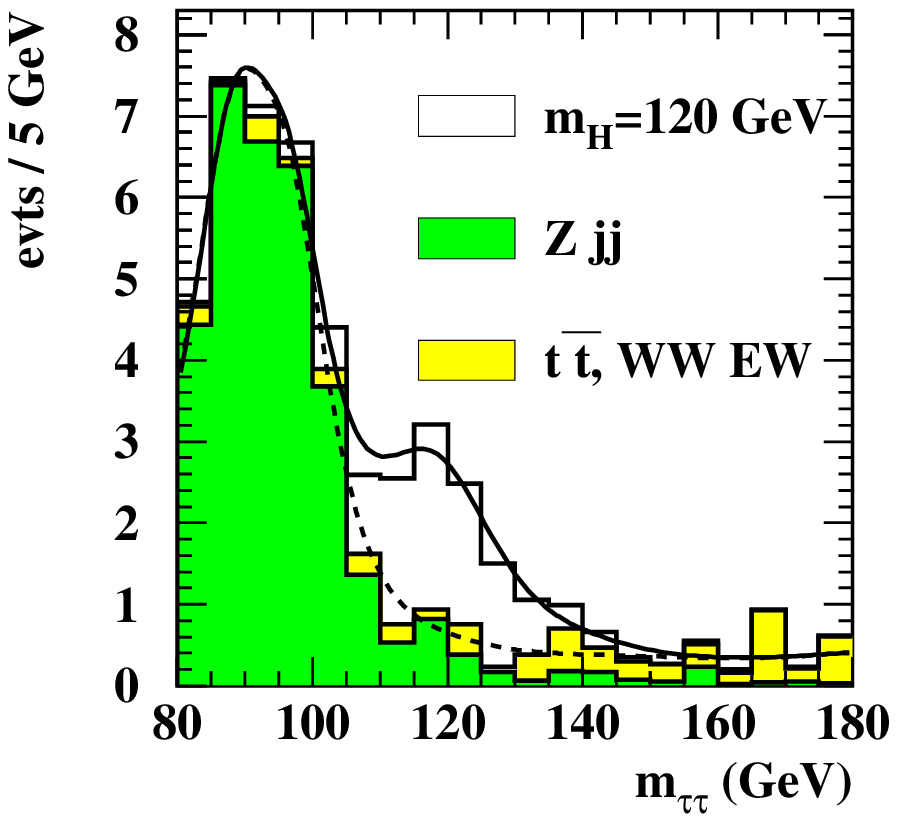,height=5.7cm}}
\end{minipage}
\begin{minipage}{6.0cm}
\hspace*{0.4cm}
\vspace*{0.2cm}
\mbox{\epsfig{file=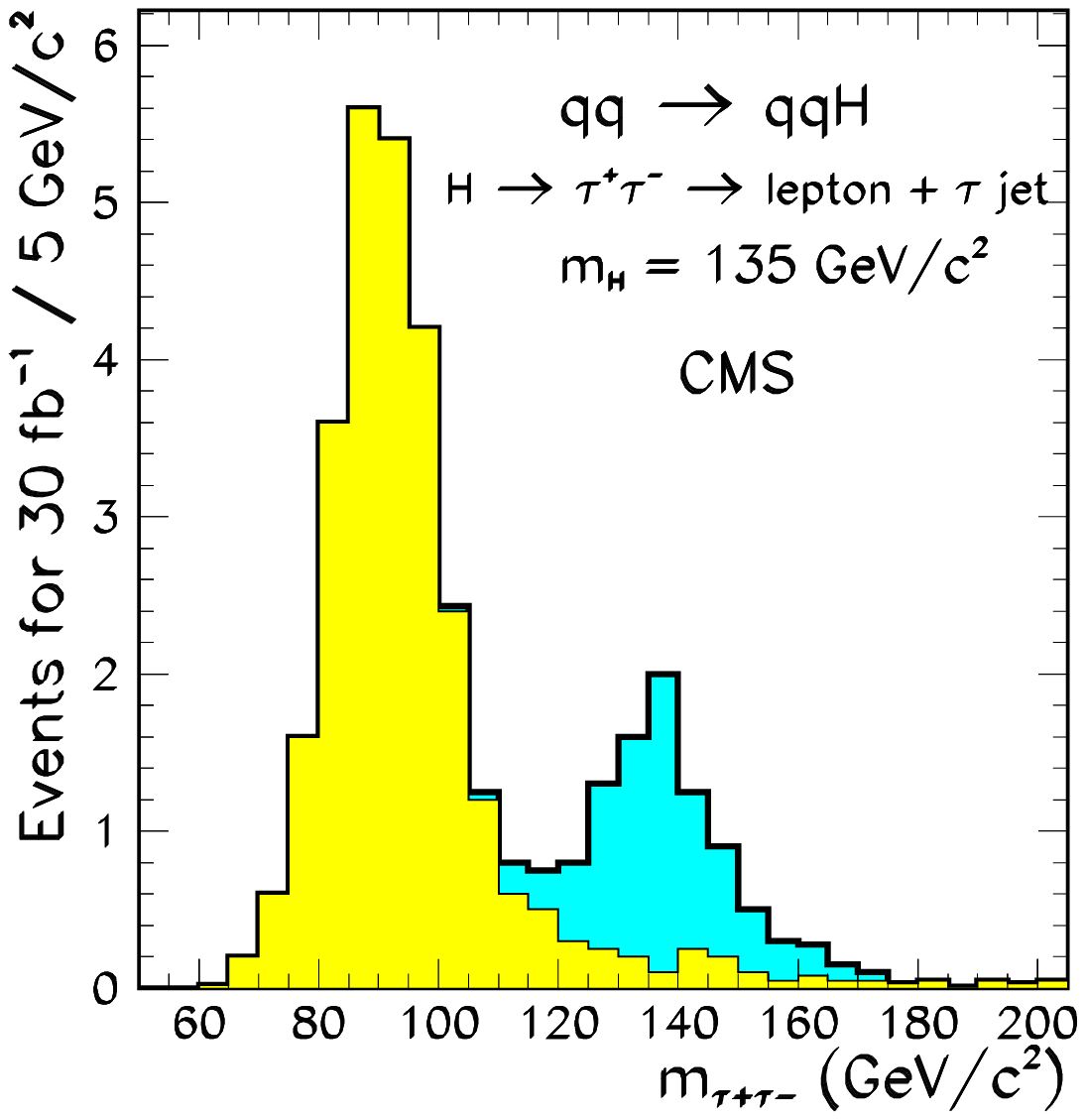,height=5.4cm}}
\end{minipage}
\caption{\small \it 
(Left) The reconstructed $\tau \tau $ invariant mass for a Higgs boson signal 
of 120~\Gcsit\ in the $e \mu$ channel  
above all backgrounds after application of all cuts except the mass window cut around the Higgs
boson mass. 
The number of signal and background events are shown for an integrated 
luminosity of 30 \fbsit\ (from Ref.~\protect\refcite{atlas-vbf}). 
(Right) The same, 
for the $\ell$-had channel and for a Higgs boson with 
m$_{\it H}$ = 135~\Gcsit\ (from Ref.~\protect\refcite{cms-higgs}). 
}
\label{f:mtautau}
\end{center}
\end{figure}

It has been shown that in the mass region 
$110 < m_H < 140$~\Gcs\ the ATLAS and CMS experiments are sensitive to 
the $\tau \tau$ decay mode of the Standard Model Higgs 
boson in the vector boson fusion channel.\cite{cms-vbf,atlas-vbf-tau} 
Searches for $H \rightarrow \tau \tau$ decays  
using the  double leptonic decay mode
$ qqH  \rightarrow qq \ \tau \tau \rightarrow qq \ \ell \nu \bar{\nu} \  
\ell \bar{\nu} \nu$ and the lepton-hadron decay mode
$ qqH  \rightarrow qq \ \tau \tau$ $\rightarrow qq$  $\ell \nu \nu \ had \ \nu$
seem to be feasible. In these analyses, $Z$+jet production with $Z \rightarrow \tau 
\tau$ constitutes the principal background.
The $\tau \tau$ invariant mass can be reconstructed using the collinear 
approximation. 
In signal and background events, the $H$ and $Z$ bosons are emitted with 
significant \PT,
which contributes to large tau boosts and causes the tau decay products 
to be nearly collinear in the laboratory frame. 
Assuming that the tau directions 
are given by the directions of the visible tau decay products
(leptons or hadrons from tau decays respectively), 
the tau momenta and therefore the  
$\tau  \tau$ invariant mass can be reconstructed. 
As an example, distributions of the reconstructed $\tau \tau$ invariant 
mass for the $e \mu$ and for the $\ell$-had final states are shown 
in Fig.~\ref{f:mtautau}  
for Higgs boson masses of 120~\Gcs\ and 135~\Gcs. 
A discovery based on a combination of these
final states would require an integrated luminosity of about 30 $\fbs$.
It should however be stressed that this assumes that the background from 
$Z \to \tau \tau$ decays in the signal region is known with a precision of 
$\pm 10\%$. More studies are needed to establish that this precision can indeed 
be achieved. The detection of the $H\to\tau\tau$ decay mode is 
particularly important for a measurement of the Higgs boson couplings
to fermions and for Higgs boson searches in MSSM scenarios, as discussed in 
Sections \ref{s:couplings} and \ref{s:mssm-benchmarks}. 

\subsubsection{$qqH \to  qq \gamma \gamma$}
In addition, the prospects for discovering a Standard Model Higgs boson 
in the  \hgg\ decay mode have been
evaluated.\cite{cms-vbf-gg,Bruce01}
Due to the small \hgg\ decay branching ratio
the expected signal rate in the vector boson fusion mode is 
small. As in the inclusive analysis, there are significant 
backgrounds from non-resonant $\gamma \gamma$ production with jets and 
from QCD multijet and direct photon production. In the 
analysis,\cite{Bruce01} the $\gamma \gamma
jj$ (QCD and electroweak),    
$\gamma jjj$ and $jjjj$ matrix elements have been considered. After applying a
similar selection as in the inclusive case and requiring tagging jets with a 
large separation in pseudorapidity, a signal significance of 2.2$\sigma$ 
for $m_H$ = 130~\Gcs\ has been found assuming an integrated luminosity of 
30~\fbs. Although the significance is lower than in the inclusive case, 
a much larger signal-to-background ratio with values around 0.5 can be reached. 
On the other hand, this decay mode appears to 
be considerably more sensitive to the background resulting 
from QCD processes with jets misidentified as photons.

\subsubsection{$qqH \to qq ZZ^{(*)}$}
The $H \to ZZ \to \ell \ell qq$ mode has been studied for Higgs boson masses 
larger than $2 \ m_Z$.\cite{Bruce02} In that channel, the background is 
largely dominated by the $Z$+4jet production with  $Z \to \ell \ell$. For a 
Higgs boson with \mh\ = 300~\Gcs, a signal 
significance of 3.8$\sigma$ has been estimated, assuming an integrated luminosity of 
30\,\fbs. Also in this case, the signal significance appears to be weaker than 
for the standard $H \to 4 \ell$ channel and the estimated background cross 
sections are subject to large theoretical uncertainties. 

\begin{figure*}
\begin{minipage}{1.0\linewidth}
\begin{center}
\mbox{\epsfig{file=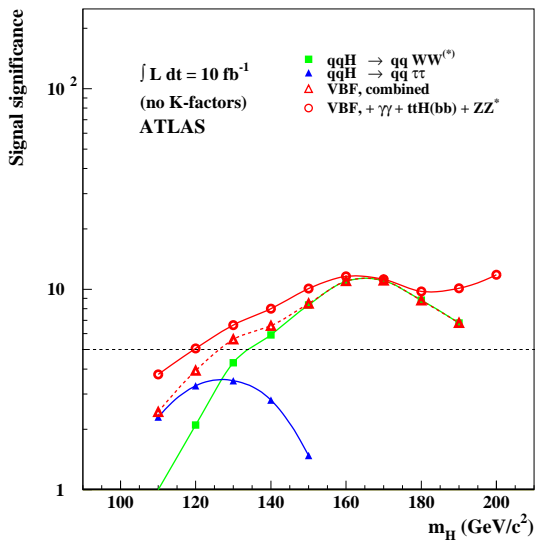,height=8.0cm}}
\caption{\small \it 
ATLAS sensitivity for the discovery of a Standard Model Higgs boson for an integrated 
luminosity of 10 \fbsit\ in the low mass region. The 
signal significances are plotted for individual channels, as well as for the
combination of several channels. A systematic uncertainty of 
$\pm$ 10\% on the background
has been included for the vector boson fusion channels
(from Ref.~\protect\refcite{atlas-vbf}).
}
\label{f:signif01}
\end{center}
\end{minipage}
\end{figure*}

\subsubsection{$qqH \to qq \bbbar$}
The detection of the fully hadronic decay mode $qq H \to qq \bbbar$ is extremely 
challenging in a hadron collider environment, given the huge background from 
QCD jet production. Already at the first trigger level the acceptance for 
signal events is low, since it is foreseen to run the experiments with 
relatively high \et\ thresholds for pure jet triggers.\cite{atlas-triggers}
Even if the trigger issue is ignored, a very low signal-to-background 
ratio is expected. First parton level studies\cite{roeck,vbfbb}
have shown that the dominant 
$bbjj$ background is about 300 times larger than the signal. 
To claim a signal,
the background must be known with accuracies at the per mille level, which 
will be very hard to achieve. Studies using a
realistic trigger simulation are still in progress. 

\subsubsection{Discovery potential in the vector boson fusion channels}
The vector boson fusion channels provide a large discovery 
potential even for small integrated luminosities. The expected 
signal significance in the mass region 110~$< m_H <$~190~\Gcs\ 
is shown in Fig.~\ref{f:signif01} for the ATLAS experiment for the 
two main channels for an integrated 
luminosity of only 10~\fbs. Combining the two channels, a 
Standard Model Higgs boson can be discovered with a 
significance above 5$\sigma$ in the mass range 130 to 190~\Gcs, assuming 
a systematic uncertainty of 10\% on the background. 
If the vector boson fusion channels are combined 
with the Higgs boson discovery channels discussed in Sections \ref{s:higgs-incl} and 
\ref{s:higgs-ass}, the 5$\sigma$ discovery range, for that value of the integrated 
luminosity, can be extended down to $\sim$120~\Gcs. 
According to these expectations, a Higgs boson discovery at the LHC should be 
possible in the low mass region in each experiment after about one 
year of running at low luminosity, provided the detectors are well understood
in terms of lepton identification, \met-resolution and forward-jet tagging.

\subsection{Higgs boson searches using the associated $WH$, $ZH$ and $\ttbar H$ production 
\label{s:higgs-ass}}

\subsubsection{\hgg\ decays}
The $\gamma \gamma$ decay mode has also been studied for the associated production 
of a Higgs boson with a $W$ or $Z$ boson or a  \ttbar\ 
pair.\cite{physics-tdr,cms-higgs} 
In both production modes at least one additional lepton is required from the 
vector boson or top-quark decay. 
Those channels have been found to have a better signal-to-background 
ratio than the inclusive \hgg\ channel, 
{\em e.g.}, assuming an integrated luminosity of 100~\fbs, 
13.2 signal events for $m_H$ = 120~\Gcs\ are expected in the ATLAS experiment  
from the sum of $WH$,  $ZH$ and $\ttbar H$ production
above a total background of 5.7 events.\cite{physics-tdr} The background 
is dominated by irreducible contributions from   
$W \gamma \gamma$, $Z \gamma \gamma$ and $\ttbar \gamma \gamma$ production,
which sum up to an irreducible fraction of 70 -- 80\%.   
The statistical significance 
is found to be around 4.3$\sigma$ for masses in the range 
between 100 and 120~\Gcs. An observation in this channel would therefore 
represent an independent confirmation of a possible discovery of a light 
Higgs boson at the LHC. 

If the accompanying $W$ boson or one of the top quarks decays
leptonically, a neutrino leading to missing transverse energy is
present in the final state.  
In addition, missing transverse energy appears in the decay mode 
$Z H \to \nu \nu \gamma \gamma$. In a recent study it has been investigated 
whether an accompanying missing transverse energy signature (instead of an
additional lepton) can be used to establish a $\hgg$ signal.\cite{beauchemin}

\begin{figure}[hbtn]
\begin{center}
\begin{minipage}{0.44\textwidth}
\mbox{\epsfig{file=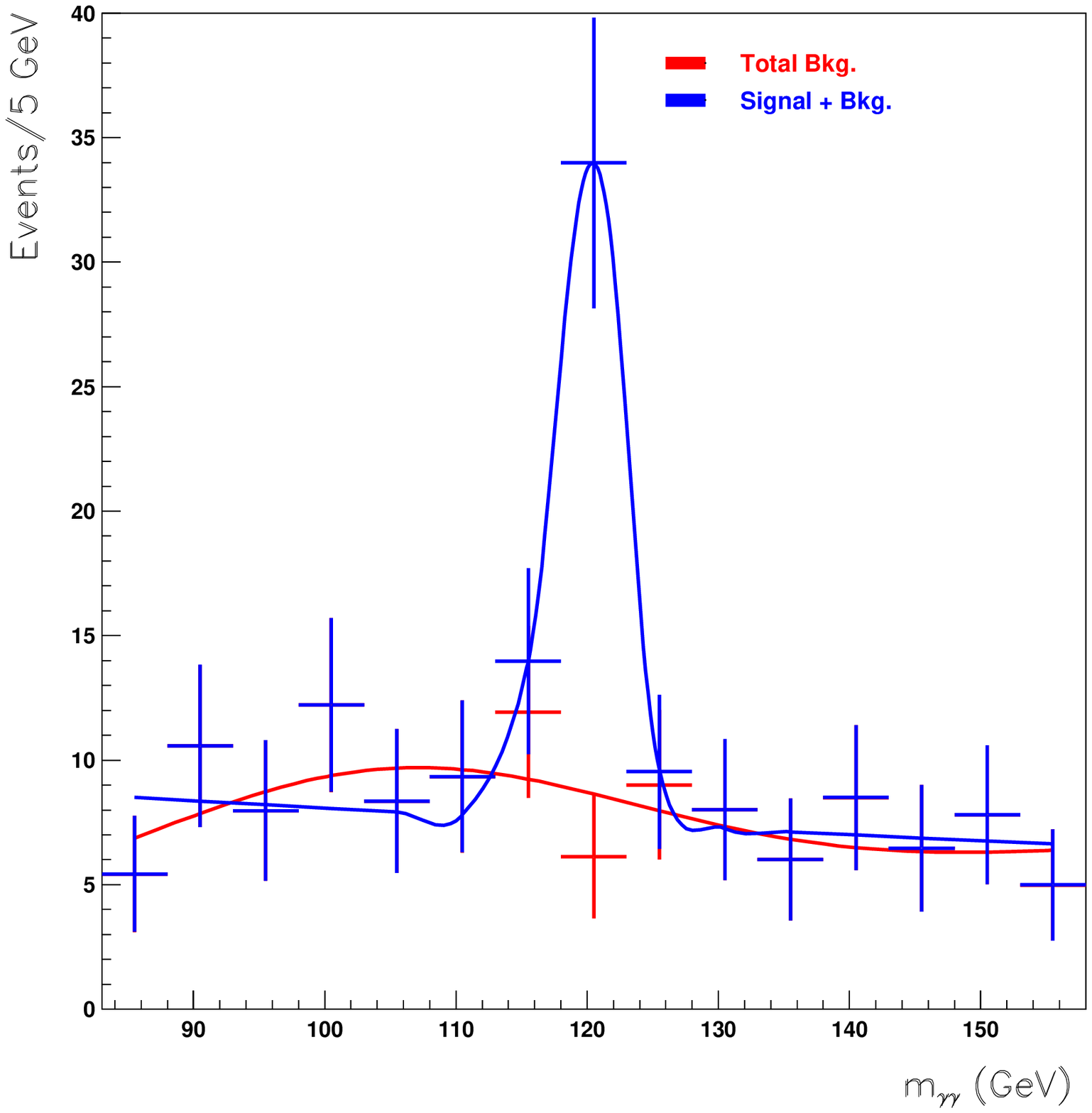,width=0.83\textwidth}}
\end{minipage}
\begin{minipage}{0.44\textwidth}
\vspace*{0.2cm}
\mbox{\epsfig{file=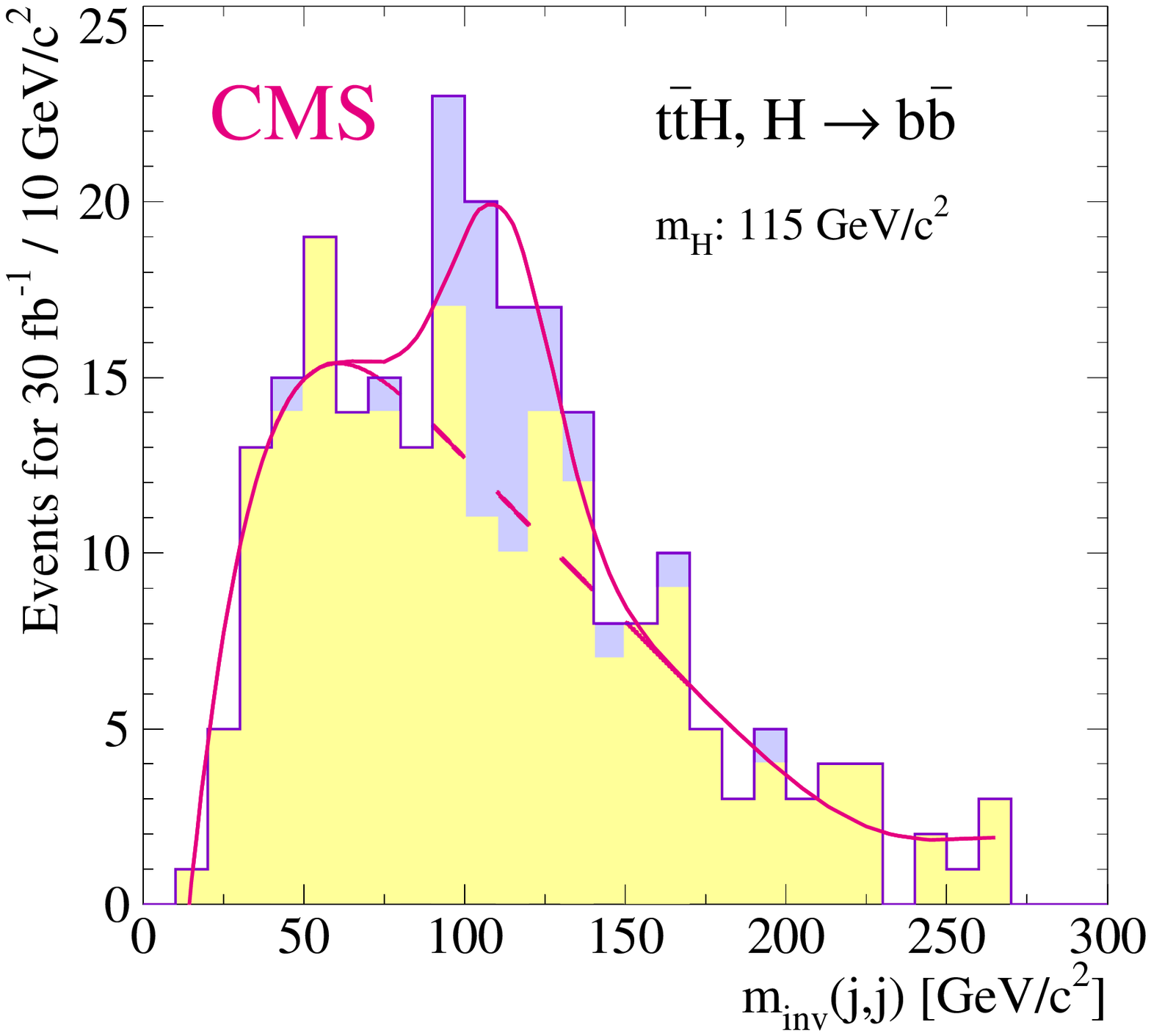,width=0.95\textwidth}}
\end{minipage}
\caption{\small \it 
(Left) Reconstructed $\gamma \gamma$ invariant mass distribution after the 
$\gamma \gamma + \met$ selection for an integrated luminosity of 100~\fbsit\
(from Ref.~\protect\refcite{beauchemin}).
(Right) 
Reconstructed \bbbar\ invariant mass distribution of a  
$tt H \to \ell \nu b \ q q b \  b b$ signal (dark) with m$_{\it H}$~=~115~\Gcsit\ 
and the background (light) for an integrated 
luminosity of 30~\fbsit\ (from Ref.~\protect\refcite{drollinger_02}).} 
\label{f:tth-bb-cms}
\end{center}
\end{figure}

In the analysis, the same photon selection as in the inclusive case is applied. 
In addition, large missing transverse energy, $\met >$ 66~\Gcs\ is required. 
For an integrated luminosity of 100\,\fbs\ and a Higgs boson with a mass of 120~\Gcs, 
a signal of 20.9 events above a background of only 
5.4 events is expected. The distribution of the invariant $\gamma \gamma$ mass is
shown in Fig.~\ref{f:tth-bb-cms}(left).  
The signal is dominated by contributions from the 
$WH + ZH$ (45\%) and $\ttbar H$ (50\%) associated production. Due to the 
\met\ requirement the $gg \to H \to \gamma \gamma$ process 
accounts for only 5\% of the 
selected signal events, despite the much larger production cross section. 
Likewise a large fraction of the irreducible $\gamma \gamma$ background is 
rejected. The residual background 
is dominated by events from $Z \gamma \gamma$ and $W \gamma \gamma$ 
production. The background has been estimated using a tree-level Monte Carlo 
generator\cite{COMPHEP} and a fast simulation of the ATLAS detector performance.\cite{ATLFAST} 
It should be stressed that in this simulation non-Gaussian tails in 
the \met\ distribution, which might arise from instrumental effects, are not taken 
into account.

Additional cuts allow to disentangle the various contributions to 
the signal, which is important for the determination of the Higgs boson couplings 
to fermions and bosons (see Section \ref{s:couplings}). 
An additional lepton veto, for example,  rejects $WH$ and $\ttbar H$ events 
and selects preferentially $ZH \to \nu \nu \gamma \gamma$ events.
An additional jet veto leads to a strong suppression of $\ttbar H$ events.\cite{beauchemin}

\subsubsection{\hbb\ decays}
At low mass, the associated \hgg\ channel is nicely complemented by a search for 
the decay mode \hbb, which has the largest branching ratio in this mass range.
Due to the huge backgrounds from QCD jet production in this decay mode, 
only the associated production modes have sensitivity. It has been demonstrated that  
the discovery potential for a Standard Model Higgs boson 
in the $WH$ production mode at the LHC is 
marginal.\cite{physics-tdr,richter-was-wh,wh-king} It is limited by  
large backgrounds from $Wb\bar{b}$, $Wq\bar{q}$, and \ttbar\ production. For small integrated 
luminosities, the extraction of a signal appears to be very difficult, 
even under the most optimistic assumptions for b-tagging performance and calibration 
of the shape and magnitude of the various backgrounds from data 
itself. If backgrounds are well known, evidence for a signal in this channel 
may be reached at large integrated 
luminosities,\cite{drollinger_01} and valuable information can be
provided for the measurement of the Higgs boson coupling to vector bosons.

On the contrary, the extraction of a Higgs boson signal in the $\ttbar H$, \hbb\ 
channel appears to be feasible in the low 
Higgs boson mass region.\cite{drollinger_02,cammin,wh-king}
Here it is assumed that the two top-quark decays
can be fully reconstructed with  
an acceptable efficiency, which calls for an excellent b-tagging capability 
of the detector. Another critical item is the knowledge of the shape of the 
main residual background from $\ttbar jj$ production. If the shape can be 
accurately determined using real data from \ttbar\ production, a Higgs boson signal 
could be extracted with a significance of more than 5$\sigma$ in the mass range from 
80 to 120~\Gcs, assuming an integrated luminosity of 30\,\fbs.
As an example, the reconstructed invariant \bbbar\ mass distribution for the
$\ttbar H \to \ell \nu b \ q q b \  b b$ signal with $m_H$ = 115~\Gcs\ 
and background events is shown in Fig.~\ref{f:tth-bb-cms}(right) for the CMS experiment, 
assuming an integrated luminosity of 30~\fbs. 

In addition, the channels $ZH \to \ell \ell \bbbar$, $ZH \to \nu \nu  \bbbar$ and 
$\bbbar H \to \bbbar \bbbar$ have been suggested in the literature for Higgs
boson searches in the \bbbar\ decay mode.\cite{bb-gunion}
For $ZH \to \ell \ell  \bbbar$ a similar
signal-to-background ratio is  
expected as for the $WH$ channel.\cite{physics-tdr}
The other two have so far not been considered by the LHC
collaborations, due to  
the challenging trigger and background conditions.

\subsubsection{$H \to WW^{(*)}$ decays}

The signal significance in the $H \to WW^{(*)} \to \ell \nu \ \ell \nu$ decay 
mode can still be enhanced in the mass region around 160~\Gcs\
by searching for the associated production mode 
$WH$ with $W \to \ell \nu$, leading to a three-lepton final state accompanied by 
missing transverse momentum.\cite{baer_whlll}
The third lepton allows for a significant suppression of the background and therefore
for a better signal-to-background ratio than in the $gg\to H \to WW^{(*)}$ channel.

Furthermore, Higgs bosons decaying to $WW$ can be searched for in
the associated \ttbar\ production mode.
Experimentally this channel leads to 
striking signatures with multilepton and multijet final states. 
In a recent study, final states with two like-sign leptons
or with three leptons have been investigated.\cite{jessica} 
In order to reject the large backgrounds from heavy flavour 
decays into leptons, strict isolation criteria have been applied. After final cuts, 
the signal-to-background ratio is of the order of one for both channels and for 
the most favourable Higgs boson mass of 160~\Gcs. Assuming an integrated luminosity
of 30~\fbs, about 21 and 13 signal events are expected for $\mh$ = 160~\Gcs\ in the  
two and in the three-lepton channel, respectively, compared 
to an expectation of about 20 background events from Standard Model processes 
in each channel. 
Significant contributions to the background arise from the associated $\ttbar W$, 
$\ttbar Z$ and $\ttbar \ttbar$ processes. It should be mentioned that the 
estimates of these backgrounds suffer from large uncertainties,
and more studies are
needed to obtain a more reliable estimate. 
For both smaller and larger
Higgs boson masses, the signal significance decreases because of decreasing 
branching ratios and acceptance in the former and decreasing production 
cross section in the latter case.\cite{jessica} 
This channel provides important  
information for the determination of the Higgs boson couplings to top quarks, as 
discussed in Section~\ref{s:couplings}. 

\begin{figure*}
\begin{minipage}{1.0\linewidth}
\begin{center}
\mbox{\epsfig{file=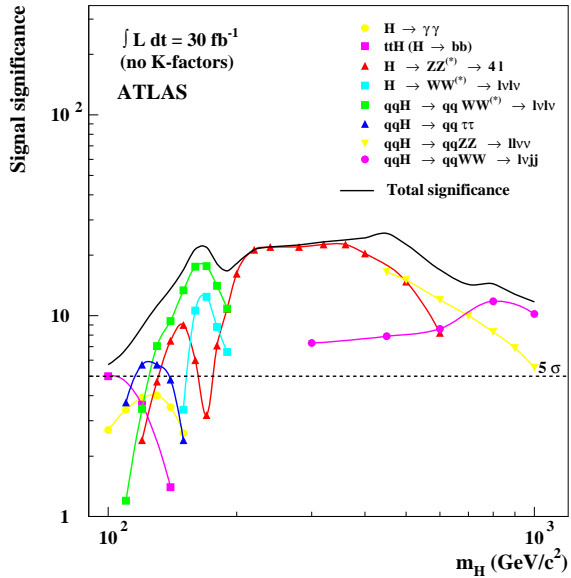,height=8.5cm}}
\caption{\small \it 
ATLAS sensitivity for the discovery of a Standard Model Higgs boson for an integrated 
luminosity of 30 \fbsit\ 
over the full mass region. The 
signal significances are plotted for individual channels, as well as for the
combination of channels. Systematic uncertainties on the background
have been included for the vector boson fusion channels ($\pm$ 10\%) and for the 
$H \to WW^{(*)} \to \ell \nu \ \ell \nu$ channel ($\pm$ 5\%).}
\label{f:signif02}
\end{center}
\end{minipage}
\end{figure*}

\begin{figure*}
\begin{minipage}{1.0\linewidth}
\begin{center}
\mbox{\epsfig{file=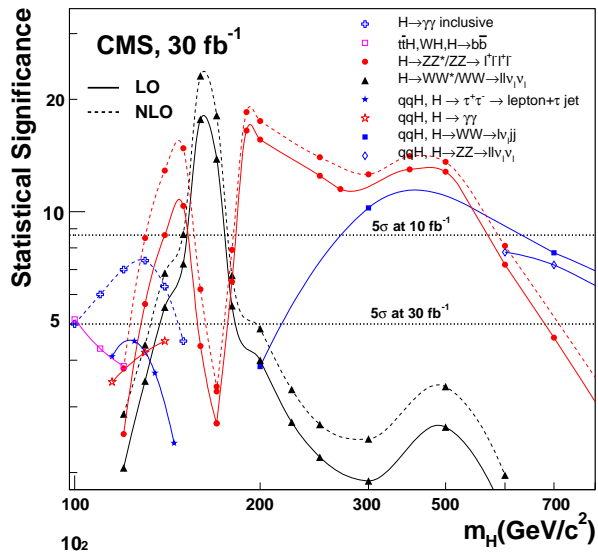,height=8.0cm}}
\caption{\small \it 
CMS sensitivity for the discovery of a Standard Model Higgs boson for an integrated 
luminosity of 30 \fbsit\ 
over the full mass region. The 
signal significances are plotted for individual channels, as well as for the
combination of channels. A systematic uncertainty of $\pm$ 5\% on the background
has been included for the $H \to WW^{(*)} \to \ell \nu \ \ell \nu$ channel. 
For the channels \hgg, $H \to ZZ^{(*)} \to 4 \ell$ and 
$H \to WW^{(*)} \to \ell \nu \ \ell \nu$ results are shown for both leading order 
and next-to-leading order predictions for signals and backgrounds 
(from Ref.~\protect\refcite{cms-higgs}).}
\label{f:signif-cms}
\end{center}
\end{minipage}
\end{figure*}

\subsection{Combined signal significance}

The combined ATLAS Higgs boson discovery potential over the full mass range, 
100 $< m_H <$ 1000~\Gcs, assuming an integrated luminosity of 
30 \fbs\ is shown in Fig.~\ref{f:signif02}.  
The full mass range up to $\sim$1~\Tcs\ can be covered with a signal 
significance of more than 5$\sigma$ with several discovery channels
available at the same time. In this evaluation  
no K-factors have been included. 

A similar performance has also been established
for the CMS experiment.\cite{cms-higgs} The corresponding discovery potential is 
shown in Fig.~\ref{f:signif-cms}. It should be noted that for the 
\hgg, $H \to ZZ^{(*)} \to 4 \ell$ and 
$H \to WW^{(*)} \to \ell \nu \ \ell \nu$ channels the analyses have also been 
performed by including NLO K-factors for the signal and 
background processes. The results obtained are included in Fig.~\ref{f:signif-cms}.

The various discovery channels available at the LHC are complementary both from physics and 
detector aspects. 
The different channels test three different production mechanisms,
the gluon fusion, the vector boson fusion and the associated $\ttbar H$ 
production.  
This complementarity also provides sensitivity to non-standard Higgs models, 
such as fermiophobic models.\cite{fermiophobic} 

Also from the experimental point of view the Higgs boson discovery potential 
at the LHC is robust. The searches exploiting the various production
and decay modes 
are complementary in the sense that different detector components are important 
for different channels. 
The $H \rightarrow \gamgam$ decays require excellent electromagnetic 
calorimetry. In the identification of vector boson fusion the measurement
of jets, in particular the reconstruction of the forward tag jets, is essential. 
The Higgs detection in \bbbar\ decays via the associated $\ttbar H$ production   
relies to a large extent on an excellent b-tagging performance.

\subsection{Outlook}

Given the recent progress in the calculation of higher order QCD corrections
for both signal and background processes 
and their implementation in form of 
Monte Carlo programs,\cite{mcfm,mc-nlo,nlo-mc-vbf} activities have started to
perform the experimental simulation using these new programs. 
In addition, new methods to match 
matrix element and parton shower calculations have been developed.\cite{sherpa}

In parallel, the simulation of the detector response is
performed in more and more detail and the reconstruction 
algorithms are further developed and improved. 
It can therefore be expected that the uncertainties on the 
estimates of the signal significance will decrease in the near future. 

In that context it is important to note that the steadily 
increasing data samples collected at the Tevatron allow 
to test the various background Monte Carlo predictions.
 This is an essential step to give confidence in 
background estimates for the LHC, such that a potential
Higgs boson signature can be identified reliably.

\pagebreak

\section{Measurement of Higgs Boson Parameters at the LHC }
After a possible observation of a Higgs boson at the LHC it will  
be important to establish its nature. Besides a precise measurement of 
its mass, which enters electroweak precision tests, a determination of 
the spin and CP-quantum numbers is important. To establish that the  
Higgs mechanism is at work, measurements of 
the couplings of the Higgs boson to fermions and bosons as well as a 
demonstration of the Higgs boson self coupling are vital.
The LHC potential for a measurement of these parameters is discussed 
in the following.

\subsection{Mass and width \label{s:SM-mass}}
A precise measurement of the Higgs boson mass can be 
extracted from those channels where the invariant mass can be 
reconstructed from electromagnetic calorimeter objects, as in the case 
of $\hgg$ and $\hzzsfourl$ decays.\cite{physics-tdr,drollinger-mass} 
With an integrated luminosity of 
300~\fbs, each experiment would measure the Higgs boson mass with a precision of 
$\sim$0.1\% over the mass range 100 -- 400~\Gcs. The measurement error is 
determined by the absolute knowledge of the lepton energy scale, which is 
assumed to be $\pm$0.1\%. The precision could be slightly improved in the mass 
range between $\sim$150 and 300~\Gcs\ if a  
scale uncertainty of $\pm$0.02\% could be achieved. For larger masses, the 
precision deteriorates because the Higgs boson width becomes large and the 
statistical error increases. However, even for masses around  
700~\Gcs\ a precision of about 1\% can be reached. 

The width of a Standard Model Higgs boson can be measured directly 
only for masses larger than about 200~\Gcs, where the intrinsic 
width of the resonance is comparable to or larger than the experimental 
mass resolution.
 This is the mass region covered mainly by searches for \hzzfourl\ decays. 
It has been estimated\cite{physics-tdr} that over the mass range 
300 $< m_H <$ 700~\Gcs\ the precision of the measurement of the Higgs boson
width is approximately constant and of the order of 6\%. 

In the low mass region, $m_H \lesssim$ 200~\Gcs, the 
Higgs boson width can be constrained only indirectly by using the 
visible decay modes, as proposed and discussed in 
Refs.~\refcite{zeppenfeld-width} and \refcite{duehrssen}.

\subsection{Spin and CP eigenvalue}
After finding a resonance, one of the first priorities must be a  
determination of the spin and the CP eigenvalues. Recently, an
analysis to determine these quantum numbers has been 
presented in Ref.~\refcite{spin-cp}. The decay mode
\hzzfourl\ has been used in the mass range above 200~\Gcs\ to 
extract information on the spin and CP eigenvalue by studying two distributions: 
(i) the distribution of the cosine of the polar angle $\cos \theta$ of the 
decay leptons relative to the $Z$ boson momentum and (ii) the distribution of 
the angle $\phi$ between the decay planes of the two $Z$ bosons in 
the rest frame of the Higgs boson. 
Since a heavy Higgs boson decays
mainly into longitudinally polarized vector bosons, the cross section 
$d \sigma / d cos\theta$ should show a maximum around $\cos \theta = 0$. 
For a Standard Model Higgs boson a (1 + $\beta \ \cos 2 \phi$) like behaviour 
is expected for the angle $\phi$. 
However, the small vector coupling of leptons as well as experimental 
acceptance and resolution effects flatten this distribution. 
Apart from the Standard Model CP=0$^+$ Higgs boson, also a vector,  
pseudovector and a pseudoscalar scenario have been studied.

For Higgs boson masses above 250~\Gcs, the distribution of the polar angle
provides a good measurement of spin and CP. 
All scenarios considered can be separated from the Standard Model case 
with a significance of more than 
8$\sigma$, assuming an integrated luminosity of 100 \fbs. The decay plane 
angle correlation becomes more important for lower Higgs boson masses, 
where the discrimination power of the polar angle variable 
decreases and a determination of spin and CP is more difficult. 
However, with higher integrated luminosities non-Standard Model scenarios 
could still be ruled out, {\em e.g.,}
for $m_H$ = 200~\Gcs\ the spin-1, CP-even
hypothesis can be rejected with a significance of 6.4$\sigma$, while for the 
spin-1, CP-odd case, the significance is at the level of 3.9$\sigma$. 

In Ref.~\refcite{zerwas-spin} the method has been systematically generalized 
to arbitrary spin and parity assignments ($J^P$) of the decaying particle. 
It has been shown that for $m_H > 2 \ m_Z$ any odd spin state can be ruled 
out. Even spin states with $J > 2$ may mimic the spin-0 case, but they 
could, nevertheless, be ruled out by measuring
angular correlations of the $Z$ bosons  
with the initial state.\cite{zerwas-spin} 
Below the threshold for real ZZ production, {\em i.e.}, for $H \to Z Z^*$ decays, 
the study of the threshold behaviour of the invariant mass spectrum of 
the off-shell $Z$ boson is used as a key element. The measurement of this 
distribution in 
combination with angular correlations allows to rule out 
non Standard Model
$J^P$ assignments. However, a full experimental simulation, including 
backgrounds and detector resolution effects, still needs to be carried out. 

In this context it should be noted that the $J = 1$ hypothesis
could also be ruled out by observing non-zero $H \gamma \gamma$ and 
$H g g $ couplings.

As suggested in Ref.~\refcite{gunion-spin}, the $\ttbar H$ production
channel could potentially be used to distinguish a
CP-even from a CP-odd Higgs boson. The method  
proposed requires the reconstruction of the momenta of both top quarks. 
The corresponding studies by the LHC collaborations have not yet been made.

Higgs boson production via weak boson fusion can be used to determine the 
CP properties via the azimuthal angle distribution of the two outgoing
forward tag jets.\cite{CP-zeppenfeld} The technique is independent of the 
Higgs boson mass and of the observed decay channel.

\subsection{Couplings to bosons and fermions \label{s:couplings}}
For a Higgs boson with $m_H < 2 \ m_Z$,
the experiments at the LHC will be able to observe
the dominant decays into bosons and heavy fermions over mass ranges where the 
branching ratios in question are not too small. Decays to light fermions, {\em i.e.}, 
to electrons, muons or to quarks lighter than the b-quark, will, however, 
not be observable either due to the small decay rate or due to the overwhelming 
background from QCD jet production.  

In that mass range, the total width is expected to be 
small enough such that the narrow width approximation can be used to extract 
the couplings. 
The rates $\sigma \cdot BR(H \to f\bar{f})$ measured for final 
states $f\bar{f}$ are to a good approximation proportional to 
$\Gamma_i \cdot \Gamma_f / \Gamma$, where $\Gamma_i$ and $\Gamma_f$ are
the Higgs boson partial widths involving the couplings at production and decay,
respectively, and $\Gamma$ is the total width of the Higgs boson. 

The strength of the LHC in the coupling measurements is based on the simultaneous 
information which, for a given Higgs boson mass, is available in the various 
production and decay modes. Recently, a study has
been performed where the full information of all accessible production and decay 
channels, as listed in Table~\ref{t:higgs-channels}, is used to fit the coupling 
parameters.\cite{duehrssen} In this study, the correlations among
the various channels as well as  
experimental and theoretical systematic uncertainties are taken into account.

\begin{table}
\tbl{List of all studies used in the global likelihood fit performed in 
Ref.~\protect\refcite{duehrssen}; each channel has been used in the mass
 range indicated. \label{t:higgs-channels}}
{\begin{tabular}{l  | l | c }
Production mode & Decay mode  & Mass range (\Gcs) \\
\hline
Gluon fusion    & $H \to Z Z^{(*)} \to \ell\ell \ \ell \ell $           & 110 -- 200  \\
                & $H \to W W^{(*)} \to \ell \nu \ \ell \nu$ & 110 -- 200  \\
                & $H \to \gamma \gamma                  $ & 110 -- 150  \\
\hline
Vector boson    & $ H \to  Z Z^{(*)} \to    \ell \ell \ \ell \ell$   
& 110 -- 200  \\
fusion          & $ H \to  W W^{(*)} \to   \ell \nu \ \ell \nu$ 
&110 -- 190  \\
                & $ H \to  \tau \tau \to   \ell \nu \nu \ \ell \nu \nu $ 
&110 -- 150  \\
                & $ H \to  \tau \tau \to   \ell \nu \nu \ had \ \nu $ 
&110 -- 150  \\
                & $ H \to  \gamma \gamma$ 
&110 -- 150  \\
\hline
\ttbar\ H production & $ H \to W W^{(*)} \to \ \ell \nu \ \ell \nu \ (\ell \nu)$ 
&120 -- 200  \\
                     & $ H \to \bbbar\ $       & 110 -- 140  \\
                    & $ H \to \gamma \gamma $  & 110 -- 120  \\
\hline
$W H$ production & $H \to  W W^{(*)} \to   \ell \nu \ \ell \nu \ (\ell \nu) $ & 150 -- 190  \\
                 & $H \to  \gamma \gamma $  & 110 -- 120  \\
$Z H$ production & $H \to  \gamma \gamma $  & 110 -- 120  \\
\hline
\end{tabular}}
\end{table}

Assuming that the measured values correspond to the Standard Model expectations, 
a likelihood function is formed which, for a given integrated luminosity, 
is based on the expected Poisson distribution of the event numbers and on the estimated 
systematic errors. These errors include a 5\% uncertainty on the integrated luminosity, 
uncertainties on the reconstruction and identification efficiencies of leptons (2\%), photons (2\%),
b-quarks (3\%) and on the forward jet tagging and jet veto 
efficiency (5\%). In addition, theoretical 
uncertainties on Higgs boson production (20\% for $ggH$, 15\% for $\ttbar H$, 7\% 
for $WH$ and $ZH$ and 4\% for vector boson fusion) and on
branching fractions (1\%) have been taken into account. 

\begin{figure}
\center{
\begin{minipage}{6.0cm}
\mbox{{\epsfig{file=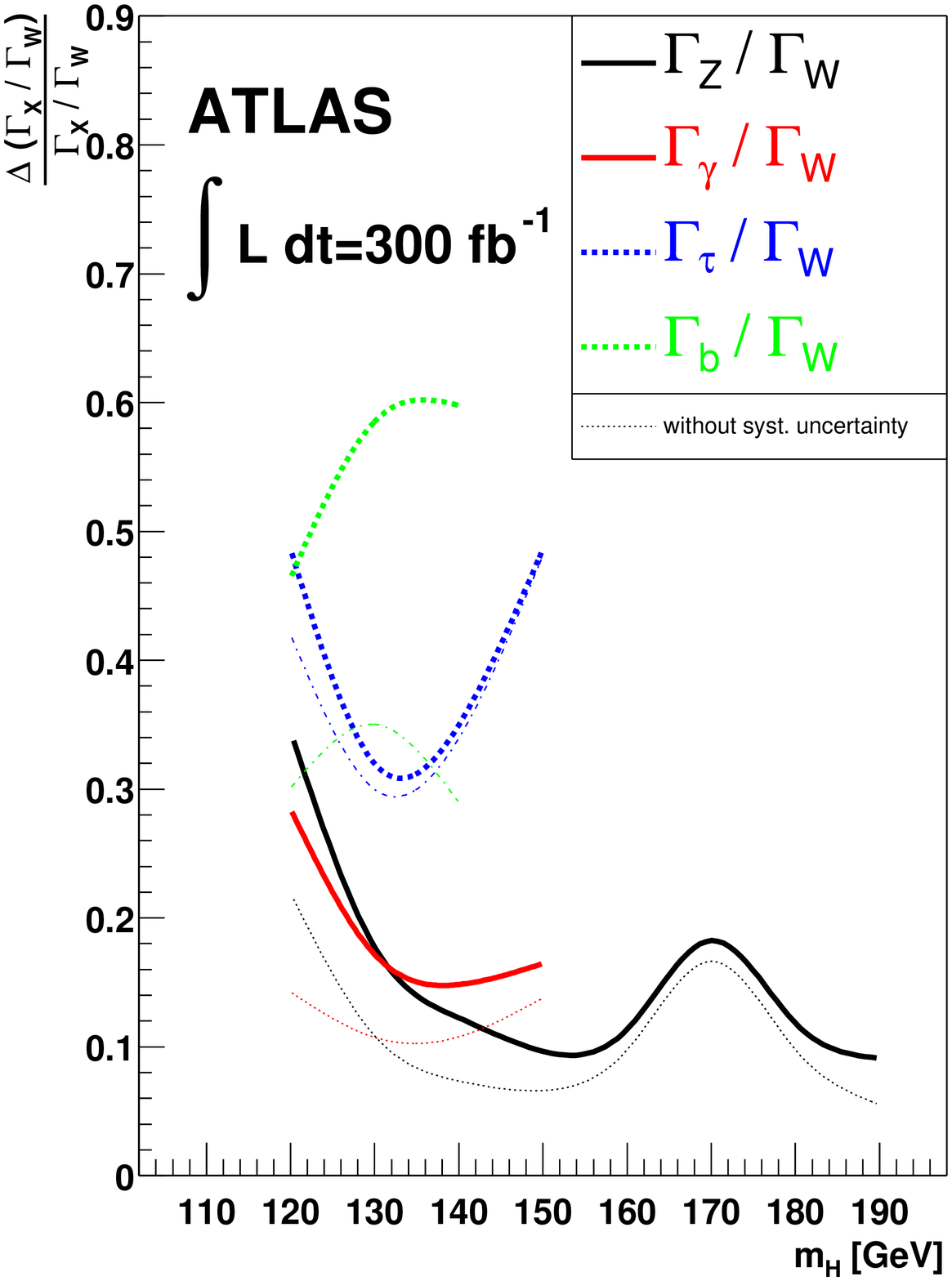,height=7.0cm}}}
\end{minipage}
\begin{minipage}{6.0cm}
\mbox{{\epsfig{file=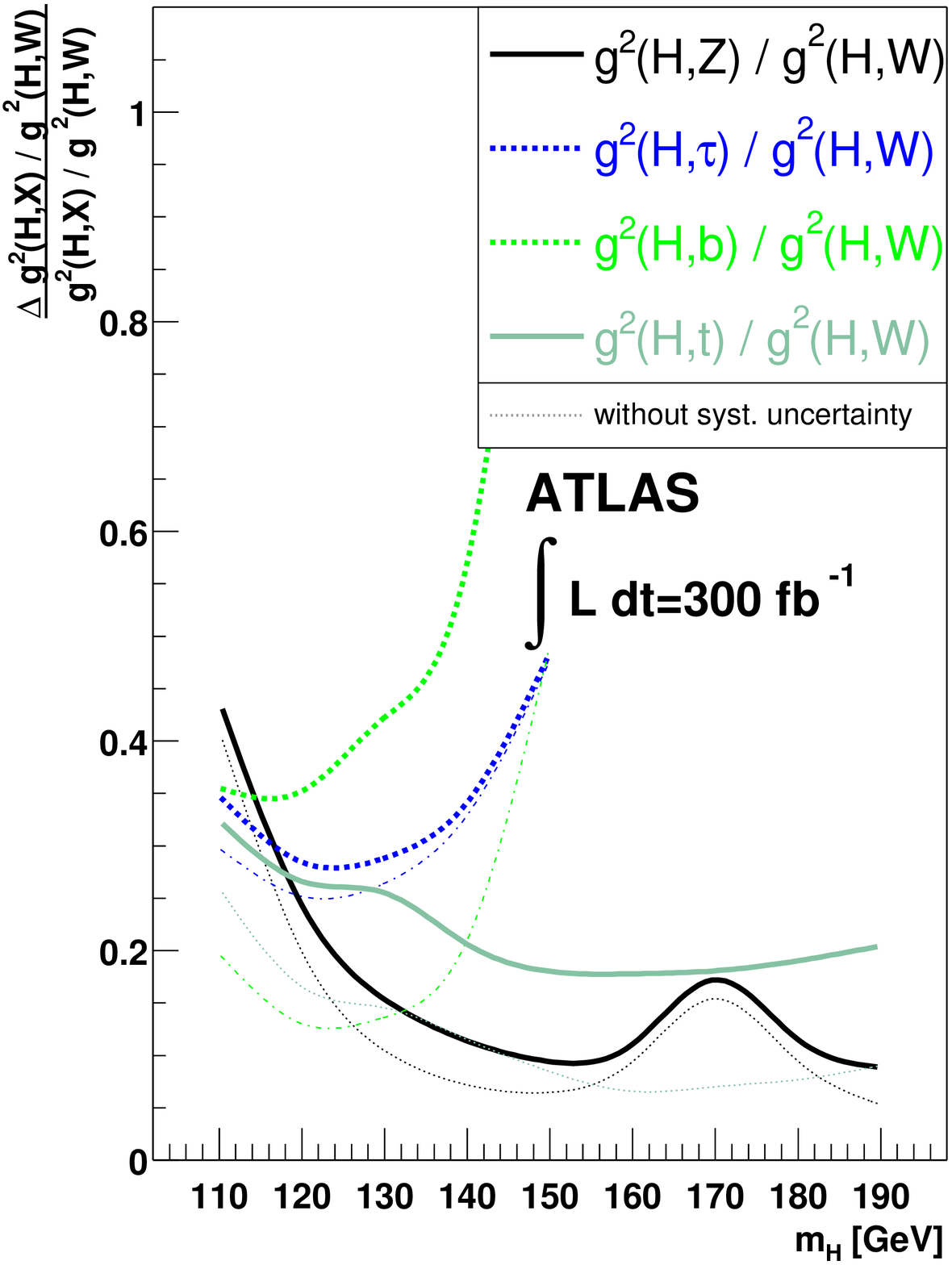,height=7.0cm}}}
\end{minipage}
}
\caption{\it 
The expected relative errors for the measurement of relative
branching ratios (left) and relative couplings (right), 
normalized to those of the $\hww$ decay and assuming an integrated luminosity of 300~\fbsit\ 
for one experiment. The dashed 
lines give the expected relative error without systematic uncertainties
(from Ref.~\protect\refcite{duehrssen}).
}
\label{f:higgs_coupl}
\end{figure}

Under the assumption that only one scalar CP-even Higgs boson exists,
relative branching ratios, which are identical to ratios of 
partial decay widths, can be measured. The decay \hww\ is used as 
normalization since it can be measured over the full intermediate mass range with 
a relatively small error. In Figure~\ref{f:higgs_coupl}(left) the expected relative errors
on the measurement of ratios of Higgs boson branching ratios are shown, assuming an
integrated luminosity of 300 \fbs. 
In particular the 
ratios $\Gamma_Z / \Gamma_W$ can be measured with an accuracy of the order of 
10 - 20\% for Higgs boson masses above 130~\Gcs. The ratios  $\Gamma_{\tau} / \Gamma_W$
and $\Gamma_b / \Gamma_W$ are less constrained, with errors expected
to be of the order of 30 to 60\%. 
It should be noted that these results are based on information from the decay only
and do not exploit any information from the production.

If additional theoretical assumptions are made, information from the production 
can be used as well, and in particular a measurement of the
top-Yukawa coupling becomes possible via the strong dependence of the gluon fusion and 
$\ttbar H$ production cross sections on this coupling. 
Assuming that only the known particles of the
Standard Model couple to the Higgs 
boson and that no couplings to light fermions are extremely
enhanced, all accessible  Higgs boson 
production and decay modes at the LHC can be expressed in terms of the Higgs boson
couplings $g_W, g_Z, g_t, g_b$ and $g_{\tau}$. The production cross sections depend on 
the square of these couplings. 
The exact dependence has to be calculated theoretically and put into the fit with 
corresponding systematic uncertainties. The gluon
fusion production cross section, for example, is not strictly 
proportional to the top-Yukawa coupling squared, but receives
additional contributions from the interference with diagrams containing a b-loop. 
Using the program of Ref.~\refcite{b-contributions}, they have been found to be at the 
level of 7\% for \mh\ = 110~\Gcs\ and 4\% for \mh\ = 190~\Gcs. These additional
contributions are ignored and it is assumed that the $b$-coupling is not  
enhanced by a factor of 10 or more compared to the Standard
Model. Furthermore, the interference 
effect between $W$- and $Z$-fusion in the vector 
boson fusion is expected to be small ($<$ 1\%)\cite{spira-xsect} and is ignored.

Similarly, the Higgs boson branching ratios are 
proportional to $g^2/ \Gamma$, and again, the 
proportionality factors are taken from theory, assuming a relative uncertainty of about 
1\%. The decay
\hgg\ proceeds either via a $W$ or a top-quark loop, with destructive interference between
the two. The relative contributions are taken from Ref.~\refcite{gammagamma-loop}. 

In Figure~\ref{f:higgs_coupl}(right) the
relative errors on the measurement of relative couplings are  
shown for an integrated luminosity of 300\,\fbs\ for one experiment.
Due to the large contributions of the gluon-fusion
and  $\ttbar H$ production modes,  
the ratio of the top-Yukawa coupling to the Higgs boson coupling to $W$
bosons can be well constrained with an estimated uncertainty of the order of 10 to 20\%. 

Recently, the determination of Higgs boson couplings at the LHC has been discussed 
for general multi-Higgs doublet models.\cite{coupl:les-houches-04} 
Imposing the additional constraint that the $HWW$ and $HZZ$ couplings are bound 
from above by their Standard Model values ($g^2_{W/Z}(SM)$), 
{\em i.e.}, $g^2_{W/Z} < 1.05 \cdot g^2_{W/Z}(SM)$, an absolute measurement of 
the Higgs boson couplings to the vector bosons and the $\tau$ lepton, $b$ and $t$ quark 
becomes possible. The constraints imposed are theoretically motivated and 
are valid in particular for the MSSM. Any model that contains only Higgs doublets and/or 
singlets satisfies the relation $g^2_{W/Z} < g^2_{W/Z}(SM)$. 
The extra 5\% margin allows for theoretical uncertainties in the translation between 
couplings-squared and partial widths and also for small admixtures of exotic Higgs
states, as for example SU(2)
triplets.\cite{coupl:les-houches-04} Contributions of additional   
particles running in the loops for \hgg\ and $gg \to H$ are allowed for and their 
contribution to the partial width is fitted. Assuming an integrated luminosity of 
300 \fbs\ (for all channels, except vector boson fusion, for which 
100 \fbs\ have been assumed)
and a combination of two experiments, typical accuracies for the absolute 
measurement of the couplings of about 20 to 30\% can be achieved for Higgs boson masses 
below 160~\Gcs. For masses above the $W$-pair threshold the measurement of the 
$W$ and $Z$ partial width can be performed with an accuracy at the level of $\pm$10\%.

\subsection{Higgs boson self coupling}
To fully establish the Higgs mechanism, it must be demonstrated
that the shape of the Higgs  potential $V_H$ has the form required for
electroweak symmetry breaking.
The potential can be expressed in terms of the physical Higgs field $H$ as
$V_H = \frac{m_H^2}{2} H^2 + \frac{m_H^2}{2v} H^3  + \frac{m_H^2}{8v^2} H^4$, 
where $v = (\sqrt{2} G_F)^{-1/2} $ = 246~\Gcs\ is the Higgs vacuum expectation value. The 
coefficients of the second and third term are proportional to the strength of the Higgs 
trilinear ($\lambda_{HHH} = 3 m_H^2 / v$) and quartic ($\lambda_{HHHH}' = 3 m_H^2 / v^2$)
Higgs boson self coupling, respectively. In order to extract information on these 
couplings, multiple Higgs boson production must be measured.\cite{higgs-self}
Since the quartic coupling is about two orders of magnitude smaller than the 
trilinear coupling, present studies have focussed on the determination
of $\lambda_{HHH}$.

The coupling strength $\lambda_{HHH}$ enters the production rate of Higgs  
boson pairs. At LHC energies, the inclusive Higgs boson pair production is dominated by gluon 
fusion.\cite{hpair-prod} Other processes have cross sections which are factors of
10 to 30 smaller. 
For $\mh >$ 140~\Gcs, $H \to W W$ is the dominant decay mode and  
$W W \ W W$ final states are produced with the largest branching ratio. In order to 
suppress the large backgrounds from \ttbar\ and multiple gauge boson production, 
like-sign lepton final states offer
the best sensitivity.\cite{self-slhc,self-baur,self-federica} 
Studies have shown that the $(\ell^{\pm} \nu jj) \ (\ell^{\pm} \nu jj)$ final state 
has the highest sensitivity for extracting information on the self coupling parameter 
$\lambda_{HHH}$. 

The studies conclude that with data corresponding to 
the ultimate luminosity expected at the LHC of 300\,\fbs\ per experiment, a determination of the  
Higgs boson self coupling will not be possible. 
A measurement with reasonable errors will require a luminosity upgrade, 
{\em i.e.}, the realization of the so called {\em Super LHC} (SLHC).

In the analysis of Ref.~\refcite{self-baur} it is concluded that the LHC may rule out 
the case of a non-vanishing $\lambda_{HHH}$ using the four $W$ final state for a mass range 
of 150 $< \mh <$ 200~\Gcs\ with a confidence level of 95\%. For the SLHC, assuming 
an integrated luminosity of 3000\,\fbs\ per experiment, it is claimed that a measurement 
of $\lambda_{HHH}$ with a precision of 20\% will be possible. In this study NLO K-factors have 
been used.

These conclusions still need to be confirmed in a full simulation of the 
detector performance. First preliminary studies\cite{self-federica} confirm the 
sensitivity at the SLHC, however, 
some background contributions might have been underestimated. Further
studies to clarify these issues are currently in progress.

The mass region $m_H \lesssim$ 140~\Gcs\ is considered to be very challenging for 
a measurement of the Higgs boson self coupling, even at the SLHC.\cite{self-baur}
Recently, it has been proposed to use the rare decay mode $ HH \to \bbbar \ \gamma \gamma$ 
to investigate the self coupling in that mass region.\cite{baur-bbgg}
At the LHC, assuming an integrated luminosity of 300\,\fbs\ per experiment
and including NLO K-factors, only six signal events are expected for
the most optimistic case of $m_H$ = 120~\Gcs.
Using this decay mode would require a luminosity upgrade to rule 
out $\lambda_{HHH} = 0$ at the 90\% confidence level.\cite{baur-bbgg} 
Also in this case, a detailed simulation of the 
detector performance has not yet been completed.

\pagebreak
\section{Search for MSSM Higgs Bosons at the LHC}

In addition to the excellent prospects for Standard Model Higgs searches, the 
LHC experiments have a large potential in the investigation of 
the MSSM Higgs sector. The experimental searches carried out at LEP 
and presently continued at the Tevatron can be 
extended to much larger Higgs boson masses. 
In the search for the light, Standard Model-like Higgs boson 
$h$ the same channels as in the search for the Standard Model Higgs 
boson will be used. Heavier Higgs bosons will be searched for 
in additional decay channels which become accessible in certain 
regions of the MSSM parameter space due to enhanced couplings, {\em e.g.},   
the decay mode $H/A \to \tau \tau$ at large \tanb.
Decays into $\tau$ leptons also contribute to the search for 
charged Higgs bosons, which at the LHC can
be extended to masses beyond the top-quark-mass. 
A comprehensive and complete 
study of the LHC discovery potential in the MSSM has first been 
presented in Ref.~\refcite{mssm-ela}. 
In that study, the discovery potential had been determined for two 
benchmark scenarios with different assumptions on mixing 
in the stop sector, as defined at LEP.\cite{LEP-MSSM}
In a so-called {\em no-mixing scenario} the trilinear coupling in the stop sector 
$A_t$ is set to zero,
whereas in the so-called {\em maximal mixing scenario} the 
value $A_t = \sqrt{6} \ m_{SUSY}$, with 
$m_{SUSY}$ = 1 \Tcs, is used. 
In the following, the search strategies for MSSM Higgs bosons at the 
LHC are briefly discussed and the discovery potential is presented for different 
MSSM benchmark scenarios. 
Throughout this chapter, all discovery contours for MSSM Higgs bosons
in the \matb-parameter space are given for a signal significance at
the 5$\sigma$ level. In most analyses  
SUSY particles are considered to be heavy enough  
to play a negligible r{$\hat{\rm{o}}$}le in the phenomenology of Higgs boson 
decays. This assumption is abandoned in Section \ref{s:susydecays}, where the interplay of 
the Higgs sector and SUSY particles is discussed.

\subsection{Search for the light CP-even Higgs boson $h$ \label{s:MSSM-light}}
\begin{figure}
\center{
\mbox{{\epsfig{file=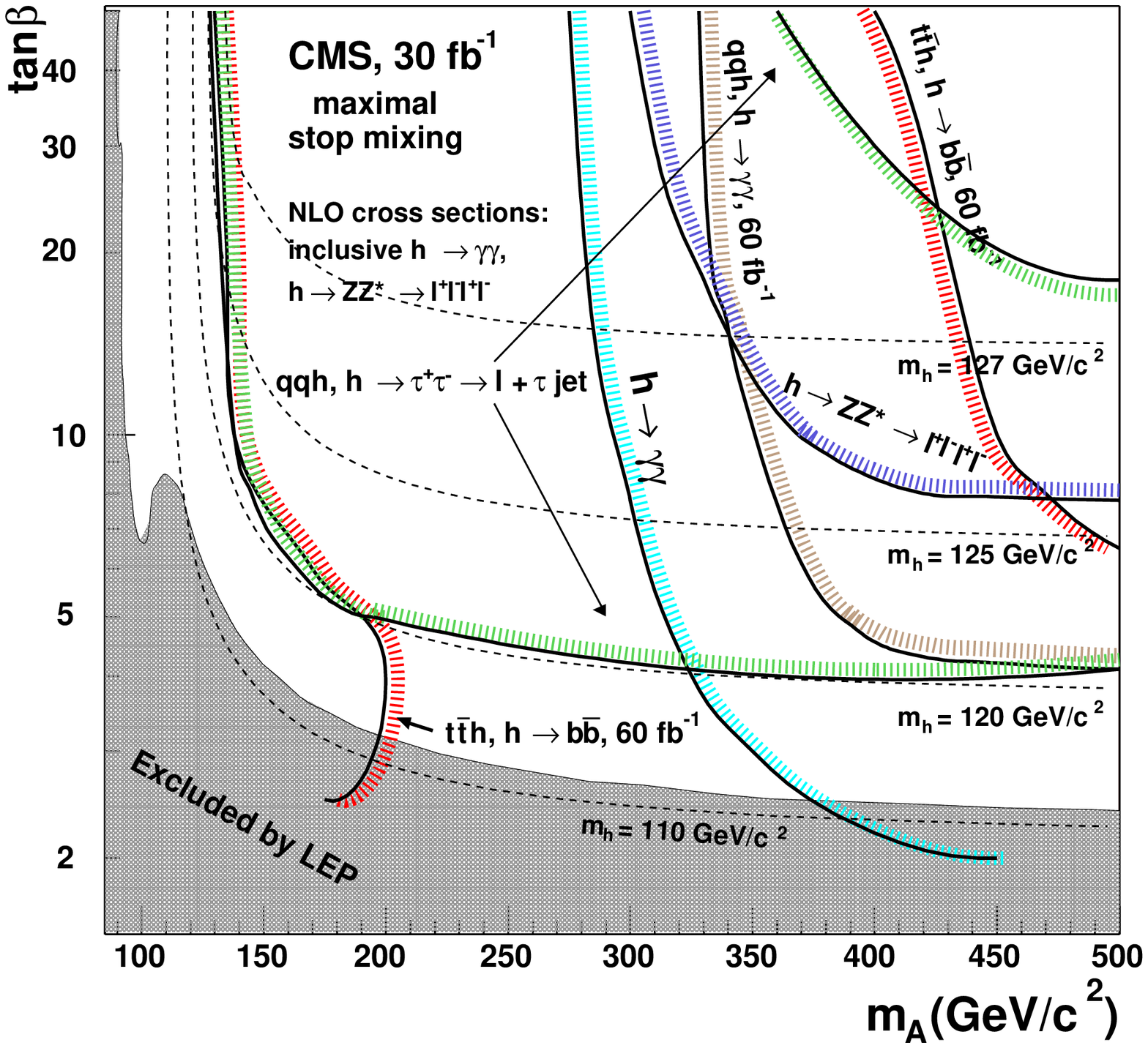,height=5.7cm}}}
\mbox{{\epsfig{file=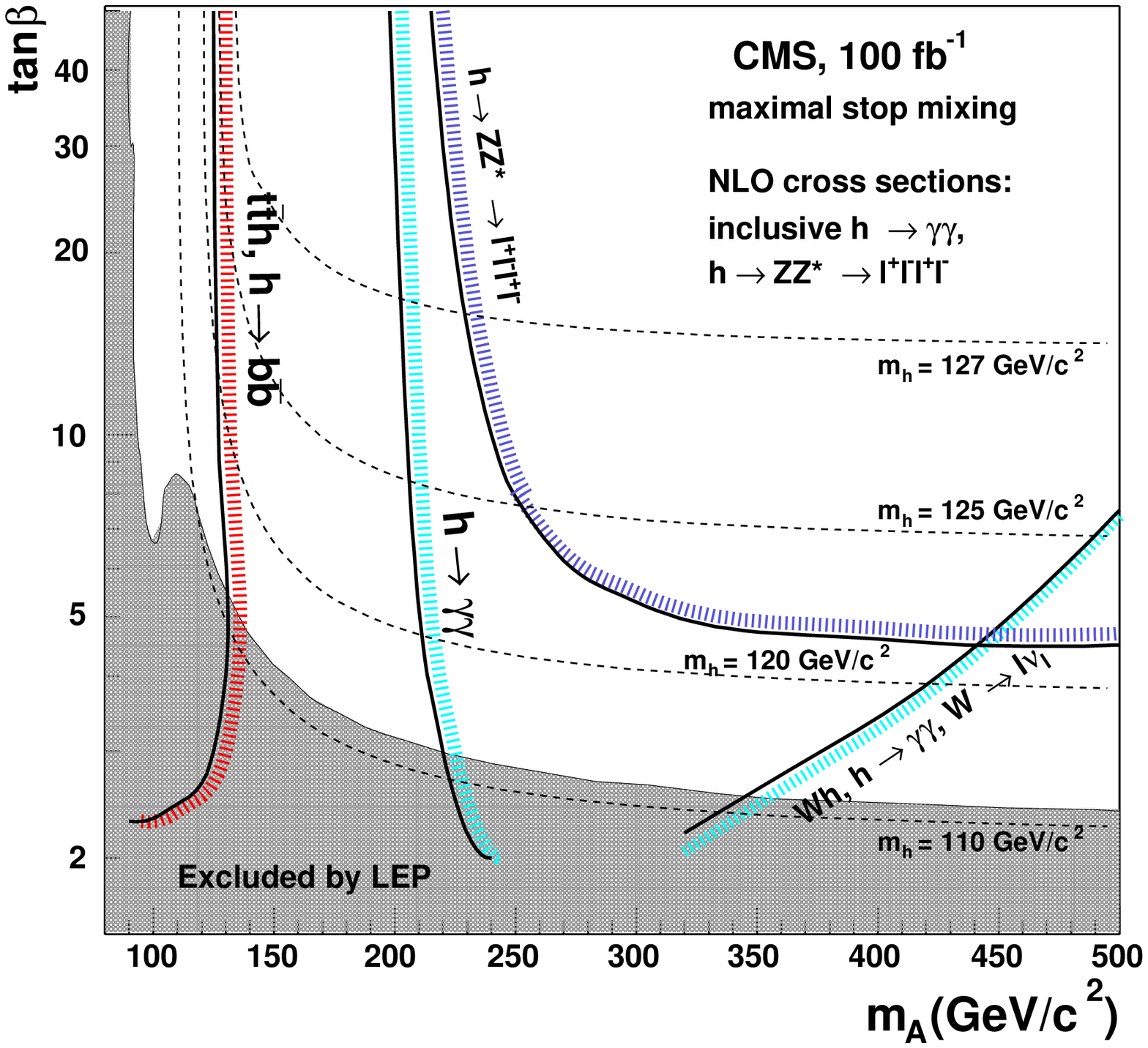,height=5.7cm}}}
}
\caption{\it 
(Left): The 5$\sigma$ discovery potential for the light CP-even MSSM-Higgs boson h in 
the \matbit-plane assuming maximal stop mixing. Integrated luminosities 
of 30 and 60\,\fbsit\ are assumed for the gluon-fusion channels (\hgg\ and \hzzsfourl)
and for the associated $\ttbar h$ and vector boson fusion channels ($qqh$), respectively. 
For the gluon-fusion channels NLO cross sections are used.
(Right): The same, assuming an integrated luminosity of 100\,\fbsit\
for all channels (from Ref.~\protect\refcite{cms-higgs}).
}
\label{f:mssm-sig-h}
\end{figure}

The channels discussed in Section \ref{s:higgs-lhc} can be used to 
search for the lightest MSSM CP-even Higgs boson $h$. Results for the 
discovery potential   
obtained in studies carried out by the CMS collaboration  
are shown in the  \matb-plane in Fig.~\ref{f:mssm-sig-h}.
In these studies, maximal stop mixing, a top-quark mass of 175~\Gcs,
m$_{SUSY}$ = 1~\Tcs\ and integrated luminosities of 30, 60 and 100 \fbs\
have been assumed.\cite{cms-mssm} 
In addition to the \hgg, \hzzsfourl\ and $ttH$ with $ H \to \bbbar$ channels  
the vector boson fusion channel $qqH \to qq \tau \tau$ contributes 
significantly to the discovery potential. For an integrated luminosity 
of 100~\fbs\ the whole parameter space, except the region 
90 $< m_A <$ 130~\Gcs, can be covered with a significance of more than 5$\sigma$
by a single experiment. 

In the maximal mixing scenario, the Higgs bosons $h$ and $A$  are
 nearly degenerate in mass in the uncovered region up to 
$m_A \sim$125~\Gcs. At large \tanb\ the enhanced 
coupling to $b$ quarks can be used to search for these Higgs bosons in the 
$pp \to \bbbar \ h/A \to \bbbar \ \mu \mu$ channel. It has been shown that 
signals can be extracted with a significance above 5$\sigma$ for 
105 $< m_A <$ 125~\Gcs\ and $\tanb \gtrsim$15.\cite{gonzales-ros,gentile} 
The signal extraction is challenging due to the overwhelming background from 
$Z \to \mu \mu$ production.
It will therefore be essential to have a good  b-tagging
performance and a good dimuon mass resolution as well as to control
the background from data, using $Z \to \ell \ell$ control samples.
It should be noted that this particular part of the parameter space can also be covered 
with a significance exceeding 5$\sigma$ by   
searches for charged Higgs bosons (see Sections \ref{s:mssm-hplus} and  \ref{s:mssm-benchmarks}).

\subsection{Search for the heavy MSSM Higgs bosons H and A \label{s:mssm-HA}} 

\subsubsection{The large \tanbit\ region}
At large \tanb\ where the couplings of the heavy Higgs bosons $H$ and $A$
to down-type fermions are enhanced, the associated Higgs boson production 
with a \bbbar\ pair is the dominant production mode. The  
decay into $\tau$ pairs has a significant branching ratio over
a large region of parameter space, such that the 
process $\bbbar \ H/A \to \bbbar \ \tau \tau$ plays a key  role. 
This channel can be complemented by the $H/A \to \mu \mu$ decay mode, 
for which the much smaller branching ratio is compensated by the 
better signal-to-background ratio.
In addition, a search in the $\bbbar \bbbar$ final state 
has been proposed in the literature,\cite{gunion-bbbb}
but it suffers from sizeable background from QCD 
jet production and therefore has only a limited discovery potential.\cite{physics-tdr} 

\begin{figure}
\center{
\mbox{{\epsfig{file=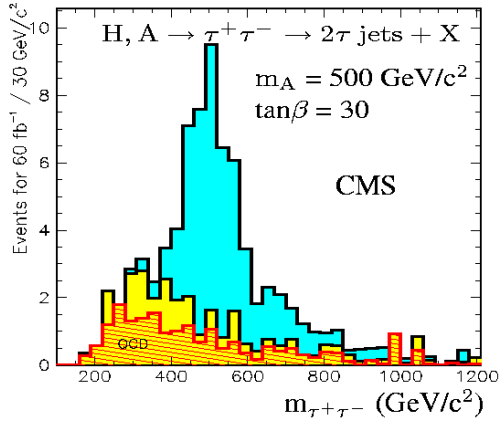,height=6.0cm}}}
}
\caption{\it 
The reconstructed $\tau \tau$ invariant mass distribution in the fully hadronic  
final state ($\tau_{had} \ - \ \tau_{had}$) from the process
$gg \to \bbbar \ H/A \to \bbbar \ \tau \tau $. 
The H/A signal (dark) for $m_A$ = 500~\Gcsit\ and \tanbit=30 
is shown on top of the background for an integrated luminosity of 
60 \fbsit\ (from Ref.~\protect\refcite{cms-higgs}).
}
\label{f:mssm-signals}
\end{figure}

\begin{figure}
\begin{center}
\mbox{\epsfig{file=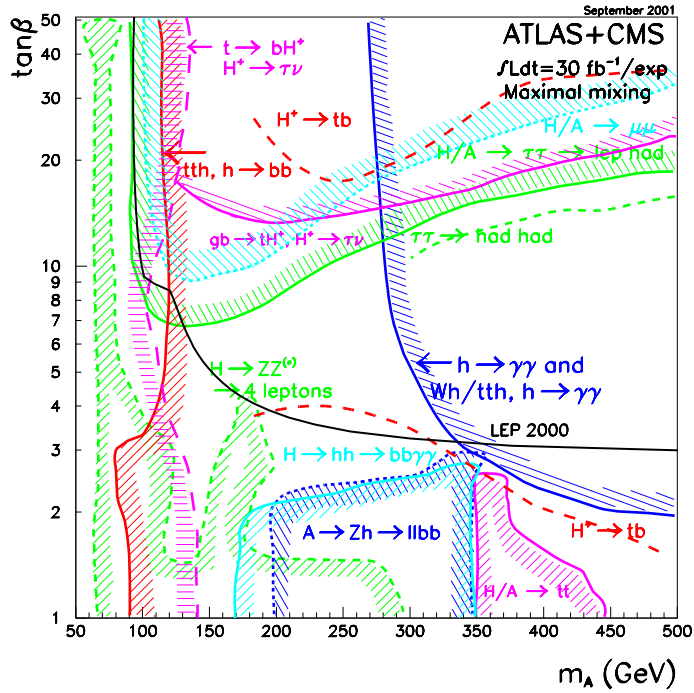,width=10.0cm}}
\end{center}
\caption{\it 
The combined sensitivity of the ATLAS and CMS experiments for the discovery of 
MSSM Higgs bosons in the maximal mixing scenario for an integrated luminosity of 
30\,\fbsit. The 5$\sigma$ discovery contour curves are shown in the \matbit-plane 
for individual channels not including the vector boson fusion channels. 
The limit from the LEP experiments is superimposed
(from Ref.~\protect\refcite{ela-mssm}).}
\label{f:mssm-contours}
\end{figure}

For the $\bbbar \ \tau \tau$ final states both excellent b-tagging
and tau identification performance are essential to suppress the 
large backgrounds from $\gamma^*/Z$, QCD multijet and $W$+jet production. 
Detailed studies have shown that both leptonic and
hadronic tau decays ($\tau_{had}$) can be used to isolate signals 
from heavy Higgs boson production.\cite{donatella-juergen,cms-mssm-taus} 
For  
$m_A  \lesssim 400$~\Gcs\ the final state with one leptonic and 
one hadronic tau decay  ($\ell-\tau_{had}$) has the largest 
discovery reach. At large $m_A$ the double hadronic decay mode contributes 
significantly, while for
small Higgs boson masses its reach is limited by the low trigger efficiency.
In all decay modes the $\tau \tau$ invariant mass can be reconstructed using 
the collinear approximation, as discussed in Section \ref{s:lhc-vbf}.  
An example of a reconstructed $H/A$ signal with $m_{H/A}$ = 500~\Gcs,
based on a simulation of the CMS detector performance in the 
$\tau_{had}-\tau_{had}$ decay mode, is  
shown in Fig.~\ref{f:mssm-signals}. An integrated luminosity of 60 \fbs\ and
\tanb\ = 30 have been assumed.  
The discovery contours for the various $\tau \tau$ decay modes are 
shown in the \matb-plane in Fig.~\ref{f:mssm-contours} for the combination
of both LHC experiments assuming an integrated luminosity of 30~\fbs.

As mentioned above, also the $\mu \mu$ decay mode 
contributes to the discovery potential. Given the good mass 
resolution for dimuon final states, 
this channel can eventually be used to separate nearby Higgs boson 
resonances using a precise measurement of the shape of the 
mass distribution. 
For small \mA\ it covers regions of parameter space not yet 
excluded by the LEP experiments (see discussion at the end of 
Section \ref{s:MSSM-light}).

\subsubsection{The small \tanbit\ region} 
In the region of small \tanb\ heavy Higgs bosons can be searched for via their 
decays into the lightest Higgs boson.  
This allows for a simultaneous observation of two Higgs bosons,
for example via the decay modes $H \to hh$ and $A \to Z h$.

The final states of interest for a search in the $H \to hh$ decay mode are
$\bbbar \ \gamma \gamma$, $\bbbar \ \tau \tau$ and  $\bbbar \ \bbbar$.
The decays into $\bbbar \ \gamma \gamma$ 
can reliably be triggered on and offer good kinematic constraints 
and mass resolution for the reconstruction of $m_H$. 
However, due to the small branching ratio into $\gamma \gamma$, the discovery potential in that 
channel is limited by the small signal rate. 
For an integrated luminosity of 300\,\fbs, 
an obervation is possible in the parameter region 
$\tanb <$ 4 and  $2 \  m_h < m_H < 2 \  m_t$.\cite{physics-tdr} The 5$\sigma$ discovery 
contour is shown in Fig.~\ref{f:mssm-contours}, assuming an integrated luminosity of 
30\,\fbs\ for each LHC experiment.

Searches in both the $\bbbar \ \tau \tau$ and $\bbbar \ \bbbar$
final states have been found to be less promising.\cite{physics-tdr}
For the $\bbbar \ \tau \tau$ case, the signal 
extraction is difficult due to large backgrounds from $W$+jets and \ttbar\ production 
and due to the poor mass resolution for the signal.
A detection of the $H\to hh$ decay 
in the  \bbbar\ \bbbar\ mode requires a four-jet trigger with as low a 
\PT\ threshold as possible and excellent b-tagging performance to control the 
overwhelming background from four-jet events. 

The $A \to Zh$ decay mode has so far been studied in the final
states $Zh \to \ell \ell \ \bbbar$ and $Zh \to \bbbar \
\bbbar$.\cite{physics-tdr,cms-higgs}  
A search in the latter decay mode is as
challenging as the $H \to hh \to \bbbar \ \bbbar$ channel discussed above.
The decays into $\ell \ell \ \bbbar$ can easily be triggered on and offer the largest rates 
apart from the dominant $\bbbar \ \bbbar$ decays. 
Given that the signal rate is rapidly falling with increasing \tanb, the 
$A \to Z h \to \ell \ell \ \bbbar$ channel can only be observed at small 
\tanb\  and for 200~\Gcs $ < m_A < 2 \ m_t$. 
The 5$\sigma$ discovery 
contour is shown in Fig.~\ref{f:mssm-contours}, assuming an integrated luminosity of 
30\,\fbs\ for each LHC experiment. 

For $m_A > 2 \ m_t$ and $\tanb \sim 1$,  the $H \to \ttbar$ and $A
\to \ttbar$ branching ratios are close to  
100\%. Since the $H$ and $A$ bosons are almost degenerate in mass in the 
relevant region of parameter space, 
the two decays cannot be distinguished experimentally.  
As discussed in the literature,\cite{HA-ttbar} a signal from 
$H/A \to \ttbar$ would appear as a peak in the \ttbar\ invariant mass spectrum above
the \ttbar\ continuum background. The interference of the signal and background 
amplitudes leads to a strong suppression of the signal at high 
masses. For a Higgs boson mass of 370 (450)~\Gcs, the 
signal-to-background ratio is expected to be of the order of 9\% (1\%) only.
Under the optimistic assumption that the
background \ttbar\ mass spectrum will be known to better  
than 1\% from experimental data and theory, a signal extraction would be possible 
in a limited region of parameter space for $ 2 m_t < m_A <$
450~\Gcs.  The 5$\sigma$ 
discovery contours are shown in Fig.~\ref{f:mssm-contours}.

\subsection{Charged Higgs bosons \label{s:mssm-hplus}}

At the LHC, a charged Higgs boson can be detected in several scenarios.
For $m_{H^{\pm}} < m_t - m_b$, top-quark production represents a copious source 
via the decay $t \to H^\pm b$.
If the charged Higgs boson is heavier than the top quark, it can be produced 
via the process $gg \to H^\pm t b$ (or $gb \to H^\pm t$).\cite{production-hplus-t,hplus-nlo}
Additional contributions come from the Drell-Yan type
process $gg, \qqbar \to H^+ H^-$ and from the associated production with a 
$W$ boson, $\qqbar \to \hplus W^{\mp}$.\cite{gunion-hplus} 
Assuming a heavy SUSY mass spectrum, the charged Higgs boson decays into 
Standard Model particles only. For small \tanb\ and $\mhplus < m_t$, 
the main decay channels are $\hplus \to \tau \nu , c s, W h$ and $t^* b$. For mass 
values above the top-quark mass, the $\hplus \to t b$ decay mode becomes dominant, 
in particular at small \tanb. At larger \tanb, also the $\hplus \to \tau \nu$ decay 
has a significant branching ratio, which, however, decreases with increasing Higgs
boson mass (see Fig.~\ref{f:br_mssm} for details).

Due to the large \ttbar\ production cross section, 
charged Higgs bosons with $m_{H^{\pm}} < m_t - m_b$ can be detected nearly 
up to the kinematic limit. 
The LHC discovery potential has been studied for both 
the dominant $H^{\pm} \to \tau \nu$ and the $H^{\pm} \to qq'$ channel, via the decay chain
$pp \to \ttbar \to H^{\pm} b \ W^{\mp} \bar{b}$.\cite{physics-tdr,cms-higgs}
In the region 5 $< \tanb <$ 10, 
the hadronic decay mode contributes significantly and increases the discovery 
potential.\cite{dosil}

Following the studies presented in Refs.~\refcite{gunion-hplus-tb} and \refcite{roy-hplus},  
both LHC collaborations have recently evaluated their discovery potential for heavier 
charged Higgs bosons, {\em i.e.}, $m_{H^\pm} > m_t - m_b$.\cite{atlas-hplus,cms-hplus}
For the process $g b \to \hplus t \to t b \ t \to \ell \nu b  b \ q \bar{q} b$,
the large backgrounds can be sufficiently suppressed
if three b-tags are required and both top quarks are reconstructed. 
In addition, it has been studied whether the discovery 
range in the $\hplus \to t b$ decay mode can be extended if four b-jets are 
required in the final state (via the process 
$gg \to \hplus t b$).\cite{atlas-gghplus} 
Although this requirement reduces the background, 
the reconstruction of the charged Higgs boson is rendered more difficult 
due to the combinatorics in the reconstruction of the 
multijet final state. Overall, this 
channel does not lead to a significant extension of the discovery 
potential provided by the three b-tag analysis.

\begin{figure}
\center{
\mbox{{\epsfig{file=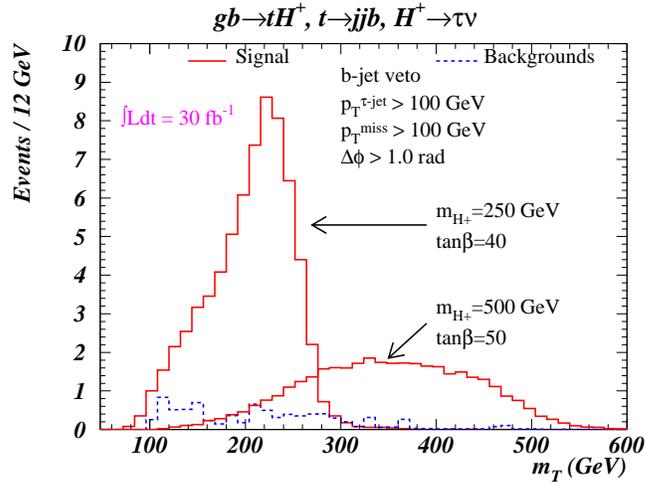,height=7cm}}}
}
\caption{\it 
The reconstructed transverse mass of the $\tau_{had}-\pet$ system from
the process $gb \to  H^+ t \to \tau \nu \ qq b $ for charged Higgs bosons with 
masses of 250~\Gcs (at \tanbit = 40) and 500~\Gcsit\ (at \tanbit\ = 50) in comparison to 
the background. The simulation has been performed for the ATLAS experiment
assuming an integrated luminosity of 30 \fbsit\ 
(from Ref.~\protect\refcite{atlas-hplus}).
}
\label{f:mssm-h+signal}
\end{figure}

More favourable signal-to-background conditions are found in the 
$\hplus \to \tau \nu$ decay mode for the production process 
$ gb \to \hplus t \to \tau \nu \ q q b$, with $\tau \to had \ \nu $.
Both the associated top quark and the tau lepton are required to 
decay hadronically. This has the advantage that the transverse mass of 
the $\tau_{had}-\pet$ system can be used to discriminate between 
signal and background. The signal distribution is expected to show 
a Jacobian peak structure at the Higgs boson mass. 
The channel must be triggered by a hadronic tau 
+ $\pet$ or hadronic tau + multijet trigger, both of which are foreseen in the
two LHC experiments.  
The backgrounds from $W$+jet, \ttbar, $Wt$ and QCD 
multijet production can be suppressed 
by requiring in addition to an identified hadronic tau exactly three high-\PT\ 
jets,  one of which must be b-tagged. The signal can be enhanced by 
exploiting the $\tau$ polarization in one prong tau decays,\cite{roy-hplus}
which leads to a harder spectrum of single pions when the tau originates
from an \hplus\ rather than from a $W$ decay. Requiring, {\em
e.g.}, that  more than 80\% of the visible tau 
energy is carried by the single charged pion, 
the \ttbar\ background can be reduced by a factor of $\sim$300 while the 
signal efficiency can be kept at the 10 -- 20\% level.\cite{cms-hplus}
The discrimination between signal and background is shown in 
Fig.~\ref{f:mssm-h+signal}, where the reconstructed transverse mass of the 
$\tau_{had}-\pet$ system is shown for two Higgs boson masses in comparison to the 
background. After a transverse mass cut, the backgrounds are small and
a charged Higgs boson can be discovered with a significance of more 
than 5$\sigma$ for $m_{H^\pm} > m_t$ in the large \tanb\ region. 
In the $\tau \nu$ decay mode,  the two b-tag final 
state (via $gg \to \hplus t b$ production) also contributes to the 
discovery potential.\cite{cms-mssm}

The discovery contours for both the $\hplus \to \tau \nu$ and the 
$\hplus \to t b$ channels are superimposed  
in the \matb-plane in Fig.~\ref{f:mssm-contours}. 
Recently, the discovery potential in the transition region 
($m_{H^\pm} \sim m_t$) has been studied more carefully.\cite{ketevi-trans}
In particular, it has been shown that charged Higgs bosons can be detected 
over the full transition region using the $\hplus \to \tau \nu$ decay mode.

Within the MSSM, the sensitivity to the $\hplus \to W^{(*)} h$ decay mode
is weaker.\cite{atlas-hplus} For Higgs boson masses below the top-quark 
mass, the extraction of the signal is only possible in the small \tanb\ 
region (1.5 $< \tanb <$ 2.5). 
For Higgs boson masses above the top-quark mass, the discovery 
potential in this channel is marginal. However, in extended models, 
as for example in the NMSSM,\cite{nmssm}
where the Higgs sector of the MSSM is extended by a  
complex singlet scalar field, the $\hplus \to W^{(*)} h$ channel shows 
a viable signal.\cite{atlas-hplus}

In addition, the s-channel Drell-Yan type production, $ q \bar{q}' \to \hplus \to \tau \nu$, 
has been investigated by using the hadronic tau decay mode and exploiting
the $\tau$ polarization.\cite{cms-dy-hplus} For this production mode 
it is difficult to extract 
a signal  since the reconstructed transverse mass distribution overlaps with 
the tail of the overwhelmingly large  $ q \bar{q}'
\to W \to \tau \nu$ background.

\subsection{The MSSM discovery potential in various benchmark scenarios \label{s:mssm-benchmarks} }

Different benchmark scenarios have been proposed for the interpretation of MSSM Higgs boson 
searches.\cite{mssm-benchmarks} 
In the MSSM, the masses and couplings of the Higgs bosons depend, in
addition to \tanb\ und $m_A$, on the 
SUSY parameters through radiative corrections. In a constrained model,
where unification of the SU(2) and U(1) gaugino masses is assumed, the
most relevant parameters are $A_t$, the trilinear coupling in the stop
sector, the Higgs mass parameter $\mu$, the gaugino mass term $M_2$,
the gluino mass $m_g$ and a common scalar mass $M_{SUSY}$. Instead of
the parameter $A_t$, the stop-mixing parameter $X_t := A_t - \mu \cot
\beta$ can be used.  
In particular the phenomenology of the light Higgs
bosons $h$ depends on the SUSY scenario, for which the following have been considered
in a recent study:\cite{schumacher}  
 
\begin{enumerate}
\item {\em $m_h$-max scenario}: the SUSY parameters
are chosen such that for each point in the \matb-parameter space a Higgs 
boson mass close to the maximum possible value is 
obtained. For fixed $M_2$, $\mu$, $m_{SUSY}$ and $m_g$ this is achieved by
adjusting the value of $X_t$. This scenario is 
similar to the maximal mixing scenario discussed above.   
\item {\em No mixing scenario}: in this scenario vanishing mixing in the stop sector
is assumed, {\em i.e.}, $X_t = 0$. 
This scenario typically gives a small  
mass for the lightest CP-even Higgs boson 
$h$ and is less favourable for the LHC.
\item {\em Gluophobic scenario}: the effective coupling of the light Higgs boson 
$h$ to gluons is strongly suppressed for a large area of the \matb-plane.
This requires large mixing in the stop sector, leading to cancellations 
between top-quark and stop loops such that the production cross section 
for gluon fusion is strongly suppressed. 
\item {\em Small $\alpha$ scenario}: The parameters are chosen such that the 
effective mixing angle $\alpha$ between the CP-even Higgs bosons is small.
This results in a reduced branching ratio into \bbbar\ and $\tau \tau$
for large \tanb\ and intermediate values of $m_A$.
\end{enumerate}

The parameters for the four scenarios are summarized in 
Table~\ref{t:susy-benchmarks}.\cite{schumacher}
For a given point in parameter space, couplings and branching ratios of the 
Higgs bosons have been calculated using the program described
in Ref.~\refcite{feynhiggs},
which is based 
on a two-loop diagrammatic approach in an on-shell renormalization 
scheme.\cite{heinemeyer} In this calculation, the full one-loop 
radiative corrections and all dominant two-loop corrections are included. 
\begin{table}[hbn]
\tbl{Values of the SUSY parameters as used in 
Ref.~\protect\refcite{schumacher}
for the four benchmark scenarios. \label{t:susy-benchmarks}}
{\begin{tabular}{l  | c | c | c | c | c }
                & $m_{SUSY}$ & $\mu$ & $M_2$ & $X_t$ & $m_g$ \\
                & (\Gcs) & (\Gcs) &(\Gcs) &(\Gcs) &(\Gcs) \\
\hline
$m_h$-max & 1000 &  200 & 200 & 2000 & 800 \\
No mixing & 1000 &  200 & 200 &    0 & 800 \\
Gluophobic     & 350 & 300 & 300 & -750 & 500 \\
Small $\alpha$ & 800 & 2000& 500 & -1100& 500 \\
\hline
\end{tabular}}
\end{table}

\begin{figure}
\begin{center}
\mbox{\epsfig{file=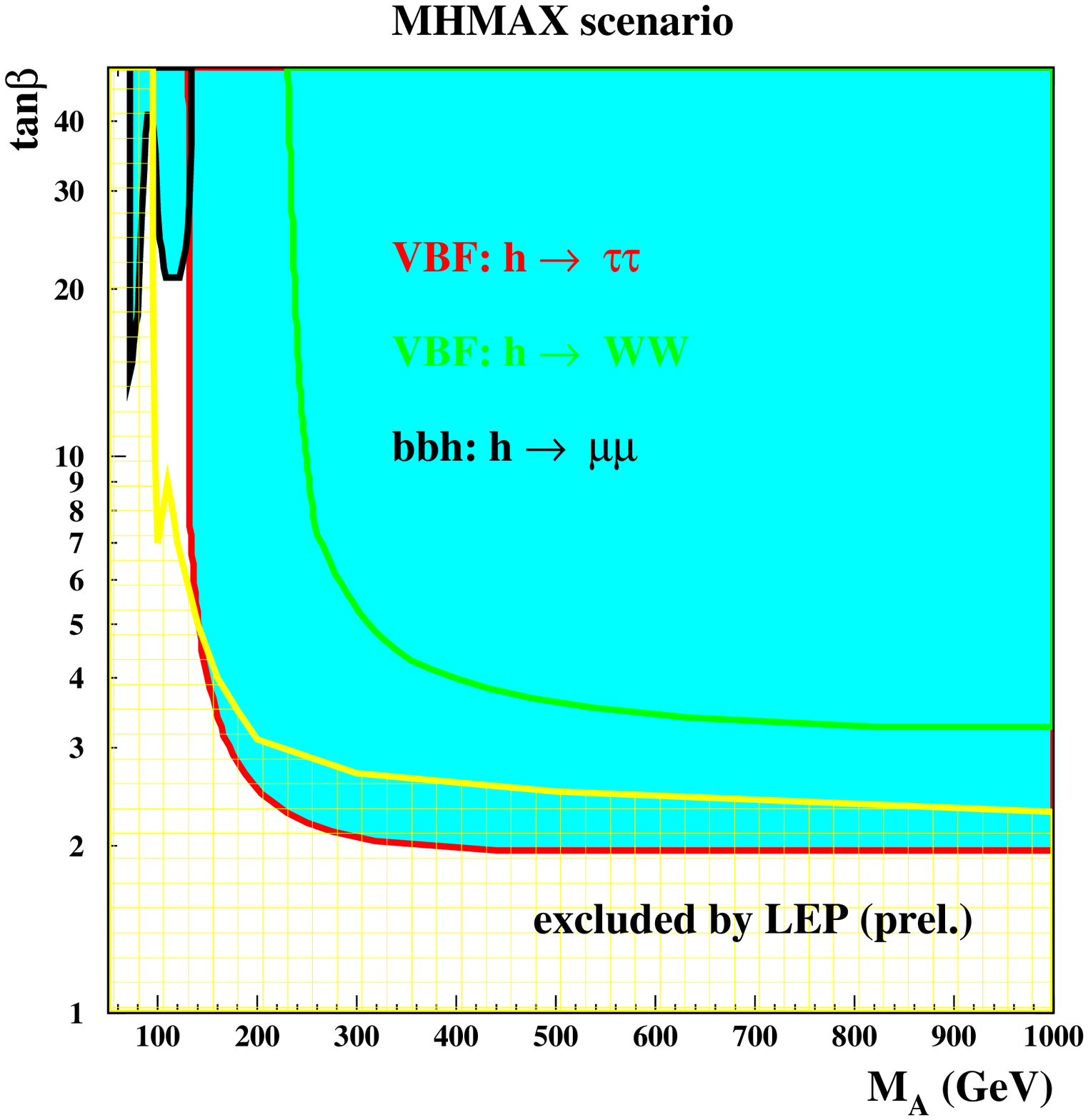,width=0.44\textwidth}}
\mbox{\epsfig{file=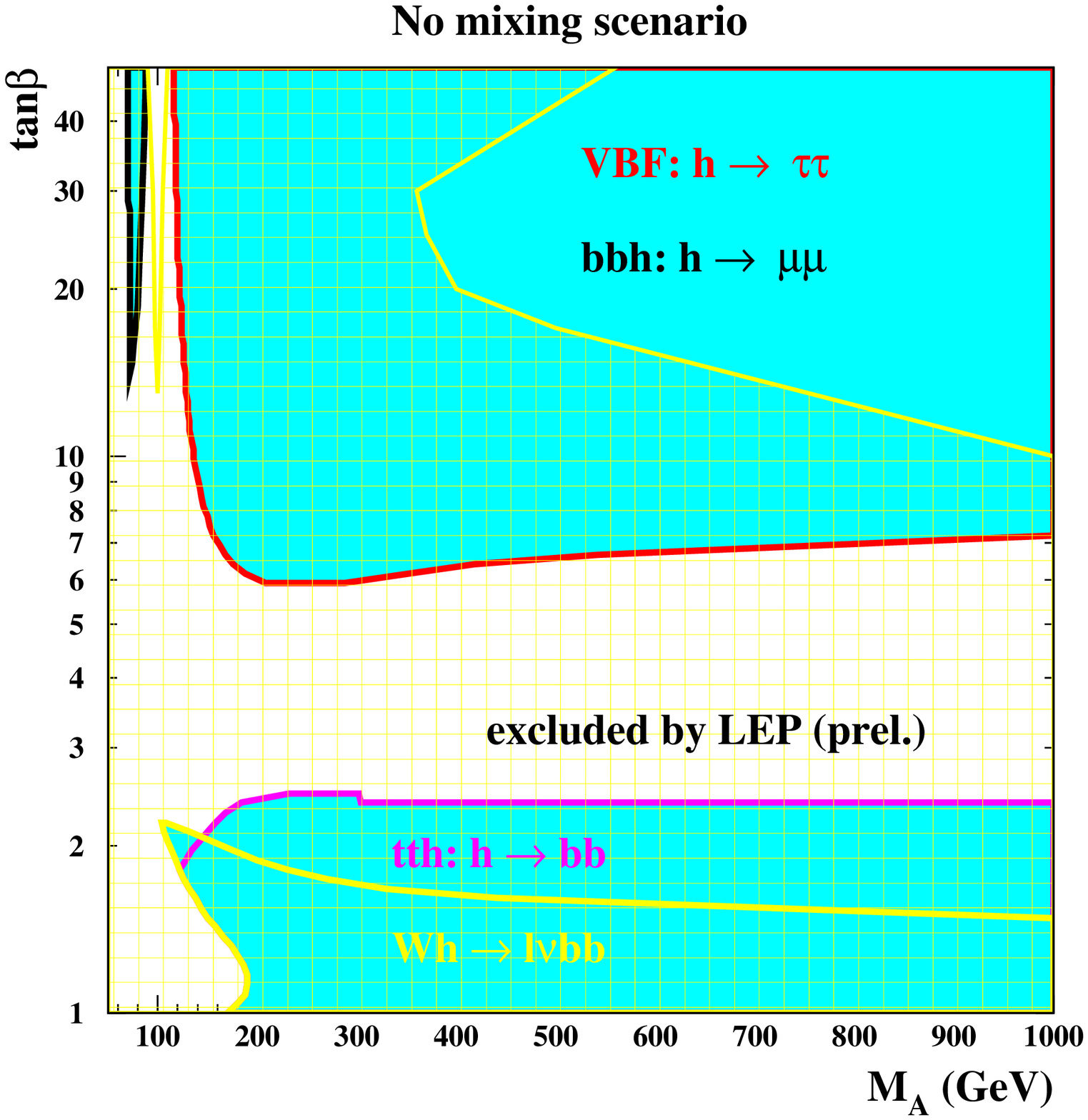,width=0.44\textwidth}}
\mbox{\epsfig{file=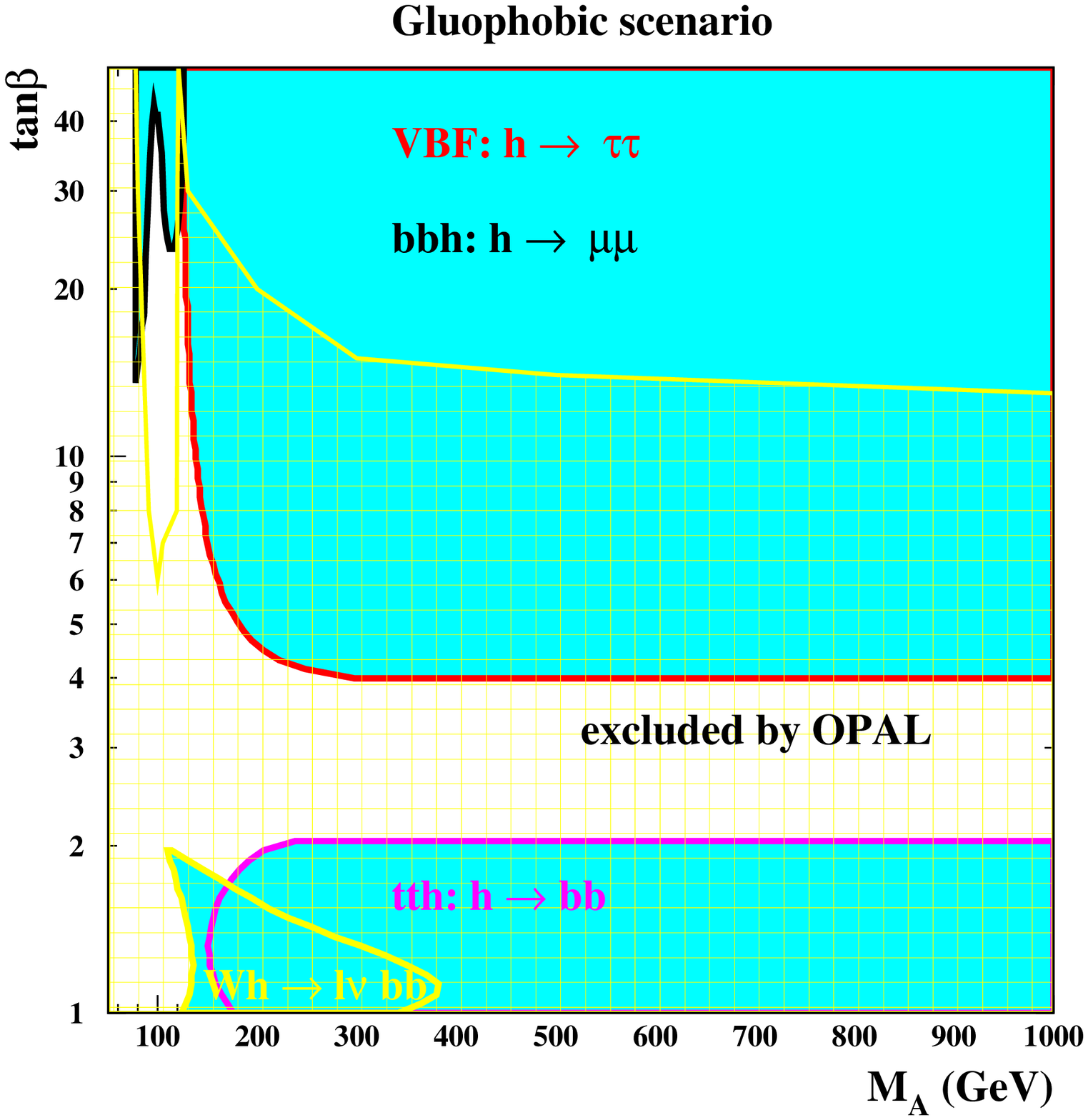,width=0.44\textwidth}}
\mbox{\epsfig{file=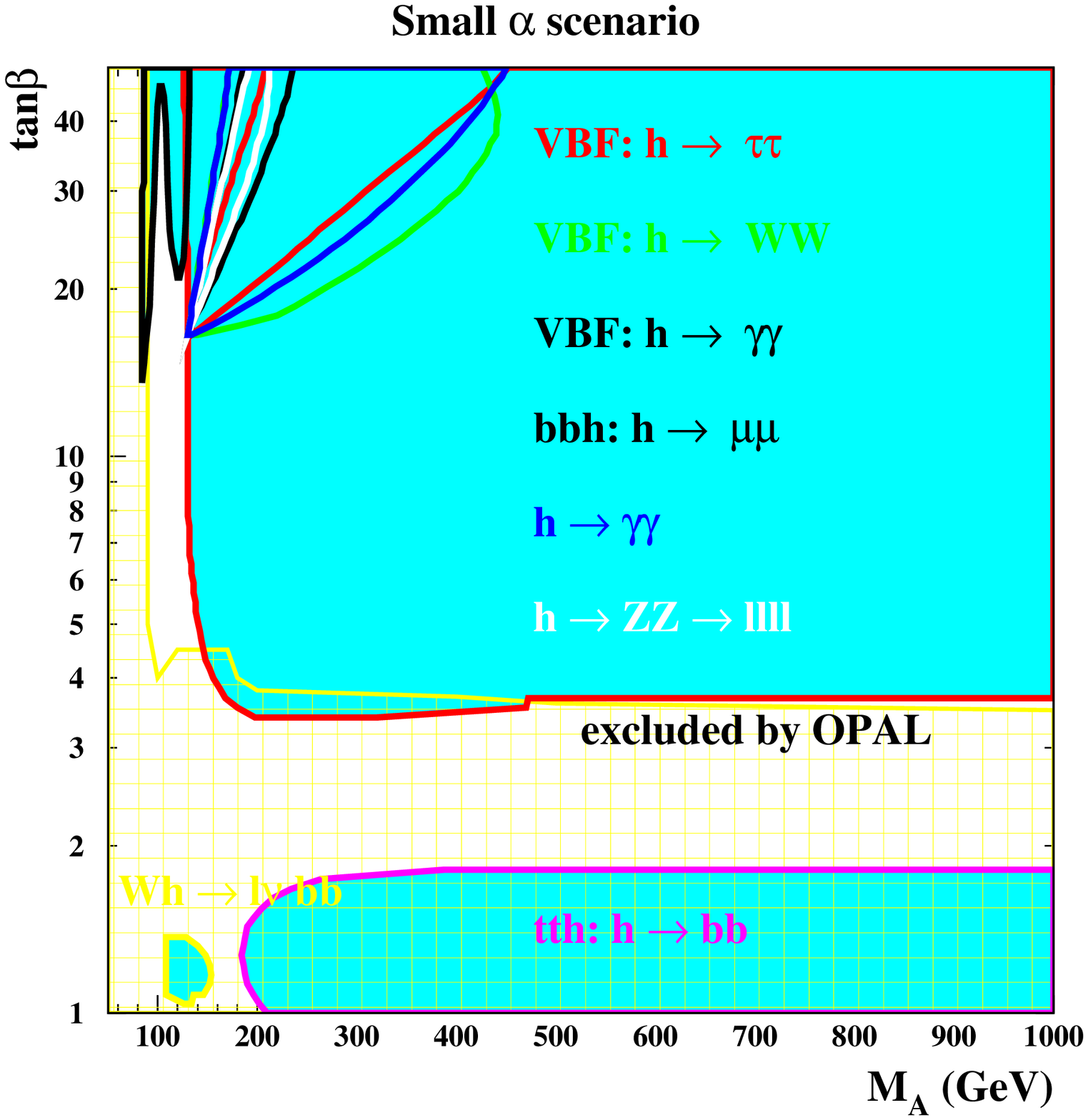,width=0.44\textwidth}}
\end{center}
\caption{\it 
The 5$\sigma$ discovery contours for the light CP even Higgs boson $h$ in the 
\matbit-plane after collecting 30\,\fbsit\ in the ATLAS experiment for the 
four benchmark scenarios considered: the $m_h$-max scenario, the no mixing
scenario, the gluophobic and small $\alpha$ scenario (see text for details). The cross-hatched 
yellow region is excluded by searches at LEP
(from Ref.~\protect\refcite{schumacher}).
}
\label{f:mssm-bench-h30}
\end{figure}

In the evaluation of the discovery potential for the light Higgs boson $h$ 
all production modes, {\em i.e.}, 
the gluon fusion, the vector boson fusion as well as the associated 
$\bbbar h$, $ \ttbar h$ and $Wh$ production have been used.
The decay modes 
$h\to \gamma \gamma$, $qqh \to qq \tau \tau$, $qqh \to qq W W$,
$\bbbar h \to bb \ \mu \mu$, $\ttbar h$ with $ h \to \bbbar$ and 
$Wh \to \ell \nu \bbbar$ have been considered.
The discovery potential for the light CP-even Higgs 
boson $h$ for the individual channels in the four benchmark 
scenarios is shown in Fig.~\ref{f:mssm-bench-h30} for the ATLAS experiment alone,
assuming an integrated luminosity of 30~\fbs. 

\begin{figure}
\begin{center}
\mbox{\epsfig{file=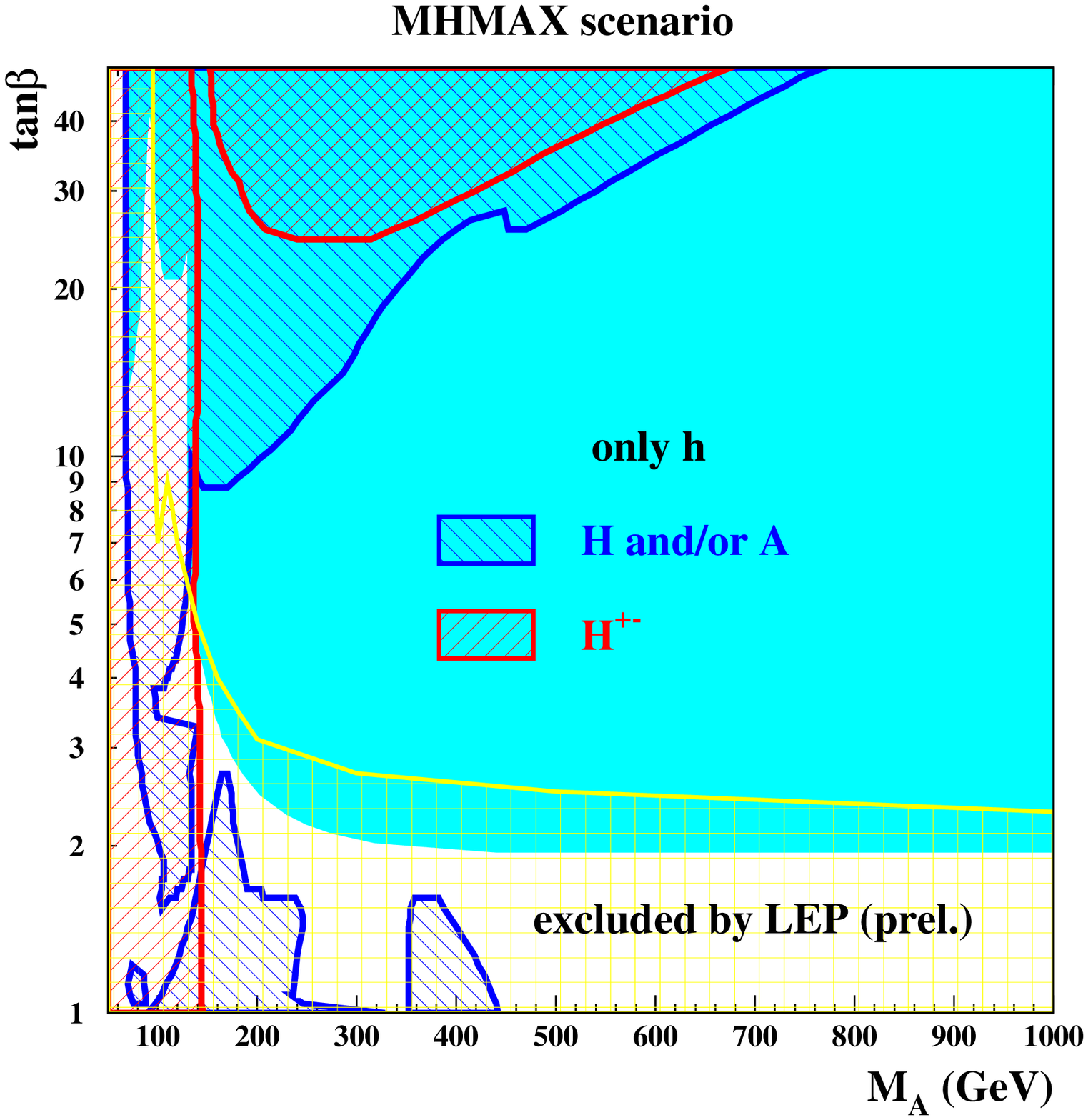,width=0.44\textwidth}}
\mbox{\epsfig{file=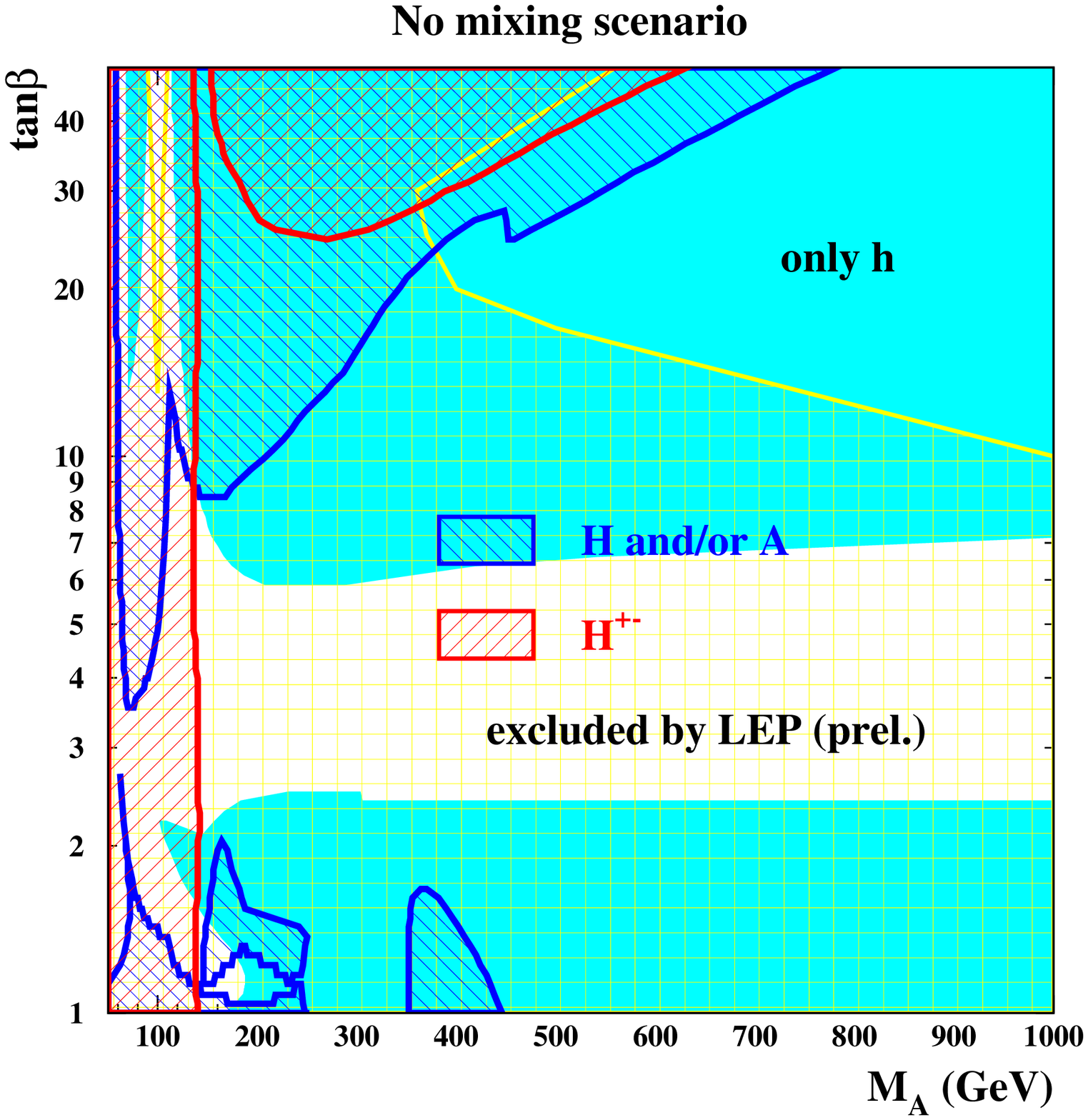,width=0.44\textwidth}}
\mbox{\epsfig{file=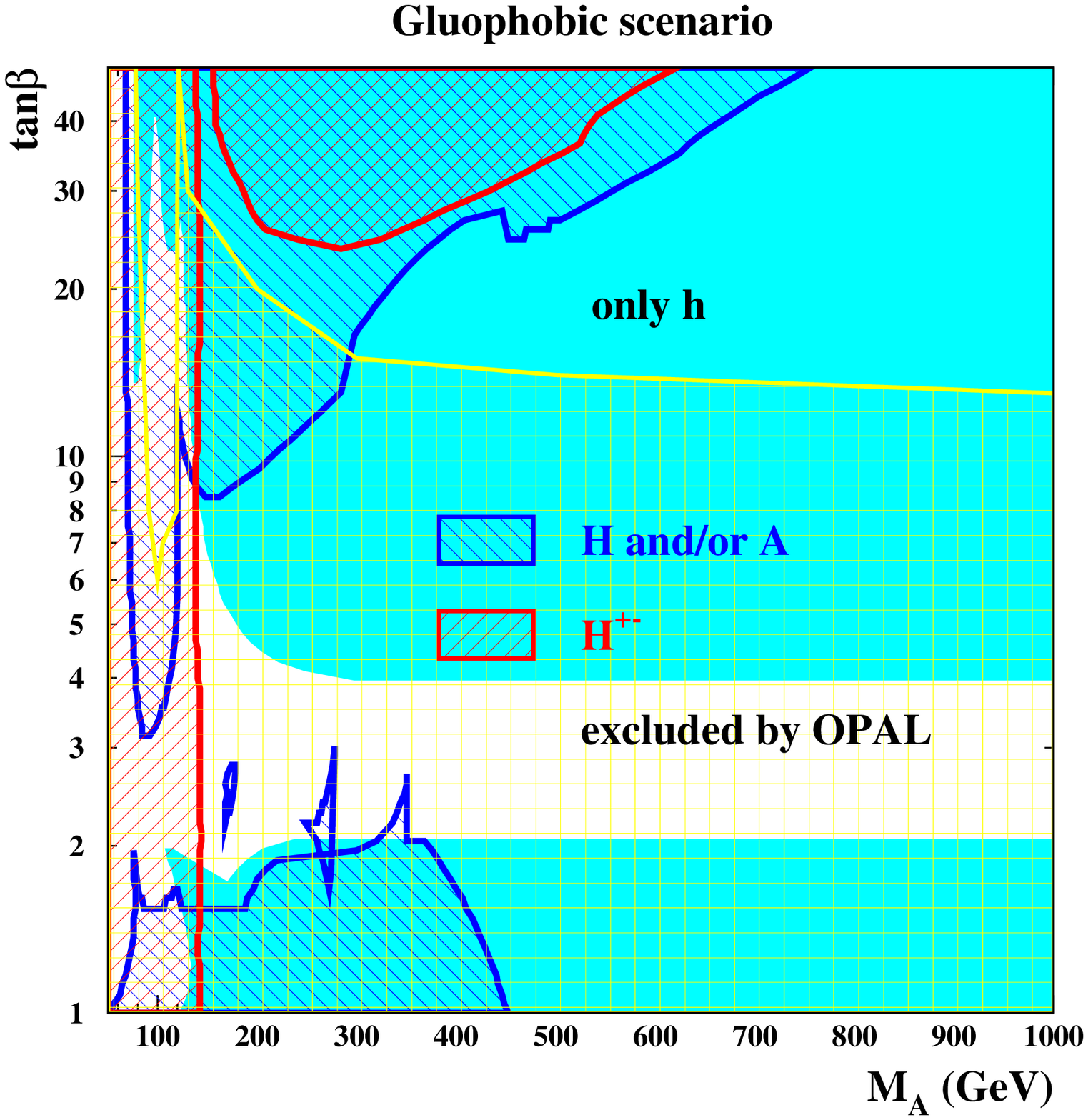,width=0.44\textwidth}}
\mbox{\epsfig{file=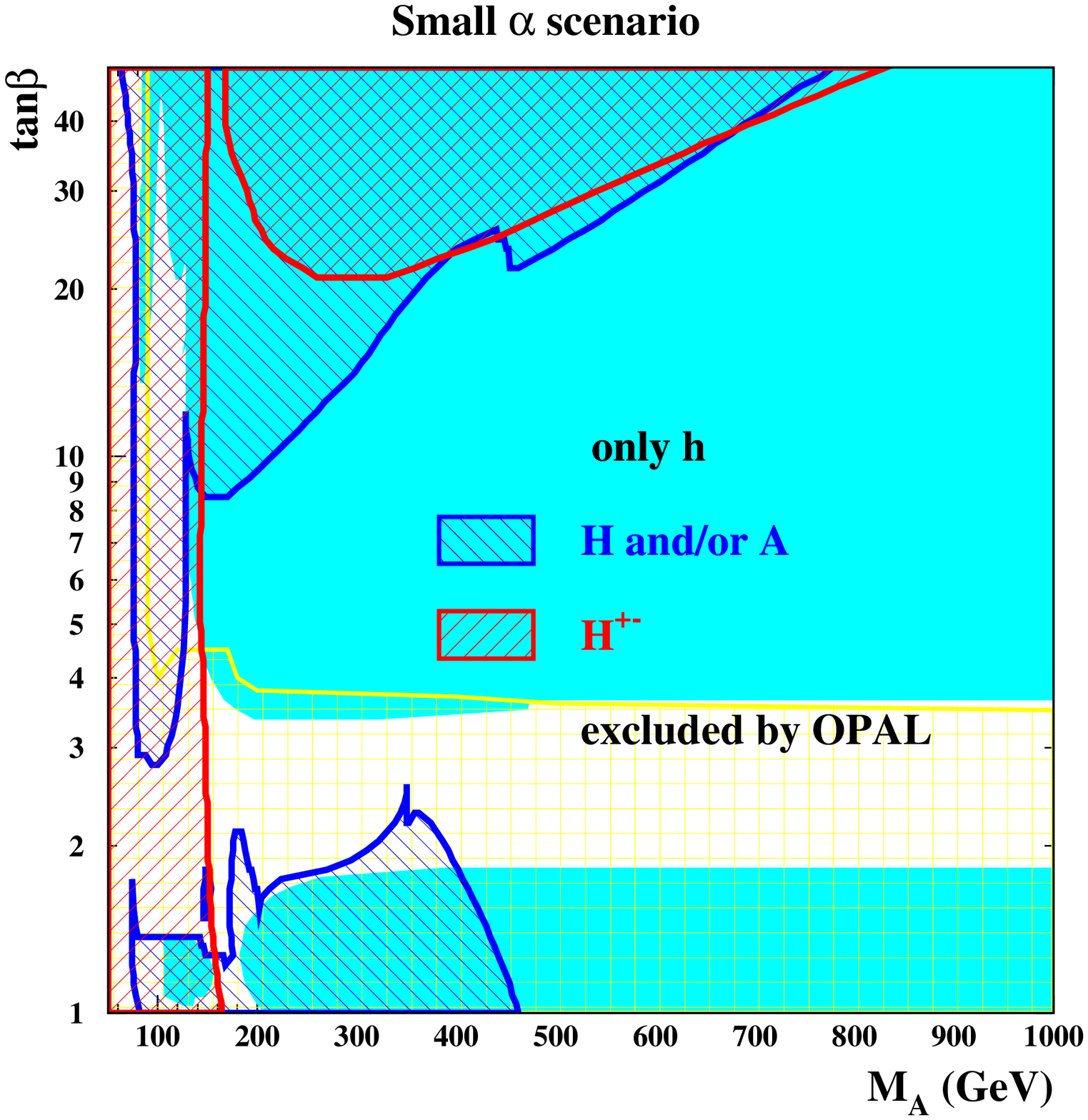,width=0.44\textwidth}}
\end{center}
\caption{\it 
The overall 5$\sigma$ discovery potential for MSSM Higgs bosons at the 
LHC for an integrated luminosity of 30\,\fbsit\ in the four benchmark 
scenarios (see text). In the shaded area (cyan) 
only the light CP-even Higgs boson $h$ can be observed. In the blue 
left-hatched area the heavy neutral Higgs bosons $H$ and/or $A$, and in the red 
right-hatched area the charged Higgs bosons $H^{\pm}$ can be detected. 
The cross hatched yellow region is excluded by searches
at LEP (from Ref.~\protect\refcite{schumacher}).
}
\label{f:mssm-bench-all30}
\end{figure}

The study shows that already 
at low luminosity the full \matb-plane can be covered 
in all benchmark scenarios considered,
apart from a region at small \mA. However, in that particular area 
the searches for heavier Higgs bosons have sensitivity, such that at least 
one MSSM Higgs boson would be discovered at the LHC already with a moderate 
integrated luminosity of 30~\fbs\ 
(see Fig.~\ref{f:mssm-bench-all30} and discussion below).  
In the area not yet excluded by the LEP data, the light CP-even Higgs
boson is almost guaranteed to be discovered in all four scenarios via the vector boson 
fusion channels.
For an integrated luminosity of 30~\fbs\  the discovery 
potential in the large $m_A$ region is dominated by the vector boson 
fusion channel with $h \to \tau \tau$. In the $m_h$-max scenario,  
the $qq h \to qq WW$ channel also contributes. 
For small $m_A$ and large \tanb, the $bbh$ associated production with $h \to \mu \mu$ dominates.
Since many complementary channels 
are available at the LHC, the loss in sensitivity due to 
suppressed couplings in certain benchmark scenarios 
can be compensated for by other channels. 
In the small $\alpha$ scenario, for example, 
the effect of the suppressed branching ratio into $\tau$ leptons 
(visible in the region $\tanb >$ 20 and $200 < \mA < 300$~\Gcs) is 
nicely compensated for by the $h \to WW$ contribution. 
For large integrated luminosities, the  $h \to \gamma \gamma$ and 
$h \to Z Z^* \to 4 \ell$ channels provide additional sensitivity.

For the heavier Higgs bosons, the production cross sections and 
decay branching ratios are similar for the various benchmark scenarios, in
particular for large values of $m_A$.  
Therefore, a similar discovery potential as 
presented in Sections \ref{s:mssm-HA} and \ref{s:mssm-hplus}
and illustrated in Fig.~\ref{f:mssm-contours} is expected for 
all scenarios considered.

\begin{figure}
\begin{center}
\mbox{\epsfig{file=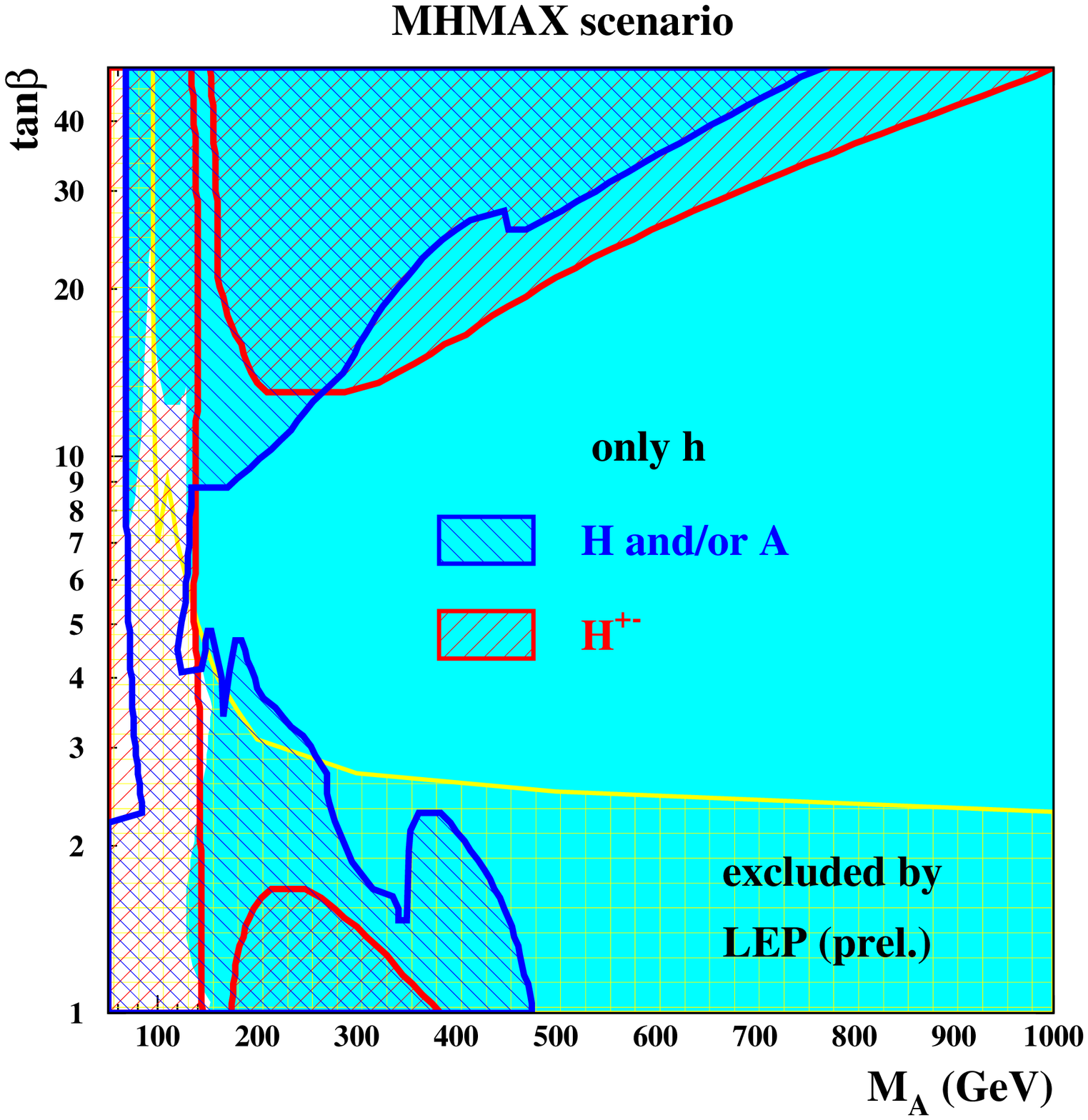,width=0.44\textwidth}}
\mbox{\epsfig{file=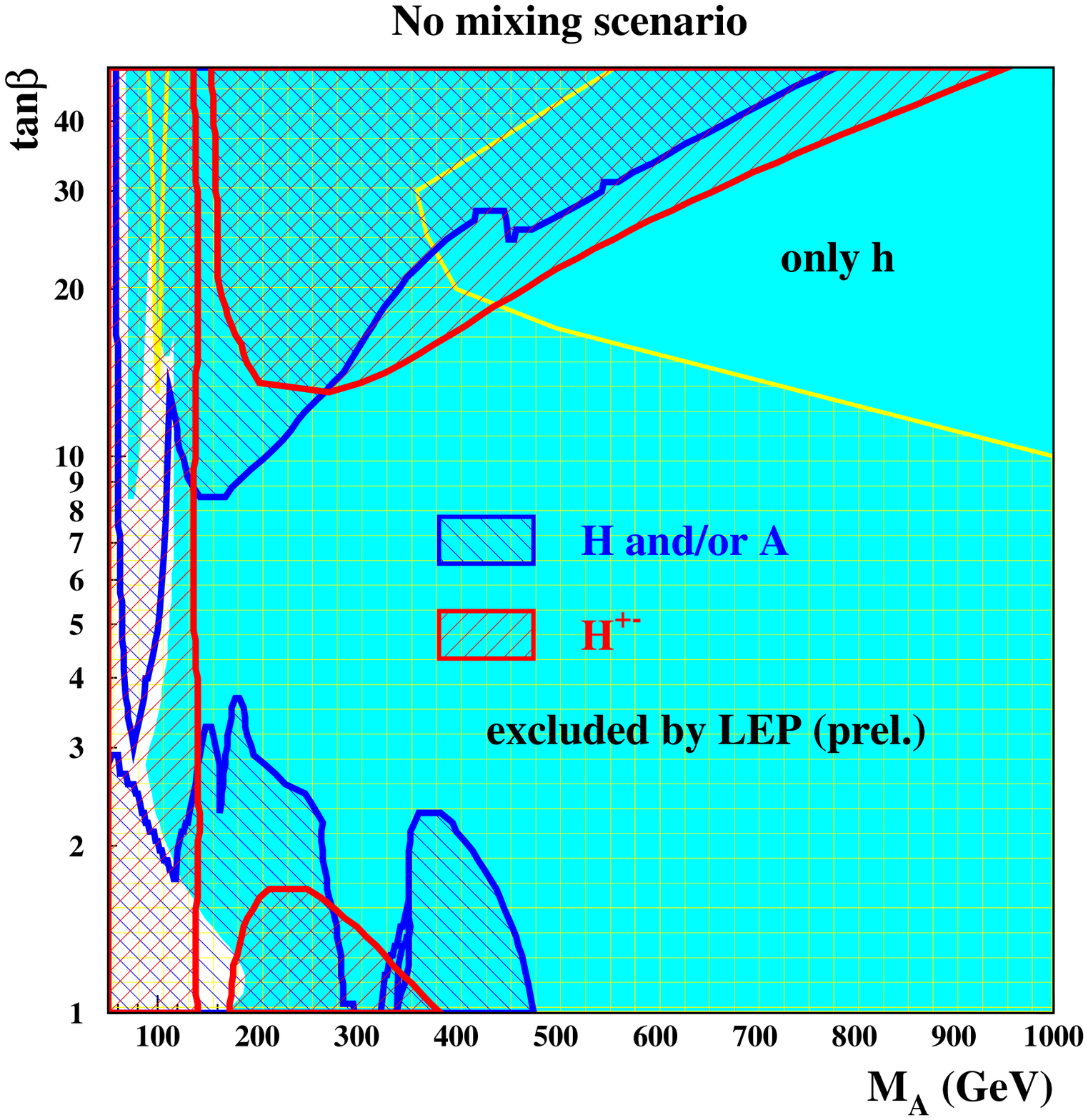,width=0.44\textwidth}}
\mbox{\epsfig{file=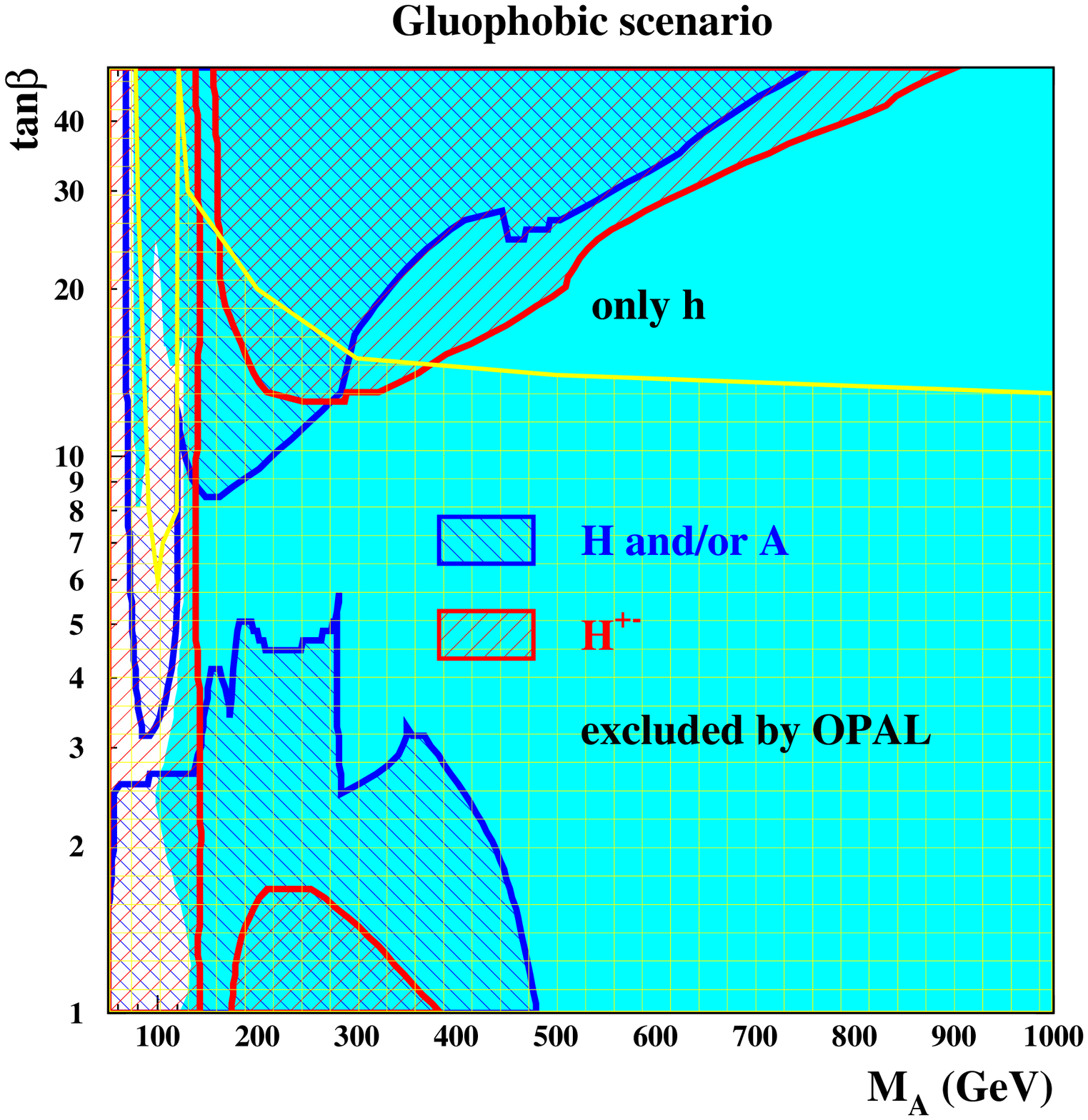,width=0.44\textwidth}}
\mbox{\epsfig{file=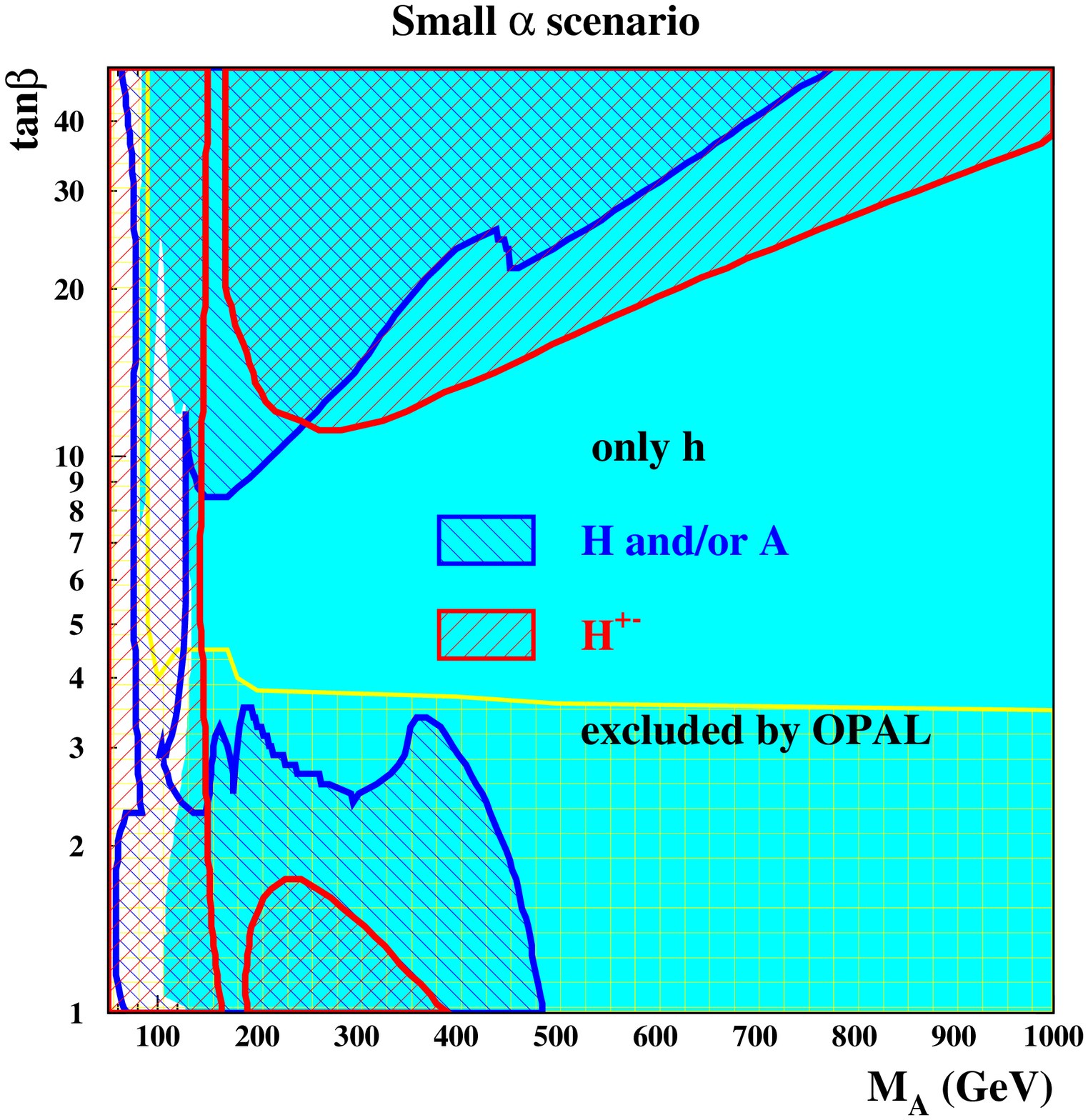,width=0.44\textwidth}}
\end{center}
\caption{\it 
The same as Fig.~\protect{\ref{f:mssm-bench-all30}}, but for 
an integrated luminosity of 300\,\fbsit\ (except for the 
$A/H \to \tau \tau$ decay mode, for which an integrated luminosity 
of 30\,\fbs\ has been assumed)
(from Ref.~\protect\refcite{schumacher}).
}
\label{f:mssm-bench-all300}
\end{figure}
The overall discovery potential for the four scenarios is presented in 
Figs. \ref{f:mssm-bench-all30} and  \ref{f:mssm-bench-all300}
for integrated luminosities of 30 and 300 \fbs, respectively. 
Already with a modest integrated luminosity of 30\,\fbs, the full parameter 
space can be covered for all benchmarks. 
However, in the region of moderate \tanb\ and large $m_A$ only one MSSM 
Higgs boson, the Standard Model-like Higgs boson $h$, can be discovered, 
even if a large integrated luminosity of 300~\fbs\ is assumed.  
The exact location of that region in the parameter plane depends on the 
details of the model considered. 
A further increase in integrated 
luminosity will only marginally reduce that region.\cite{self-slhc}
As discussed in Section \ref{s:susydecays}, 
some sensitivity to heavier Higgs bosons might, however, 
be provided via their decays into SUSY particles.

\begin{figure}
\begin{center}
\begin{minipage}{6.0cm}
\hspace*{-1.0cm}
\mbox{{\epsfig{file=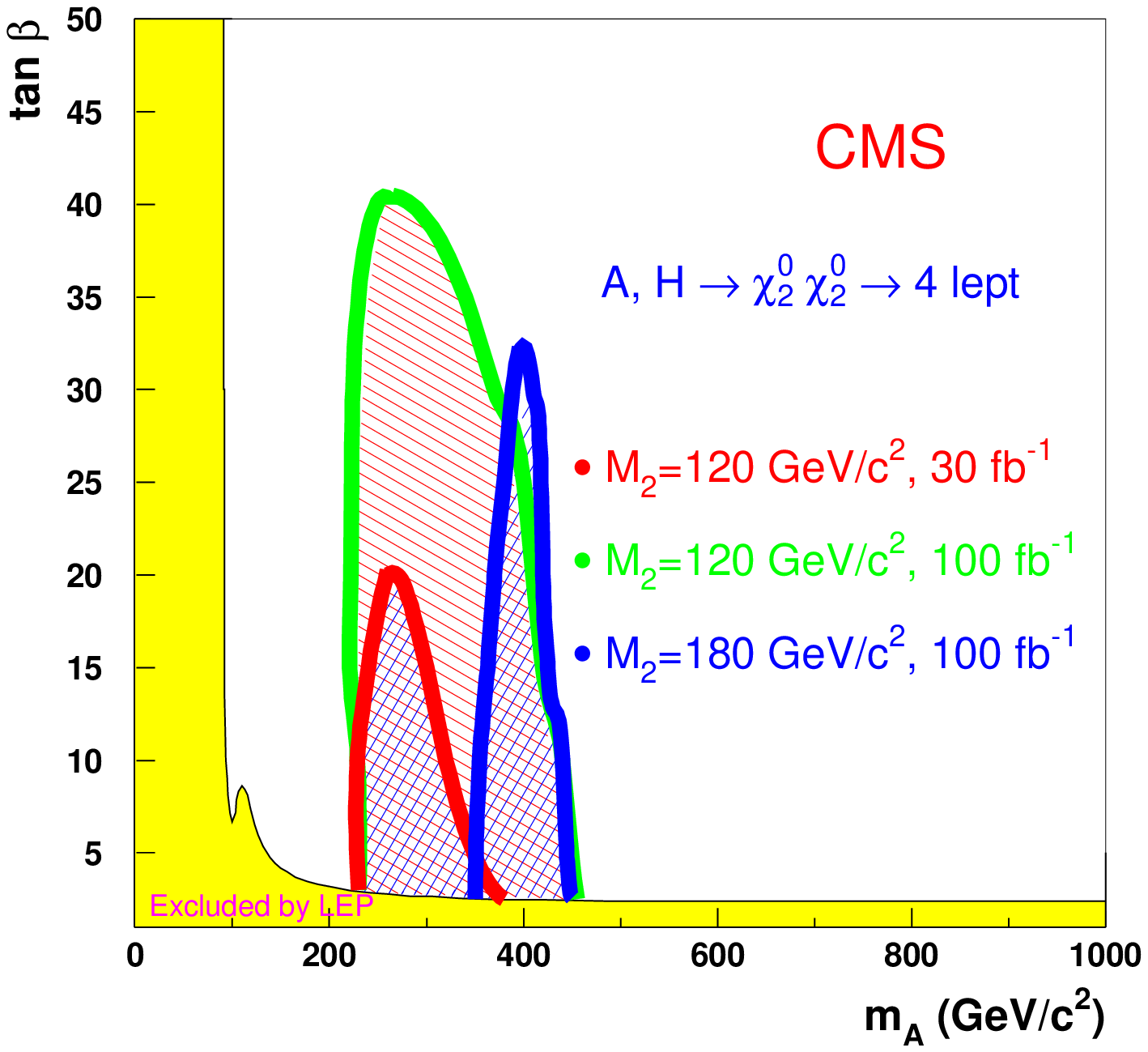,height=5.7cm}}}
\end{minipage}
\begin{minipage}{6.0cm}
\vspace*{0.0cm}
\mbox{{\epsfig{file=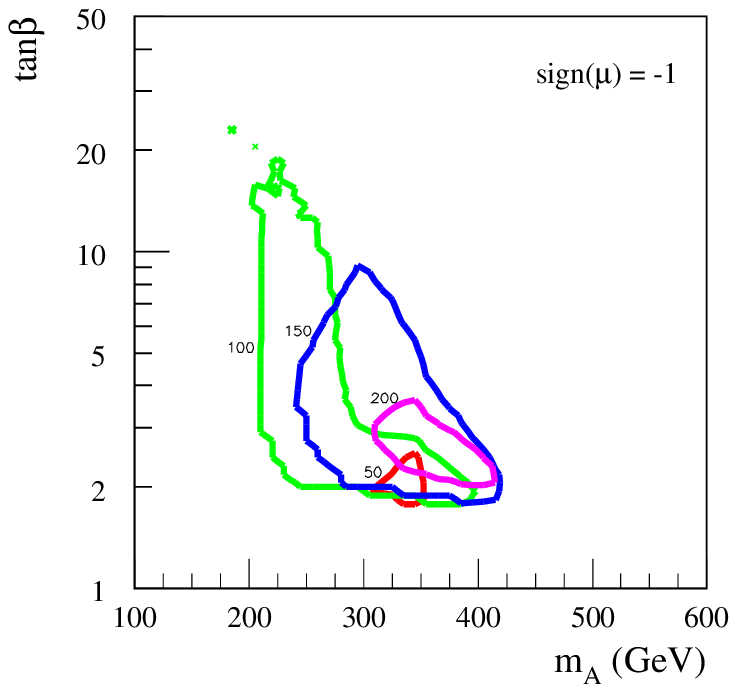,height=5.8cm}}}
\end{minipage}
\end{center}
\caption{\it 
(Left) The 5$\sigma$ discovery contours in the \matbit-plane for 
$H/A \to \chitwo \chitwo \to \chione \ell \ell \ \chione \ell \ell$ 
decays for different SUSY parameters and integrated luminosities
(from Ref.~\protect\refcite{cms-higgs}).
(Right) The 5$\sigma$ discovery contours in the \matbit-plane for 
$H/A \to \chitwo \chitwo \to \chione \ell \ell \ \chione \ell \ell$ 
in the mSUGRA scenario for fixed \mzero\ = 50, 100 and 200~\Gcsit,
$A_0$ = 0 and $\mu<0$. The parameter \mhalf\ has been 
scanned in the range between 100 and 300~\Gcsit. An integrated luminosity
of 300\,\fbsit\ is assumed 
(from Ref.~\protect\refcite{physics-tdr}). 
} 
\label{f:mssm-H/A-fourl}
\end{figure}

Finally it should be noted that Higgs boson decays into Standard Model particles 
can be strongly suppressed in certain MSSM scenarios. 
An example has been presented in Ref.~\refcite{berger-sbottom}, where 
the light Higgs boson decays predominantly into light bottom squarks leading 
to multijet final states. In such scenarios the detection of a light 
Higgs boson might be difficult at hadron colliders.

\subsection{The interplay between the Higgs sector and SUSY particles \label{s:susydecays}}
SUSY particles have an impact on the Higgs boson  
discovery potential via their appearance in the decay chain (mostly for H and A)
and in loops (mostly for production via gluon fusion and for $\hgg$ decays).\cite{baer-susy}
In particular, the heavy Higgs bosons $H$, $A$ and $\hplus$ could be detected 
via their decays into neutralinos and charginos, using multilepton
final states. 
Due to present experimental constraints, the 
decay of the light Higgs boson $h$ into the lightest SUSY particles is 
kinematically forbidden over a large fraction of the constrained MSSM parameter space. 
Instead, this Higgs boson could appear at the end of the decay 
cascade of SUSY particles, for example in the decay $\chitwo \to \chione h$,  
where \chione\ and \chitwo\ are the lightest and second lightest neutralinos.

\subsubsection{Search for the heavy MSSM Higgs bosons in SUSY decay modes}
As discussed in Section \ref{s:mssm-benchmarks}, the heavy MSSM Higgs bosons 
cannot be detected via their decays into Standard Model particles in the
parameter range of intermediate \tanb\ and large $m_A$.
It has been shown that decays into 
charginos and neutralinos can be used in some areas of SUSY parameter 
space for detection of MSSM Higgs bosons.\cite{moortgat,cms-higgs}
If the masses of these particles are small enough, the branching 
ratios for the decays $H/A \to \chitwo \chitwo$ and $\hplus \to \chi_{2,3}^0 \chi^\pm$ 
are sizeable. If in addition sleptons are light, these decays can be 
searched for in multilepton final states via the decay chain 
$\chitwo \to \slep \ell \to \chione \ell \ell$. 
Due to the invisible neutralinos, the Higgs boson mass reconstruction 
is only possible if the masses of \chione\ and \slep\ are known, 
for example from the analysis of first or second generation squark cascade 
decays.\cite{giacomo-msusy}

In the search for the decay channel 
$H/A \to \chitwo \chitwo \to \chione \ell \ell \ \chione \ell \ell$, 
lepton isolation, a large missing
transverse energy and a jet veto are used to suppress the Standard 
Model backgrounds from $ZZ$, $Z\bbbar$, $Z\ccbar$ and $\ttbar$ production
as well as the background from SUSY ($\sq, \sgl$) production.\cite{moortgat,cms-higgs} 
The discovery reach depends strongly on the choice of 
the SUSY parameters. Examples for the 5$\sigma$-discovery
reach are shown in Fig.~\ref{f:mssm-H/A-fourl}(left).

A similar analysis has been performed within the more constrained Minimal Supergravity 
(mSUGRA)\cite{mSUGRA} scenario. This model is determined by five parameters: 
common masses \mzero\ and \mhalf\ for scalars (squarks, sleptons and Higgs bosons) 
and gauginos and Higgsinos at the GUT scale, a common trilinear Higgs-sfermion-sfermion 
coupling  $A_0$, as well as \tanb\ and the sign of the Higgs mass
parameter $\mu$. 
Using renormalization group equations, the masses and mixings of all SUSY and Higgs particles 
are determined at the electroweak scale.
The parameter space has been scanned for fixed values of \mzero\ = 50, 
100, 150, 200 and 250~\Gcs\
in the range of \mhalf\ = 100 -- 300~\Gcs\ and \tanb\ = 1.5 -- 50 
with $A_0$=0.\cite{physics-tdr} The resulting 5$\sigma$ discovery 
contours are shown in Fig.~\ref{f:mssm-H/A-fourl}(right), 
projected onto the \matb-plane, for fixed values of \mzero, a negative sign of 
$\mu$ and assuming an integrated luminosity of 300\,\fbs.

In the search for charged Higgs bosons, three-lepton final states from 
the decay chain $gb \to \hplus t$, 
$\hplus \to \chi_{2,3}^0 \chi^{\pm}_{1,2}$ and $\chitwo \to \chione \ell \ell$, 
$\chi^{\pm}_{1} \to \chione \ell \nu$ can be exploited.\cite{moortgaat-hplus}
The accompanying top quark is required to decay hadronically. 
The discovery potential in this channel is limited 
to a small region of parameter space where sleptons are light and  
$ | \mu  | \leq$ 150~\Gcs, {\em i.e.}, close to the lower limit set by the 
LEP experiments.

\begin{figure}
\begin{center}
\mbox{{\epsfig{file=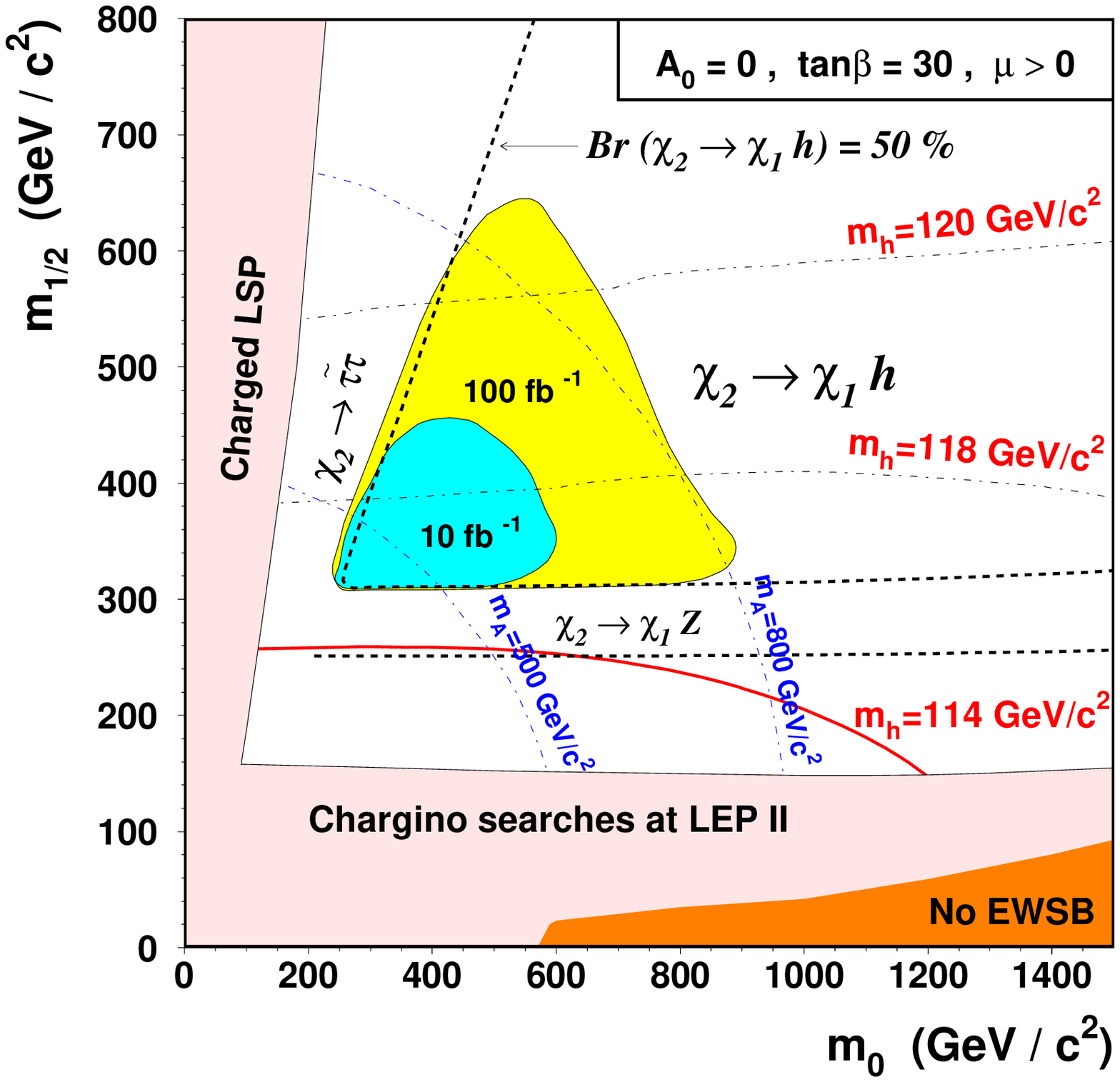,height=8.0cm}}}
\end{center}
\caption{\it 
The 5$\sigma$ discovery contours in the $(\mzero , \mhalf )$-plane 
for the CP-even Higgs bosons $h$ via 
the decay $h \to \bbbar$ from $\chitwo \to \chione h$ decays in
squark and gluino cascades for mSUGRA scenarios with \tanbit\ = 30,
$A_0$ = 0 and $\mu >$ 0 for 10 and 100\,\fbsit. The regions excluded 
by experimental and theoretical constraints are indicated as well
(from Ref.~\protect\refcite{cms-higgs}).
} 
\label{f:msugra_hbb}
\end{figure}

\subsubsection{Search for $h \to \bbbar$ in SUSY cascade decays \label{s:susy-cascades}}

The lightest Higgs boson might appear at the end of decay cascades 
of SUSY particles. One copious production source may be the decay of the 
second lightest neutralino into the lightest neutralino,  
$\chitwo \to \chione h$. The former is produced with large rates
in the decays of squarks and gluinos. 
In R-parity conserving SUSY models 
the lightest Higgs boson $h$ is in this decay mode accompanied 
by missing transverse energy, carried away by the lightest SUSY 
particle. The presence of missing transverse energy and several 
energetic jets from the squark or gluinos cascades can be used to 
obtain a sample that consists 
mainly of events containing SUSY particles. In this case, the discovery 
of the Higgs boson $h$ in its dominant decay mode $h \to \bbbar$ 
becomes possible, without requiring the presence of an additional lepton. 

Both collaborations, ATLAS and CMS, have studied the observability 
of the $h \to \bbbar$ signal in the mSUGRA parameter 
space.\cite{physics-tdr,cms-higgs,moortgat-hbb} 
As an example, the 5$\sigma$ discovery contours, as obtained in 
Ref.~\refcite{cms-higgs},
are shown in Fig.~\ref{f:msugra_hbb} in the $( \mzero , \mhalf )$-plane with $A_0$ = 0, \tanb\ = 30 
and $\mu >$ 0 for integrated luminosities of 10 and 100\,\fbs.

\subsubsection{Search for heavy MSSM Higgs bosons in SUSY cascade decays}

\begin{figure}
\begin{center}
\mbox{{\epsfig{file=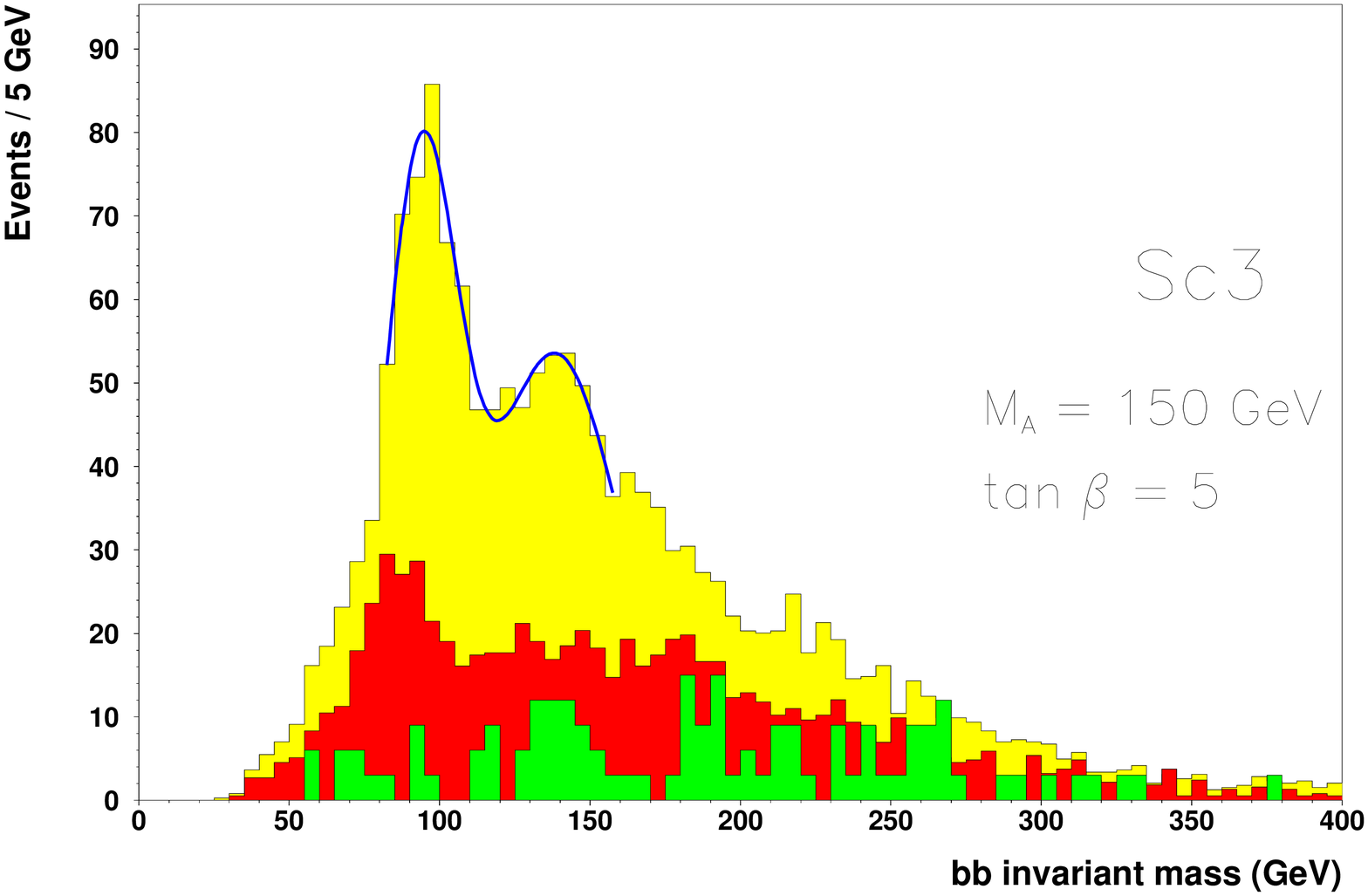,width=0.6\textwidth}}}
\end{center}
\caption{\it 
The reconstructed \bbbar\ invariant mass spectrum for a SUSY scenario 
with $M_2$ = 350~\Gcsit, $\mu$=150~\Gcsit, $m_{\sgl}$ = 1200~\Gcsit 
and $m_{\sq}$ = 800~\Gcsit\ in a simulation of the CMS experiment, assuming an
integrated luminosity of 
30\,\fbsit. The two mass peaks
result from decays of the lighter $h$ and the heavier $H/A$ Higgs bosons. They are shown
on top of the backgrounds from  SUSY processes (dark) and from 
Standard Model \ttbar\ production (light) (from Ref.~\protect\refcite{moortgat-hbb}).
} 
\label{f:cascade_susy}
\end{figure}

\begin{figure}
\begin{center}
\mbox{{\epsfig{file=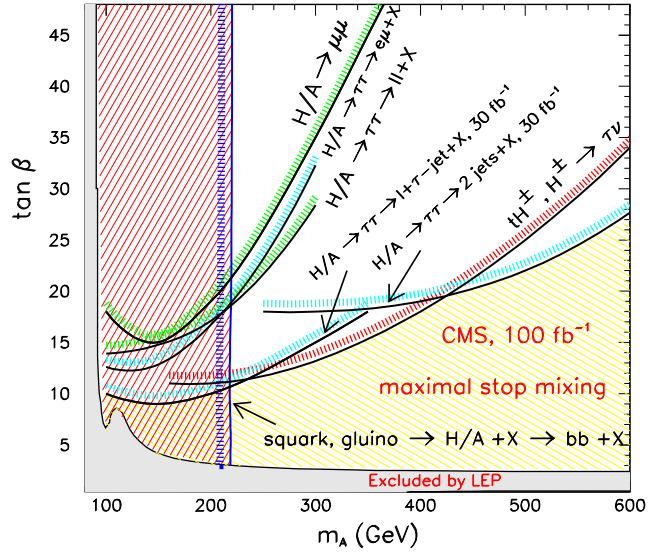,height=8.0cm}}}
\end{center}
\caption{\it 
The 5$\sigma$ discovery contours for MSSM Higgs bosons in the 
\matbit-plane for an integrated luminosity of  100\,\fbsit\
for the SUSY parameter 
point with $m_{\sgl}$= 1200~\Gcsit, $m_{\sq}$ = 800~\Gcsit, 
$M_2$ = 350~\Gcsit and $\mu$ = 150~\Gcsit.
In the region with $m_A <$ 220~\Gcsit\  the
heavier Higgs bosons $H$ and $A$ can be detected in cascade decays 
of squarks and gluinos (from Ref.~\protect\refcite{moortgat-hbb}).
} 
\label{f:cascade_susy-2}
\end{figure}

In a recent study four possible sources for detection of heavy MSSM Higgs 
bosons in SUSY cascade decays have been considered:\cite{moortgat-hbb} 
\begin{itemize}
\item Cascade decays of squarks and gluinos via the  heavy chargino
$\chi_2^\pm$ and 
neutralinos $\chi_{3,4}^0$ with subsequent decays into the 
lighter chargino $\chi_1^\pm$  or neutralinos $\chi_{1,2}^0$ and Higgs bosons: 
\begin{eqnarray}
pp \ \to \ \sgl \sgl, \sq \sq, \sq \sgl & \ \to \ & \chi_2^\pm , \chithree, \chifour \ + \ X \nonumber \\
                                    & \ \to \ & \chi_1^\pm, \chitwo, \chione \ + \ h, H, A, \hplus \ + \ X ; \nonumber 
\end{eqnarray}

\item Direct decays of squarks and gluinos into the lightest chargino and the next-to-lightest 
neutralino, with subsequent decays into the lightest neutralino and Higgs bosons: 

\begin{eqnarray}
pp \ \to \ \sgl \sgl, \sq \sq, \sq \sgl & \ \to \ & \chi_1^\pm , \chitwo  \ + \ X \nonumber \\
                                    & \ \to \ & \chione  \ + \ h, H, A, \hplus \ + \ X ; \nonumber
\end{eqnarray}

\item Direct decays of heavy stop and sbottom squarks into the lighter ones and Higgs bosons
(in the case of large enough squark mass splitting): 
\[
pp \ \to \ \st_2 \st_2, \sbot_2 \sbot_2, \rm{\ \ with \ \ } 
\st_2 (\sbot_2) \ \to \ \st_1 (\sbot_1)  +  h, H, A 
\rm{\ \ or  \ \ } \sbot_1 \ (\st_1)  + \hplus ;
\] 

\item Top quarks originating from SUSY particle cascades, decaying into \hplus\ bosons: 
\[ pp \ \to \ \sgl \sgl, \sq \sq, \sq \sgl \ \to \ t + X \ \to \ \hplus \ + \ X . \]

\end{itemize}
Four representative SUSY scenarios have been discussed: 
squarks are considered to be either lighter or heavier than gluinos 
and either light gaugino- or Higgsino-like charginos and neutralinos 
are assumed. 
Using a fast simulation of the performance of the CMS detector, signals of 
heavy neutral (charged) Higgs bosons 
have been reconstructed in the $\bbbar$ ($\tau \nu$)
decay mode. Backgrounds from both Standard Model and SUSY processes have been taken into 
account.  It has been
demonstrated that light Higgs bosons ($m_\Phi \lesssim$ 200 - 250~\Gcs)
appearing in the cascade decays can be detected with 
a significance exceeding 5$\sigma$ in some representative MSSM scenarios. 
As an example, in Fig.~\ref{f:cascade_susy} the reconstructed \bbbar\ 
invariant mass is shown for the SUSY parameter 
point with $m_{\sgl}$= 1200~\Gcs, $m_{\sq}$ = 800~\Gcs, 
$M_2$ = 350~\Gcs and $\mu$ = 150~\Gcs.
For large $M_2$ values there is sufficient phase
space for the decay of the heavier neutralinos and charginos, with
masses $\sim M_2$, into the lighter Higgsino states, with masses
$m_{\chi^\pm} \sim m_{\chi_1^0} \sim m_{\chi_2^0} \sim | \mu |$, 
and Higgs particles with masses $m_\Phi \lesssim$ 200~\Gcs. 
In the distribution two mass peaks resulting from 
the decays of the light Higgs boson $h$ and the heavier $H$ and $A$ are visible above 
the backgrounds. The discovery potential at that parameter point 
for heavy Higgs bosons is shown in Fig.~\ref{f:cascade_susy-2} for an integrated 
luminosity of 100~\fbs. In this scenario, the Higgs bosons $A, H$ (\hplus) can 
be detected via the cascade decays for masses up to $m_A \sim$ 220~\Gcs\ (200~\Gcs) 
for all \tanb.

As discussed in Ref.~\refcite{moortgat-hbb}, this study is not meant to be 
exhaustive, but should rather be considered as a preliminary investigation of a 
few representative scenarios to illustrate the discovery potential via cascade decays.  
In particular, a more detailed experimental simulation for a larger number of SUSY 
parameter points is necessary and a discussion of systematic uncertainties on  
the backgrounds and their impact on the signal extraction 
needs to be performed. In addition to the \bbbar\ decay mode, also the  
$\tau \tau$ decay mode should be considered for heavy neutral Higgs bosons. As 
discussed in Section \ref{s:susy-rad},
the ratio of the two signals could contribute to establish 
the SUSY nature of a Higgs boson signal in cascade decays.

It should be stressed that the cascade processes will be extremely useful to measure 
the couplings of supersymmetric particles to Higgs bosons, which again would be an 
essential ingredient to reconstruct the parameters of the underlying SUSY model in a  
global fit. If cascade decays via heavier neutralinos 
are kinematically allowed and the relevant branching ratios are sufficiently large, 
the decay chain $\chi_{3,4}^0 \ \to \ h, H, A + \chione \ \to \ \bbbar  + \chione$
provides important information to reconstruct the masses of the heavier neutralinos, 
which in turn helps to constrain the SUSY model significantly.

\subsection{Determination of MSSM parameters}
Assuming that non-Standard Model Higgs bosons will be
discovered at the LHC, it is important  
to extract the parameters of the underlying model. 
For this, all relevant measurements from searches for SUSY and 
Higgs particles will be taken into account in a global fit.
The masses, production rates and branching ratios of the observed Higgs bosons
are expected to contribute significantly to constrain the model. 
The accuracy which can be reached in the measurement of these input parameters 
to the global fit is discussed in the following.

\subsubsection{Measurement of the Higgs boson masses}
The precision of the mass measurement of
MSSM Higgs bosons has been determined under the same assumptions  
as described in Section \ref{s:SM-mass}.
While in addition to statistical uncertainties experimental systematic uncertainties on the 
background subtraction and on the knowledge of the absolute energy scale have been included, 
no theoretical errors have been considered. 
As in the case of the Standard Model, lepton and photon final states provide 
the highest precision for the Higgs boson mass measurements.

\subsubsection*{The mass of the lightest CP-even Higgs boson}
The light CP-even Higgs boson $h$ can be detected in the $h \to \gamma \gamma$ 
decay mode  over the full mass range of interest, {\em i.e.},
between $\sim$90 and $\sim$135~\Gcs. For an 
integrated luminosity of 300\,\fbs, a precision of the mass
measurement of the order of 0.1--0.5\% can be achieved.\cite{physics-tdr} 
The larger value is reached for   
parameter values close to the 5$\sigma$ discovery contour.    
A mass reconstruction is also possible in the vector boson fusion mode $qqh \to qq \tau \tau$ 
using the collinear approximation (see Section \ref{s:lhc-vbf}), however, the precision is 
about an order of magnitude worse.

In the $h \to \bbbar$ channel, the mass measurement is limited by the 
systematic error of $\pm$1\% on the jet energy scale once
signal rates are above a 
few hundred events. Such a signal rate will be achieved already for an integrated luminosity 
of only 60\,\fbs\ over a large region of parameter space.

\subsubsection*{The masses of the heavier neutral Higgs bosons}
For $m_A \lesssim 2 \ m_t$ and small values of \tanb, the mass of the 
$H$ boson can be determined 
in the $H \to ZZ^{(*)} \to 4 \ell$ and 
$H \to hh \to \bbbar \gamma \gamma$  
decay modes (see Fig.~\ref{f:mssm-contours}). Assuming an integrated luminosity of 300\,\fbs, 
a precision of the order of 0.1--0.3\% for the 4$\ell$ and 
of about 1\% for the $\bbbar \gamma \gamma$ channel can be achieved.\cite{physics-tdr}
While in the latter case the measurement is limited by systematic uncertainties, the range in the 
4$\ell$ channel results from the strong variation of the signal rate with \tanb\ over the 
discovery region.
 
The pseudoscalar Higgs boson $A$ can be discovered for small \tanb\ in the 
$A \to Z h \to \ell \ell \bbbar$ and in the $A \to \gamma \gamma$ decay mode. 
The expected precision of the mass measurement is 
of the order of 1--2\% for  $\ell \ell \bbbar$ 
and of the order of 0.1\% for $\gamma \gamma$ final states. 

For large values of \tanb, the heavy Higgs bosons $H$ and $A$ will be discovered in the 
$\tau \tau$ and $\mu \mu$ decay modes.  Over a large part of the parameter space they
cannot be disentangled from each other, being almost
degenerate in mass and having 
almost identical decay modes. In the region of parameter space where both $H/A \to \tau \tau$ 
and $H/A \to \mu \mu$ decays are observable,
the precision of the mass measurement is determined by the $\mu \mu$ decay mode
and is estimated to be at the level of 0.1\%. For moderate values of \tanb, where 
the discovery reach of the $\tau \tau$ channel extends further than that of the $\mu \mu$ 
channel, the precision is degraded to 1-7\%.

\subsubsection*{The mass of the charged Higgs boson}

Both the  
$\hplus \to \tau \nu$ and $\hplus \to t b$ decay modes can be used to extract information
on the mass of the charged Higgs boson. In $\tau \nu$ decays the mass 
is determined from the shape of the transverse mass 
distribution (see Fig.~\ref{f:mssm-h+signal}) using a
likelihood technique.\cite{hohlfeld,atlas-hplus}
For an integrated luminosity of 300\,\fbs, the precision of
the mass measurement varies  
between 0.8 and 1.8\% for charged Higgs boson masses in the range between 
200 to 500~\Gcs.\cite{atlas-hplus,atlas-tanb} 
In this estimate, uncertainties on the shape and 
on the rate of the background and on the energy scale are taken into account.
Due to the larger backgrounds, the precision of the mass measurement 
is found to be worse in the $tb$ decay mode.\cite{atlas-hplus}

\subsubsection{Measurement of \tanbit}
It has been suggested to determine the MSSM parameter \tanb\ from 
the measurement of the production cross section of heavy neutral and charged MSSM 
Higgs bosons.\cite{gunion-tanb}
At large \tanb\ the cross sections are approximately
proportional to $\tan^2 \beta$, however, 
loop corrections involving SUSY particles modify this behaviour. 
Due to potentially large radiative corrections to the bottom-Yukawa
coupling, the results obtained from cross section measurements in this way
correspond to a measurement of an effective parameter 
$\tan \beta_{eff}$.\cite{susy-corrections} The ex\-trac\-tion 
of the fundamental \tanb\ parameter requires additional knowledge of the 
sbottom and gluino masses as well as of the Higgs boson mass parameter $\mu$. 

The method has been applied by the ATLAS\cite{atlas-tanb} and more recently by the CMS 
collaboration.\cite{cms-tanb} In the latter analysis, 
the precision of the \tanb\ measurement has been determined by taking into account 
a systematic uncertainty of $\pm$20\% on the next-to-leading order cross sections 
and of $\pm$3\% on the branching ratios. Since the cross section depends on 
the Higgs boson mass, the uncertainty on the mass measurement has also been included. 
In order to determine  
the sensitivity of the \tanb\ measurement to SUSY parameters, they have been varied in the 
range of $\pm$20\% around their nominal values ($M_2$ = 200~\Gcs, $\mu$ = 300~\Gcs, 
$M_{SUSY}$ = 1~\Tcs\ and $A_t$ = 2450~\Gcs). Over the region of parameter space where a 
discovery is possible, such a variation changes the cross section by at 
most 11\%, which leads to a $\pm$6\% uncertainty on \tanb.\cite{cms-tanb} 
For the determination of the experimental numbers,  a 2--3\%  uncertainty on 
signal efficiencies, mainly related to uncertainties in $\tau$- and b-tagging,
and a $\pm$5\% uncertainty on the luminosity have been assumed. 

\begin{figure}
\begin{center}
\mbox{{\epsfig{file=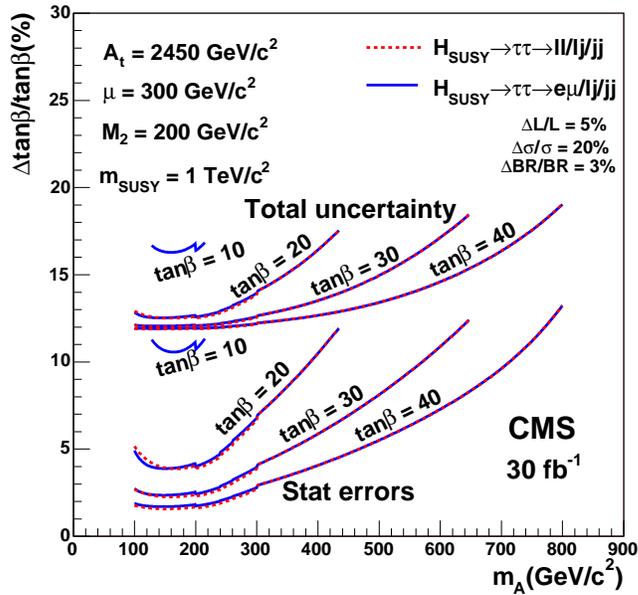,width=0.6\textwidth}}}
\end{center}
\caption{\it 
Relative uncertainty on the \tanbit\ measurement in the CMS experiment assuming an integrated 
luminosity of 30~\fbsit. The uncertainty is given for fixed values of \tanbit\ as a 
function of $m_A$ including statistical (lower curves) and statistical plus systematic 
uncertainties (upper curves)
(from Ref.~\protect\refcite{cms-tanb}).
} 
\label{f:tanb-cms}
\end{figure}

In the analysis, all combinations of tau decay channels
($\ell \ell$, $\ell   \tau_{had}$ and $\tau_{had} \tau_{had}$) 
have been used to extract the final measurement. 
In the parameter range close 
to the 5$\sigma$ discovery contour, where the signal rate is smallest, a statistical 
uncertainty of the order of 11--12\% for \tanb\ is found. 
In Fig.~\ref{f:tanb-cms} the statistical and
the combined statistical plus systematic uncertainty on \tanb\ is shown 
as a function of $m_A$ for \tanb\ = 10, 20, 30 and 40, assuming 
an integrated luminosity of 30\,\fbs. The total uncertainty ranges from 12 to 19\% 
depending on \tanb\ and $m_A$. 

As mentioned above, also the cross section measurement
will be used in a global fit together with other relevant measurements to determine the  
SUSY parameters simultaneously.

\subsubsection{Sensitivity to SUSY corrections via Higgs boson decay rates \label{s:susy-rad}}

It has been proposed to use the ratio of charged
Higgs boson decay rates, 
\mbox{$R = \rm{BR} (\hplus \to \tau \nu) / \rm{BR} (\hplus \to t b )$}, to discriminate  
between supersymmetric and non-supersymmetric 2-Higgs-doublet models.\cite{ketevi-R}
In the MSSM, Higgs boson couplings to down-type fermions receive large \tanb-dependent 
quantum corrections. Extensive theoretical analyses of one-loop corrections to both 
neutral and charged Higgs boson decay widths have been
performed.\cite{susy-corrections,susy-rad-corr} 
Depending on the parameters of the SUSY model, the value of $R$ might be changed 
significantly by supersymmetric radiative corrections.\cite{ketevi-R}
Using a simulation of the ATLAS performance it has been shown that $R$ 
can be measured in the process $g b \to H^\pm t$ (see Section \ref{s:mssm-hplus}). Assuming 
an integrated luminosity of 300\,\fbs, the accuracy on $R$ has been estimated to be
of the order of 12--14\% for \tanb\ = 50 and 300 $< m_{H^\pm} <$ 500~\Gcs.  
A full detector simulation and a study of systematic uncertainties on the 
background shape for the $\hplus \to tb$ channel still
need to be carried out. 

In the context of a global SUSY fit, a measurement of $R$
constitutes another important contribution from measurements in the Higgs sector.

\subsection{Search for an invisibly decaying Higgs boson}

Some extensions of the Standard Model predict Higgs bosons that decay
into stable neutral weakly interacting particles. 
In supersymmetric models, for example, 
Higgs bosons can decay, in some regions of the parameter space, with a 
large branching ratio into the lightest neutralinos or 
gravitinos.\cite{inv-susy,belanger}
In models with an enlarged symmetry breaking sector, Higgs bosons can decay into 
light weakly interacting scalars.\cite{inv-majoron,inv-majoron-2,inv-majoron-3} 
Invisible Higgs boson decays can also appear in models with large extra 
dimensions\cite{inv-extradim,inv-wells} or if massive neutrinos of a fourth generation 
exist.\cite{inv-neutrinos}

In a collider detector, such decays would lead
to invisible final states and triggering  
and detection would only be possible 
if the Higgs boson is produced in association with other particles.
Searches have been performed at the \epemit\ collider 
LEP in the $ZH$ associated production mode.\cite{LEP-inv-higgs} 
Since the full beam energy is absorbed in the collision, energy and momentum
conservation can be used to calculate the mass of the invisibly
decaying object (missing mass).
Since no evidence for an invisible Higgs boson has been found, a lower 
limit on its mass of 114.4~\Gcs\ has been placed, assuming Standard Model couplings in 
the production and a branching ratio into invisible final states of
100\% ($BR (H \to inv)$ = 1).\cite{LEP-inv-higgs}
Although a large fraction of the SUSY parameter space with
invisible Higgs decays is excluded in the constrained MSSM 
(given the mass limits on the lightest neutralino from LEP experiments),  invisible decays
could be enhanced if gaugino mass unification is abandoned.\cite{belanger}

At the LHC, the search for invisibly decaying Higgs bosons is much more difficult, since 
the missing mass technique cannot be applied. 
Higgs bosons can be searched for in the associated production modes with 
vector bosons ($WH, ZH$), \ttbar-pairs ($\ttbar H$) and jets (vector boson fusion mode $qq H$).
All three processes have already been suggested in the literature 
and feasibility studies have been performed.\cite{inv-roy,inv-gunion,inv-zeppenfeld} 
In a very recent paper, the discovery potential in the $ZH$ and $qqH$ channels has been re-assessed 
and the associated production of a Higgs boson with a high-\PT\ jet has been 
considered in addition.\cite{inv-han}
In all cases missing transverse energy is used as a key
signature for the presence of an invisible Higgs boson decay.

\subsubsection{Search in $WH$ and $ZH$ associated production}
In the associated production with a vector boson, the $ZH$ process with $Z \to \ell \ell$ 
has been identified to be the most promising one, since the search in the $WH$ production 
mode is plagued with large backgrounds from inclusive $W$ production.\cite{inv-roy} 
The dominant backgrounds to the $ZH$ process come from vector boson pair production, 
{\em i.e.}, $ZZ \to \ell \ell \ \nu \nu$, $WZ \to \ell \nu \ \ell \ell$ (where the lepton 
from the $W$ decay is not identified or is outside of the detector acceptance) and $WW \to \ell \nu \ \ell \nu$ 
production. After basic lepton requirements, the missing transverse energy spectrum is still 
dominated by the background processes, however, the signal-to-background ratio improves 
with increasing \met. Requiring that \met\ is larger than 75 GeV leads to a 
signal-to-background ratio of 0.25.\cite{inv-han} It is claimed that an invisibly 
decaying Higgs boson could be discovered with a 5$\sigma$ significance with an 
integrated luminosity of 30\,\fbs\ for Higgs boson masses up to $\sim$160~\Gcs, assuming 
a Standard Model-like $hZZ$ coupling and $BR (H \to inv)$ = 1.  
The simulation of this process is currently being performed by the experimental 
collaborations. 

Since evidence for a signal can only be claimed 
from an excess of events above the Standard Model backgrounds, their uncertainty needs
to be taken into account in the determination of the signal significance (see also discussion
at the end of Section~\ref{s:inv-ttH}).

\subsubsection{Search in $\ttbar H$ associated production \label{s:inv-ttH}}
In the search for an invisibly decaying Higgs boson in the associated $\ttbar H$ production, 
one of the top quarks is required to decay leptonically. This 
ensures that the events can be reliably triggered using the standard lepton triggers. The dominant
backgrounds have been identified to be \ttbar\ production and the irreducible $\ttbar Z$ production, with 
$Z \to \nu \nu$.\cite{inv-gunion} Due to the large production 
cross section at the LHC, the \ttbar\ background contributes significantly. 
In the signal selection the hadronically decaying top quark is reconstructed and two b-tags 
are required. In addition, the transverse mass of the ($\ell
- \pet$)-system is required to be large,  
typically larger than 150~\Gcs. This requirement rejects a large fraction of the  
\ttbar\ background for which the transverse mass is expected to show a peak at the 
W mass.\cite{inv-gunion} 
The residual \ttbar\ background results mainly from double leptonic decays
of the \ttbar\ system, where the second lepton ($e, \mu, \tau)$ is not identified or is outside 
the detector acceptance. The $\tau$ contribution in the residual \ttbar\ background sample, to which 
both leptonic and hadronic tau decays contribute,
amounts to about 70\%.\cite{inv-richter}
As in the $ZH$ search, the \met\ distribution is used to
establish evidence for an  
invisibly decaying Higgs boson. Requiring $\met >$ 150~GeV, a signal-to-background ratio
varying between 0.39 (for $m_H$ = 120~\Gcs) and 0.09 (for $m_H$ = 200~\Gcs) has 
been found.\cite{inv-richter} Assuming an integrated luminosity of 30\,\fbs, about 45 signal events 
are expected for $m_H$ = 120~\Gcs\ with a \met\ cut at 150~GeV. 
Based on these studies, it has been
concluded that a detection of an invisibly decaying Higgs boson  
in the associated \ttbar\ production
should be possible for Higgs boson masses up to $\sim$200~\Gcs,
assuming an integrated luminosity of 100--200\,\fbs. 
For an integrated luminosity of 30\,\fbs, the reach is expected 
to be comparable to that of the $ZH$ channel.

 However, it must be stressed that in both the 
$ZH$ and the $\ttbar H$ search no signal peak is observable. A signal is claimed from an 
excess of events above the expectations from Standard Model processes. No systematic 
uncertainties on the knowledge of these backgrounds in the extreme phase space region 
with large \met\ have been considered. In addition, detector effects might lead to 
non-Gaussian tails in the \met\ distribution such that
the backgrounds might be larger than   
anticipated so far. The experimental collaborations are studying these channels and 
addressing in particular the question of the uncertainty on the backgrounds and on the 
\met\ measurement.

\subsubsection{Search in vector boson fusion}
More recently, it has been proposed to use the vector boson fusion process 
$pp \to qq H \to qq + inv$  to search for invisibly decaying Higgs bosons.\cite{inv-zeppenfeld} 
In contrast to the two processes discussed above, the production cross section is larger, 
however, triggering these events is more difficult. 

\begin{figure}
\begin{center}
\mbox{{\epsfig{file=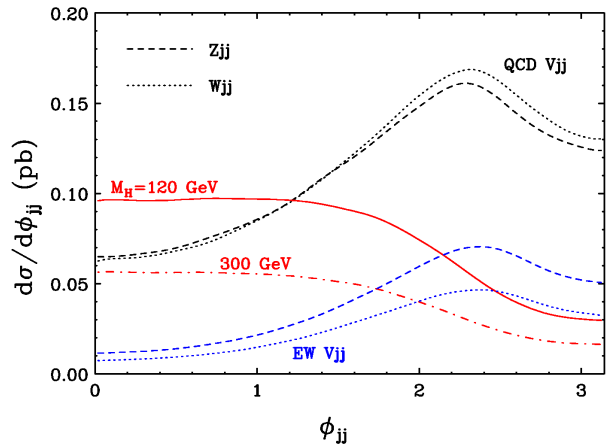,width=0.6\textwidth}}}
\end{center}
\caption{\it 
Distribution of the azimuthal angle separation $( \phi_{jj} )$ between the two tagging jets for the 
$Wjj$ and $Zjj$ backgrounds from QCD and electroweak production
and Higgs boson signals with  $m_H$ = 120 and 300~\Gcsit.
Results are shown after applying the jet tag requirements and the cut $\met >$ 100\,GeV
(from Ref.~\protect\refcite{inv-zeppenfeld}).
} 
\label{f:inv-delphi-jj}
\end{figure}

The experimental signature consists of two high-\PT\ forward jets with a large rapidity 
separation and a high dijet invariant mass, accompanied by large missing transverse 
energy.
In the analysis of Ref.~\refcite{inv-zeppenfeld}, the forward jets are selected using similar 
criteria as in the vector boson fusion analyses discussed in Section \ref{s:lhc-vbf}. 
The \PT\ thresholds have been raised to 40~\Gc\ and the cut on
the invariant mass of the two jets to 
1200~\Gcs. The missing transverse energy is required to be larger than 100~GeV. 
The Standard Model backgrounds are dominated by the $Zjj$, with $Z \to \nu \nu$, and the $Wjj$, with 
$W \to \ell \nu$, processes. 
In order to reach a better discrimination between signal and the residual background, 
an additional cut on the azimuthal separation $\Delta \phi_{jj} <$ 1 between the two 
tag jets has been suggested.\cite{inv-zeppenfeld} The coupling
structure of the Higgs boson to vector bosons favours Higgs boson emission opposite in 
azimuth to both tagging jets, which leads to small values of $\Delta \phi_{jj}$.
The expected $\Delta \phi_{jj}$ distributions for 
signal and background processes are shown in Fig.~\ref{f:inv-delphi-jj}.

In Ref.~\refcite{inv-zeppenfeld} it has been proposed to use this 
distribution to constrain the $Wjj$ and $Zjj$ backgrounds. At the LHC, sizeable 
samples of $Wjj$ and $Zjj$ samples with identified leptonic vector boson decays will be 
available. Applying the same jet tag
requirements will allow to normalize the backgrounds 
in the region $\Delta \phi_{jj} >$ 1 and to predict the background in the signal region 
($\Delta \phi_{jj} <$1).\cite{inv-zeppenfeld}
Taking into account the normalization uncertainty, it has been concluded
that with data corresponding to integrated luminosities of  
10\,\fbs\ (100\,\fbs) an invisibly decaying Higgs boson with $BR ( H \to inv)$ = 1
can be detected with a 5$\sigma$ significance for masses up to
480~\Gcs\ (770~\Gcs).

These promising results motivated a study using a more detailed simulation 
of the ATLAS detector.\cite{inv-neukermans}
Similar cuts as used in the parton level analysis have been applied. 
In addition to the tag-jet requirements, jet veto and lepton veto cuts are applied. Events are rejected if 
jets with $\PT >$20~\Gc\ in the rapidity region between the tag jets 
or identified leptons are found. 
After applying all cuts, a signal cross section of the order of 60 fb is expected for 
an invisibly decaying Higgs boson with $m_H$ = 130~\Gcs\ and $BR ( H \to inv)$ = 1.
For both the $Zjj$ and $Wjj$ backgrounds, cross sections of the order of 120 fb are expected. 
In comparison to these backgrounds, the contribution from QCD jet production with $\met >$ 
100~GeV is expected to be small. 
Performing the background normalization as described above, 
it is expected that the $Zjj$ background in the signal region
can be predicted with a total uncertainty  
of 6\% (4\%) with data corresponding to an integrated luminosity of 10\,\fbs\ 
(30\,\fbs).\cite{inv-neukermans} Similarly, the $Wjj$ background
can be predicted with an accuracy of $\pm$3\% (2.5\%) for 10\,\fbs\ (30\,\fbs). 

\begin{figure}
\begin{center}
\mbox{{\epsfig{file=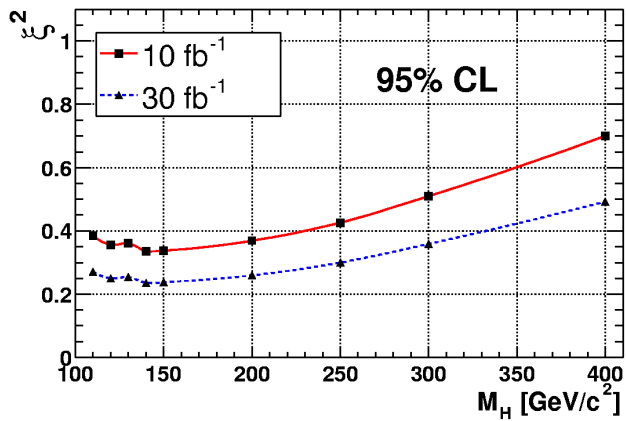,width=0.6\textwidth}}}
\end{center}
\caption{\it 
$\xi^2$ values (see text) which can be excluded with a confidence level of 95\% for integrated 
luminosities of 10 and 30\,\fbsit\ as a function of the Higgs boson mass 
(from Ref.~\protect\refcite{inv-neukermans}). 
} 
\label{f:inv-xi}
\end{figure}

To quantify the ATLAS potential for discovery of
an invisibly decaying Higgs boson,  the variable 
$ \xi^2 \ = \ BR (H \to inv) \cdot \sigma_{qq \to qqH} / (\sigma_{qq \to qqH})_{SM} $
has been introduced.\cite{inv-neukermans}
The second term accounts for a possible suppression of the production 
cross section as compared to the Standard Model value. 
In Fig.~\ref{f:inv-xi} the $\xi^2$ values which can be excluded with a confidence level of 
95\% are shown for various integrated luminosities as a function of the Higgs boson 
mass. For an integrated 
luminosity of 10~\fbs, it seems possible to probe $\xi^2$ values down to 35\% for a light 
Higgs boson and down to 70\% for a Higgs boson with $m_H$ = 400~\Gcs. 
This confirms the claim of Ref.~\refcite{inv-zeppenfeld} that the vector boson fusion mode 
extends considerably the sensitivity of the search for an invisibly decaying
Higgs boson at the LHC. 

As stressed in Ref.~\refcite{inv-neukermans}, these very encouraging results have been obtained with a 
fast detector simulation. 
Detector effects leading to non-Gaussian tails 
in the \met\ distribution cannot be estimated reliably using this simulation.
Therefore, the contributions from 
QCD jet production might be underestimated and a more reliable estimate is needed.

The largest disadvantage of the detection of invisible Higgs boson decays in the vector boson 
fusion mode is the challenge to trigger on these events. 
An invisibly decaying Higgs boson produced via the 
fusion of vector bosons 
must be triggered exploiting the two jet plus \met\ signature. Since the QCD 
jet production rate is huge at the LHC, high thresholds will have to be used for the 
jet triggers.\cite{atlas-triggers}
 At present, studies are ongoing to evaluate the design and the rates of 
a possible (2-jet + \met)-trigger.\cite{inv-leshouches} 

\subsubsection{Summary}
The analyses presented show that the experiments at the LHC have the 
potential to detect invisibly decaying Higgs bosons. Different 
production modes can be exploited which is important to understand whether 
a possible excess of events with large missing 
transverse energy originates from Higgs boson production.
In all cases evidence for a signal will be extracted from an excess of 
events with large \met\ above the background.  For a 
reliable measurement a normalization of the backgrounds in the 
experiment will be necessary. In the associated $ZH$ and $\ttbar H$ production 
modes the expected signal cross sections are small, in the range of 
a few fb, and the normalization might be affected by large uncertainties.  
However, those channels are nevertheless of interest since they can be reliably triggered, 
and in addition, the $\ttbar H$ channel does not rely on the vector boson coupling, which  
might be suppressed in certain scenarios. 
It should also be mentioned that all studies performed only consider backgrounds 
from Standard Model processes. The impact of non-Standard Model backgrounds, 
which might be present if Higgs bosons decay invisibly, remains to be studied.

\section{Conclusions}

It has been demonstrated in numerous experimental studies that the 
experiments at the Tevatron \ppbar\ collider and at the CERN Large 
Hadron Collider have a huge discovery potential for the Standard 
Model Higgs boson. If the Higgs mechanism is 
realized in nature the corresponding Higgs boson 
should not escape detection at the LHC. The full mass range can 
be explored and the Higgs boson can be detected in several decay 
modes. In addition, the parameters
of the resonance can be measured with adequate precision to 
establish Higgs-boson like couplings to bosons and heavy fermions. 
For both the discovery and the parameter measurements the 
recently studied vector boson fusion mode plays an important role. 
If the Standard Model Higgs boson is light enough, the experiments 
at the Tevatron should already be able to observe first evidence of a signal.
The significance level that can be reached at the Tevatron will depend 
strongly on how much luminosity can be accumulated before the LHC startup. 

For the Minimal Supersymmetric Standard Model, Higgs bosons can be
detected across the entire parameter space.
The Tevatron experiments will be able to exclude the maximal mixing
scenarios at 95\% confidence level if no signal is present. 
At the LHC, MSSM higgs bosons can be discovered with a significance of 
more than 5$\sigma$ for established benchmark scenarios.

\subsection*{Acknowledgements} 
The authors would like to thank the numerous colleagues from both the
theoretical and experimental community for providing such an impressive
wealth of results to review.
In particular, gratitude is expressed to
Gregorio Bernardi, Klaus Desch, Beate Heinemann, Filip Moortgat, Sacha Nikitenko,
Giacomo Polesello, Michael Spira, Guillaume Unal and Dieter Zeppenfeld for 
very useful discussions and comments. KJ wishes to thank his colleagues from the 
ATLAS Higgs working group for the excellent collaboration over many years. Particular thanks goes 
to Donatella Cavalli, Daniel Froidevaux, Fabiola Gianotti 
and Elzbieta Richter-Was for their contributions and stimulating discussions.

\end{document}